# Arquitectura de Rutas de Vuelo para Drones

**Trabajo de proyecto de grado presentado al departamento de Ingeniería de Sistemas y Computación**

**Universidad de los Andes, Colombia**

Presentado por:
**Andrés Forero Osorio**
**Carlos Andrés Torres Echeverría**

Asesorados por:
**Mario Fernando De la Rosa Rosero**
**Nelson Andrés Sánchez Otálora**

# Tabla de contenido

# Índice de Figuras

## Resumen

En este proyecto se partió de una arquitectura preexistente la cual facilita la planificación y ejecución automática de rutas para dron en un espacio conocido mediante un ambiente 3D de realidad virtual. Nuestro trabajo consistió en extender dicha arquitectura mediante la integración de un nuevo componente web, haciendo uso de un API de mapas 3D, para facilitar a las personas que no tienen acceso al hardware de realidad virtual la posibilidad de planear rutas de vuelo que tienen como parámetros la <latitud, longitud> con respecto al globo terráqueo y también un componente en metros que representa la altura a la que se eleva el dron en un determinado punto. Adicionalmente se ampliaron las posibilidades de configuración de una ruta con el fin sacar partido a uno de los componentes más valiosos de este tipo de aeronaves no tripuladas: la cámara y su potencial de uso en múltiples contextos y escenarios. La extensión de esta solución permite al usuario planificar diversas tareas de cámara a lo largo de la ruta, ver en tiempo real lo que la cámara está captando y posterior al vuelo recuperar el contenido multimedia que se creó.

## Abstract

This project was built from a pre-existing architecture that facilitates the planning and automatic execution of drone routes in a known space through a 3D virtual reality environment. Our work consisted in extending this architecture by integrating a new web component, making use of a 3D map API, to facilitate to people who do not have access to virtual reality hardware, the possibility of planning flight routes that have as parameters the <latitude, longitude> with respect to the globe and also a component in meters that represents the height at which the drone rises in a certain point. Additionally, the configuration possibilities of a route were extended in order to take advantage of one of the components that gives more value and potential to unmanned aircrafts: the use of the camera in multiple contexts and scenarios. The extension of this solution allows the user to assign different camera tasks along the route, see in real time what the camera is capturing and, after the flight, retrieve the multimedia content that was created.

# 1. Introducción

El uso de aeronaves no tripuladas controladas remotamente ha venido en aumento en los últimos años. Esta tecnología, que antes era costosa y exclusiva de operaciones militares o industrias especializadas, ahora está al alcance de cualquier persona. Dicha persona podría estar interesada en darle todo tipo de usos a esta aeronave, desde la simple recreación hasta la filmación de video profesional, vigilancia, arqueología, agricultura, cartografía etc. Solo por mencionar algunos casos en la interminable lista de contextos en los que estas aeronaves pueden ser de utilidad. Paralelamente, en un campo de investigación tan amplio como lo es la robótica, se han venido haciendo avances en un área de estudio conocida como interacción humano-robot, que como su nombre lo indica busca alternativas para facilitar la forma en la que un ser humano puede controlar total o parcialmente las funciones de un robot, ya sea en tiempo real o no. En el contexto anteriormente descrito, se ha encontrado la necesidad y la oportunidad de mejorar la experiencia de controlar un dron para que usuario sin entrenamiento previo pueda maniobrar con facilidad este tipo de aeronaves y sacar provecho de esta tecnología.

Desde el grupo de investigación IMAGINE de la Universidad de los Andes, se ha venido trabajando en una arquitectura que soporta la planeación de rutas 3D para drones con integración de ambientes de realidad virtual. Esta misma solución utiliza un aplicativo móvil donde se establece la comunicación con la aeronave y se ejecuta una ruta preestablecida, se hacen ajustes a esta y se visualiza el estado del recorrido en un mapa 2D. Con base en este trabajo anterior, adelantado por Nelson Sánchez y Santiago Munera, este proyecto busca extender las funcionalidades de esta arquitectura al integrar un nuevo componente que soporte la creación y simulación de rutas aun cuando no se cuente con el hardware requerido para interactuar con el escenario virtual o cuando el escenario de interés no ha sido recreado en un ambiente 3D. Adicionalmente se busca agregar más posibilidades de configuración y planeación de tareas asociadas con la cámara, las cuales pueden ser usadas como base para la virtualización de recorrido de espacios o para el mapeo de estructuras arquitectónicas. Finalmente se pretende usar esta extensión de la solución en un escenario o situación real donde para pueda ser probada y evaluada.

Este documento se encuentra organizado de la siguiente forma: En el capítulo 2 se presenta un recuento de los principales trabajos relacionados con la temática de uso de drones en diferentes contextos. Posteriormente, en el capítulo 3, se presentan las motivaciones de este proyecto, y en el capítulo 4 se presentan los objetivos propuestos que guían el desarrollo. En el capítulo 5 se hace la propuesta de diseño, incluyendo requisitos funcionales y exploración de tecnologías. Los capítulos 6 y 7 exponen, respectivamente, la arquitectura de solución propuesta y el desarrollo de esta solución. En el capítulo 8 se presenta el diseño y los resultados de las pruebas que validan la propuesta de solución, y en el capítulo 9 se presentan las conclusiones del proyecto y el trabajo futuro propuesto. Finalmente, en el capítulo 10 se recopilan la bibliografía empleada en el desarrollo del proyecto y de este documento.

## 2. Estado del arte y trabajos relacionados

Investigadores de todas partes del mundo han estado haciendo avances en el desarrollo de soluciones basadas en naves no tripuladas para soportar tareas de toda índole, implementando diferentes tecnologías. Estos trabajos se resumen a continuación:

### 2.1 Survey on path and view planning for UAVs [3] (2020)

Este trabajo hace una recopilación de sobre los avances más recientes en cuanto a planeación de rutas de dron. Estos avances se agrupan de acuerdo a las 3 actividades más populares que soportan los UAVs: Reconstrucción de escenas, exploración de entornos y cinematografía aérea. Lo destacable de este trabajo es que logra condensar una gran variedad de fuentes actuales sobre varias áreas de interés centradas en la operación de drones. Es de particular interés, puesto que sirven de referencia para el desarrollo que se haga en este proyecto.

### 2.2 Drones como apoyo a la ganadería a gran escala [4] (2017)

En contextos donde se tienen aplicaciones reales a este tipo de tecnologías se tiene este trabajo en el que se integra el uso de drones, sistemas de sensores IoT, para automatizar tareas relacionadas con la ganadería. Esta solución se compone de un sensor de proximidad, una Raspberry Pi (ordenador de placa reducida), servidores MTC (middleware para comunicación máquina a máquina (M2M)) y un dron programable junto con su controlador. La arquitectura propuesta por esta solución puede apreciarse a continuación:

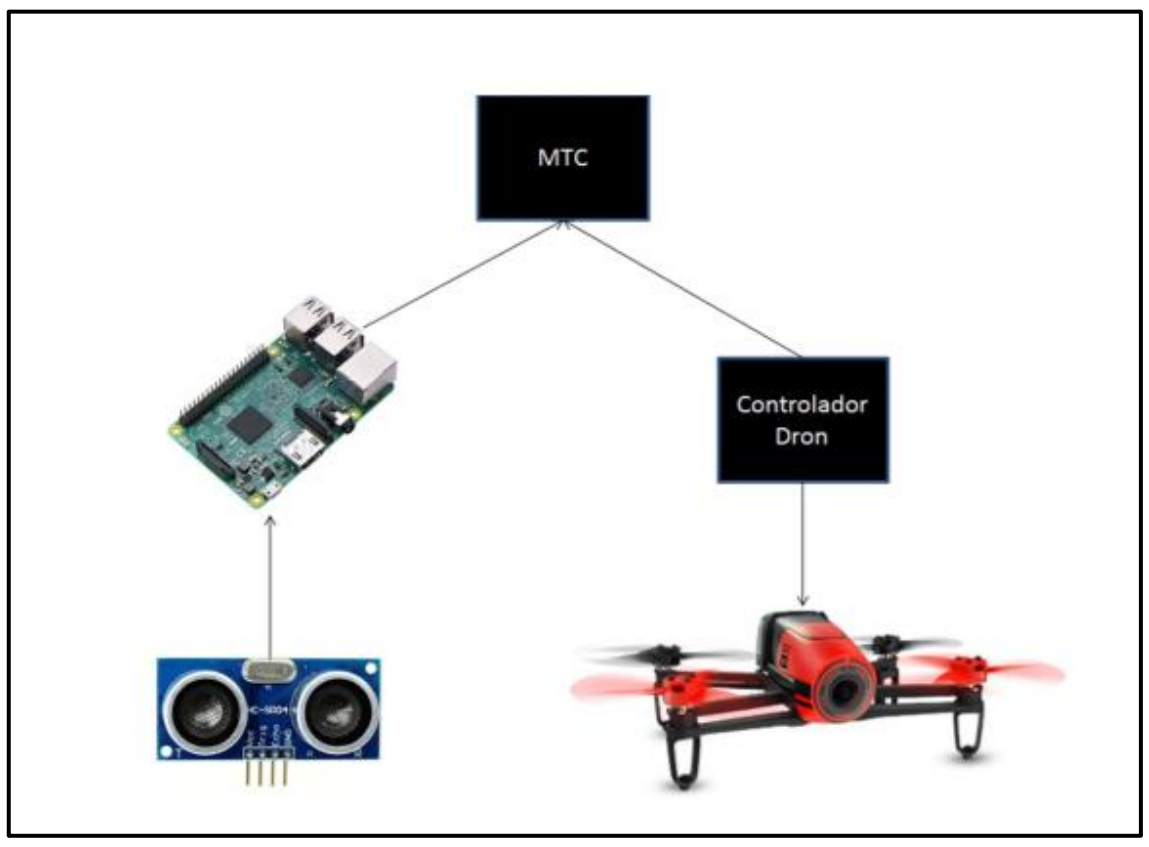

*Figura 1. Arquitectura de solución implementada en [4]*

El funcionamiento de esta solución se consigue enlazando los componentes a un servidor MTC, el cual recibe los datos cuando se detecta movimiento por el sensor y los permite ser descargados por el controlador del dron. Este envía las coordenadas del sensor al dron para que realice su vuelo y grabe el recorrido. Lo destacable de esta solución es la integración de sensores de movimiento con la toma de video de dron mediante vuelos autónomos sin GPS, lo cual es aplicable en el ámbito de la vigilancia.

### 2.3 Arquitectura de operación de drones apoyada en integración de componentes heterogéneas de planificación, control e interacción [5] (2020)

En este trabajo se busca crear una arquitectura que facilite la comunicación entre herramientas de realidad virtual, simuladores y frameworks de análisis de imagen para enriquecer la experiencia en el uso de drones. La arquitectura implementada se puede apreciar a continuación:

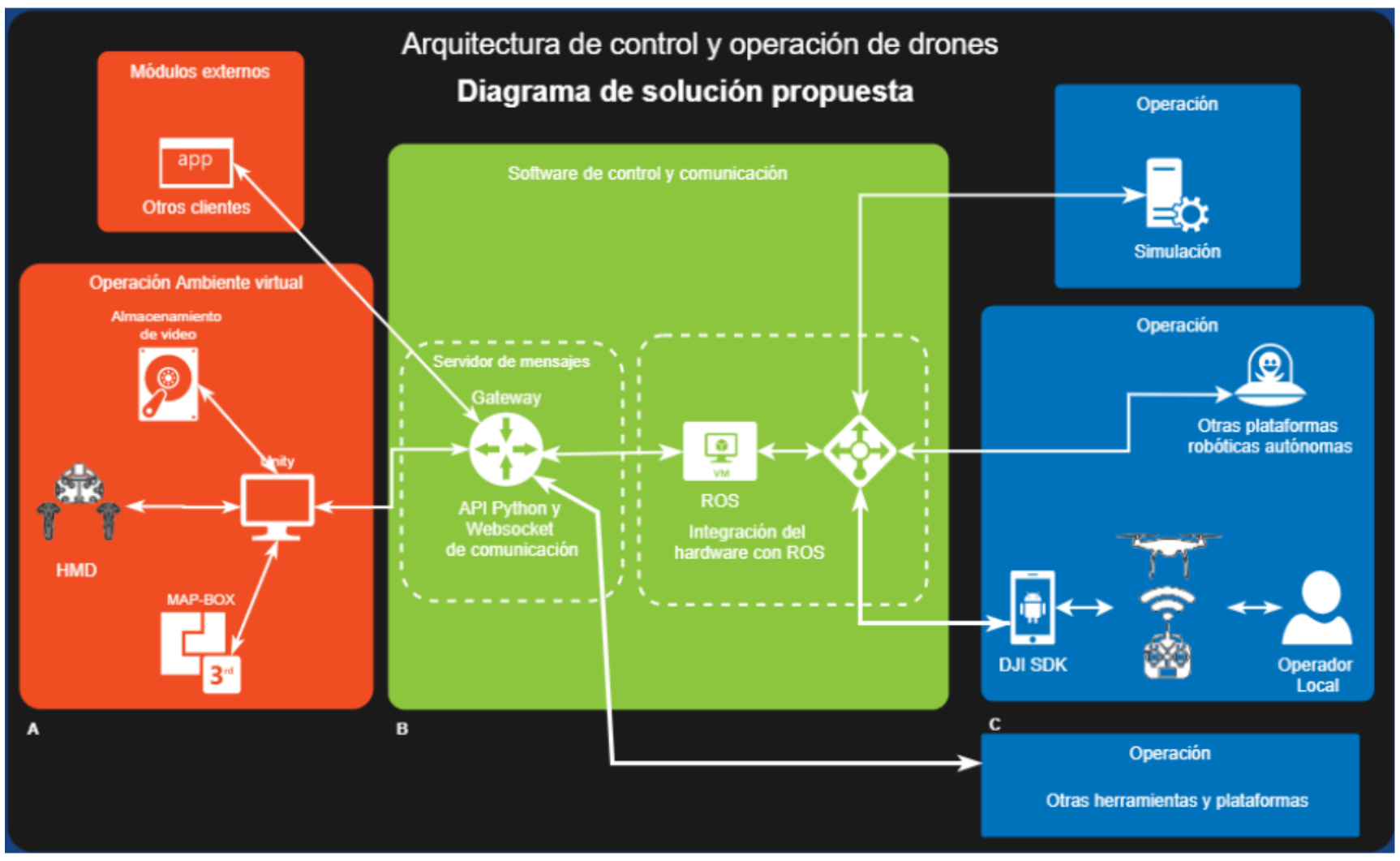

*Figura 2. Arquitectura heterogénea implementada en [5]*

Esta arquitectura se basa en el middleware ROS (Robot Operative System) para la comunicación entre los diferentes componentes. Se integra la aplicación de Realidad Virtual para planificación de rutas de vuelo y la aplicación móvil Android para el seguimiento de rutas desarrolladas por [1] y [2]. Así mismo, se introducen módulos de análisis de imágenes y componentes de simulación. El resultado destacable de esta solución es la escalabilidad de la arquitectura, puesto que facilita la integración de nuevas y diferentes tecnologías.

### 2.4 Control de Robots Aéreos en entorno Matlab-ROS utilizando el simulador V-REP [6] (2018)

En este trabajo también se hace una exploración en la integración de diferentes tecnologías para la automatización y control de drones, usando entornos y frameworks como Matlab/Simulink para visión, ROS para control y navegación, y V-REP para simulación. Lo destacable de este trabajo es el intento de aplicabilidad de hacer seguimiento de un vehículo por medio de un dron. Así mismo, las conclusiones a las que se llega son de particular interés, pues se evidencia la dificultad de integrar los diferentes componentes. Esto es de utilidad para identificar componentes compatibles entre sí para una futura arquitectura.

## 2.5 Realtime Aerial Image Panoramic by Using Drone [7] (2019)

En este trabajo se hace una implementación de planeación de rutas y toma de imágenes panorámicas aéreas. Este desarrollo involucra un dron armado con componentes específicos para la tarea. El circuito cuenta principalmente con: ArduPilot APM 2.8, como controlador de vuelo; Raspberry Pi, como controlador de la cámara; y Ublox Neo, como GPS. La arquitectura de solución propuesta en este trabajo se aprecia a continuación:

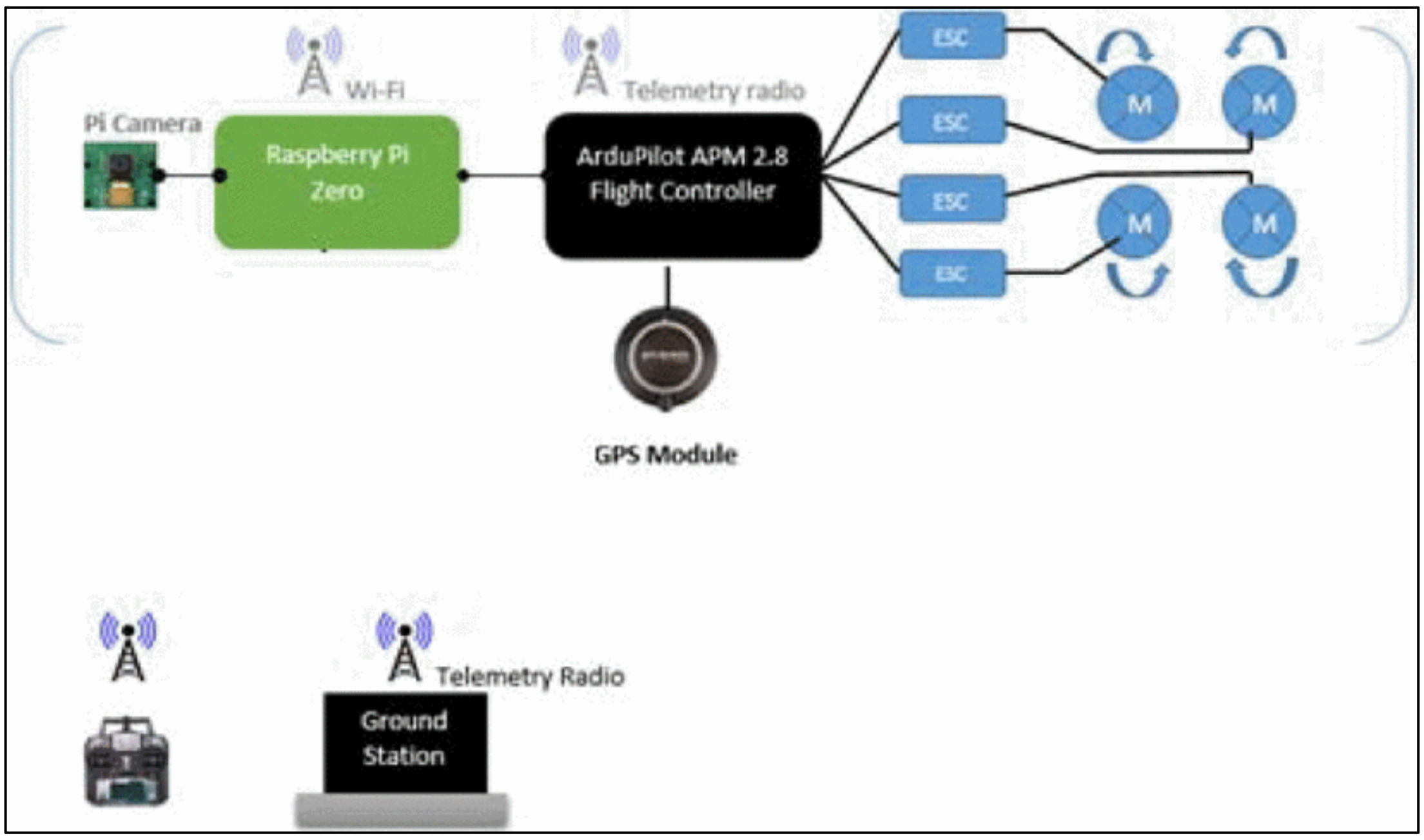

*Figura 3. Arquitectura heterogénea implementada en [7].*

El objetivo de este trabajo es que el dron ejecute de manera autónoma la ruta planeada y el usuario pueda elegir en que puntos capturar la panorámica. El resultado destacable es que se logra implementar la funcionalidad de captura de imágenes panorámicas de forma autónoma. Sin embargo, cabe resaltar que el dron fue construido específicamente para este propósito, por lo que es de interés obtener estos resultados con drones comerciales ya construidos.

## 2.6 System Design of an Open-Source Cloud-Based Framework for Internet of Drones Application [8] (2019)

En este trabajo se explora el desarrollo de una arquitectura de vuelos de drones basada en computación en la nube. Esta arquitectura busca facilitar las tareas asociadas a vuelos de drones con acceso a internet (Internet of Drones), que se ven restringidas por los recursos energéticos de la aeronave y su baja capacidad de cómputo. Esta arquitectura se puede apreciar a continuación:

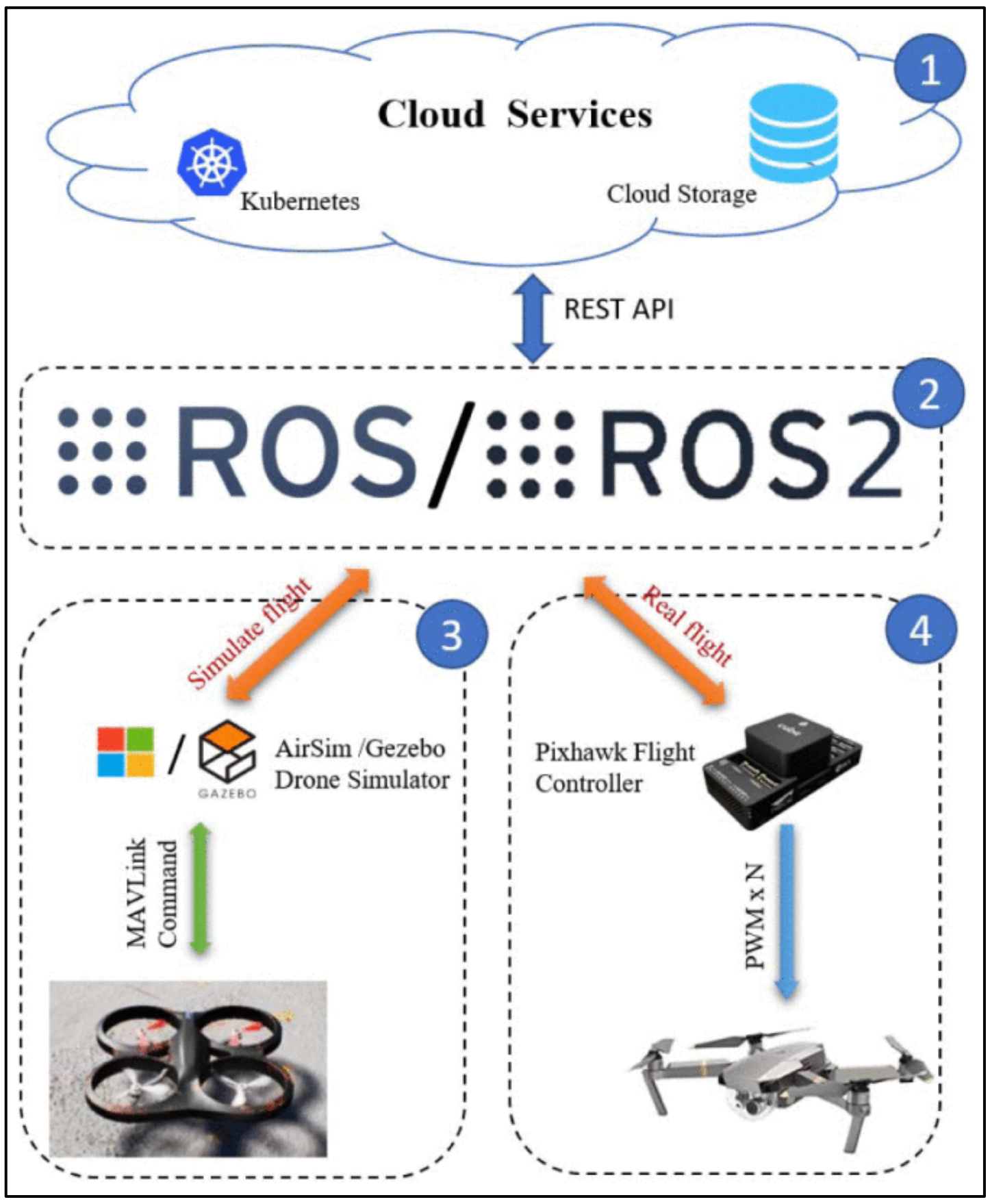

*Figura 4. Arquitectura de tres capas para la solución de [8]*

Esta propuesta consta de tres capas que se comunican entre sí para soportar las tareas de vuelo: Una capa de servicios en la nube, una capa de comunicación con ROS1/ROS2, y una capa con simuladores y herramientas de control de vuelo. Los resultados destacables de esta solución es el despliegue de los servicios en la nube que pueden extender las actividades que los drones pueden realizar en sus vuelos, sin necesidad de sobreexplotar las capacidades físicas de las aeronaves.

## 2.7 Exploring the performance of ROS2 [9] (2016)

Este trabajo no está centrado específicamente en drones, sino en la evaluación de desempeño de ROS2, usando como marco de referencia a ROS1. En este proyecto se revisan los puntos débiles de ROS1 y se explora cómo ROS2 los refuerza. Una de estas debilidades es la falta de aplicabilidad de ROS1 en sistemas de tiempo real, debido a que no logra satisfacer requerimientos de tiempo real. Para esto, ROS2 hace uso de DDS (Data Distribution Service), el cual facilita el transporte de información y la comunicación entre componentes, además de ser más escalable. Otra debilidad de ROS1 es el corto rango de sistemas operativos en los que este se puede ejecutar. Para resolver esto, ROS2 permite su ejecución en sistemas operativos Windows, macOS y Ubuntu Linux. Finalmente, se evalúa el desempeño de ambos middlewares en diferentes ambientes y con diferentes configuraciones de DDS. Lo destacable de este trabajo es el análisis que se hace sobre las diferentes configuraciones de DDS y cómo afectan el desempeño de ROS2, además de la comparación entre ROS1 y ROS2.

## 3. Motivaciones

Teniendo como plataforma el trabajo de Nelson Sánchez [1] y Santiago Munera [2] se puede ver que se logró su objetivo principal el cual es soportar y facilitar a cualquier tipo de usuarios, incluido aquellos que no tienen experiencia en el pilotaje de drones, la creación de una ruta en un espacio 3D en un ambiente de realidad virtual para su posterior comunicación, ejecución y seguimiento del plan de vuelo mediante el uso de un dispositivo Android.

Para el proyecto de Sánchez y Munera se trabajó con el dron DJI Mavic Pro 1 (Fig. 5), de la casa DJI, la cual es líder en el mercado y se caracteriza por integración de alta tecnología en sus productos, así como un conjunto de librerías y kits de desarrollo en diferentes plataformas para construir aplicaciones para el control de vuelo de drones DJI.

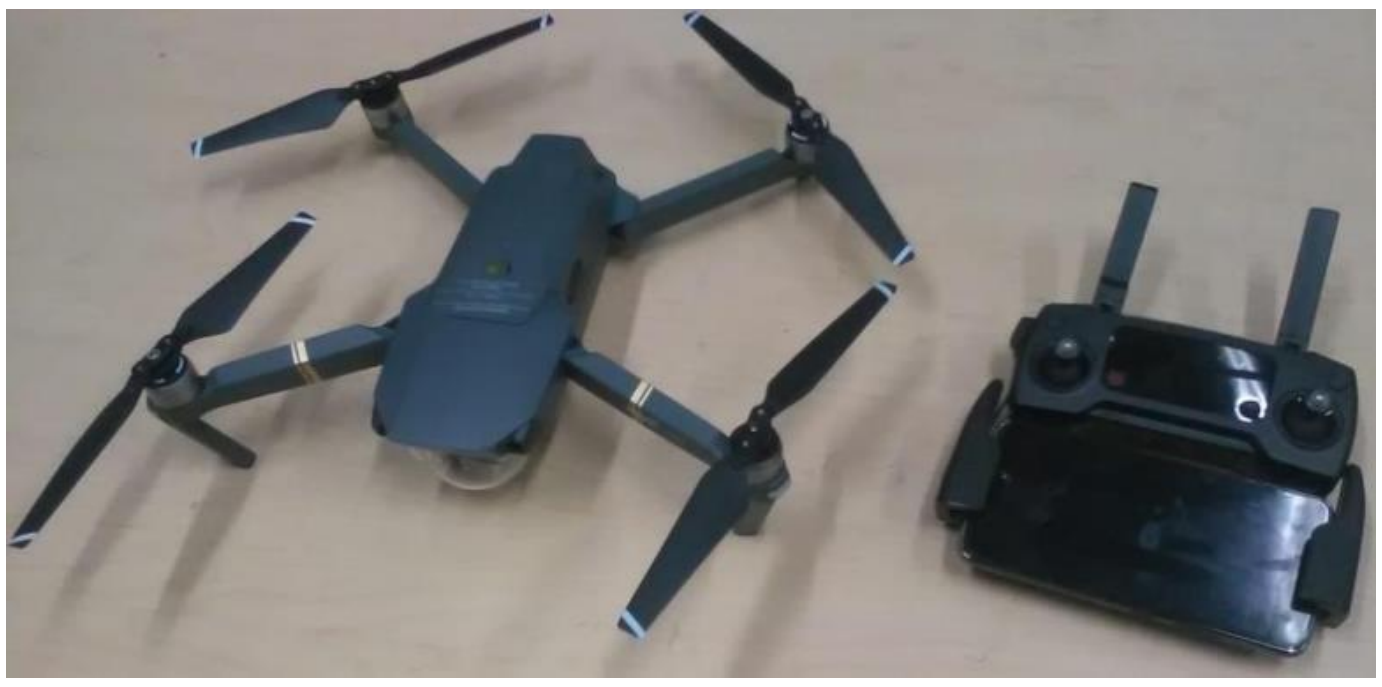

*Figura 5. Dron mavic Pro del laboratorio Colivri. tomado de [1]*

Cabe mencionar que para este modelo solo es posible usar el SDK (software development kit) para dispositivos móviles, por lo que se ha venido trabajando con una arquitectura que necesariamente utilice un dispositivo Android versión 5.0 (API Level 21) o superior para que sirva como mediador para enviar las instrucciones al dron, además de ser la interfaz de configuración adicional previo a ejecutar un plan de vuelo.

En la figura 6 se puede ver el esquema de comunicación de la arquitectura solución de Sánchez la cual es la base para el actual proyecto. La comunicación entre componentes consiste en transmitir una serie de rutas en formato JSON (JavaScript Object Notation), que fueron definidas en un ambiente de realidad virtual, a una base de datos no relacional de tiempo real en la plataforma de Firebase, cuyos datos serán consumidos por la aplicación móvil.

Dicha aplicación móvil, estará encargada de conectarse a la aeronave y convertirá las rutas del formato origen en datos que las librerías de control del SDK puedan usar como parámetro para automatizar instrucciones al dron.

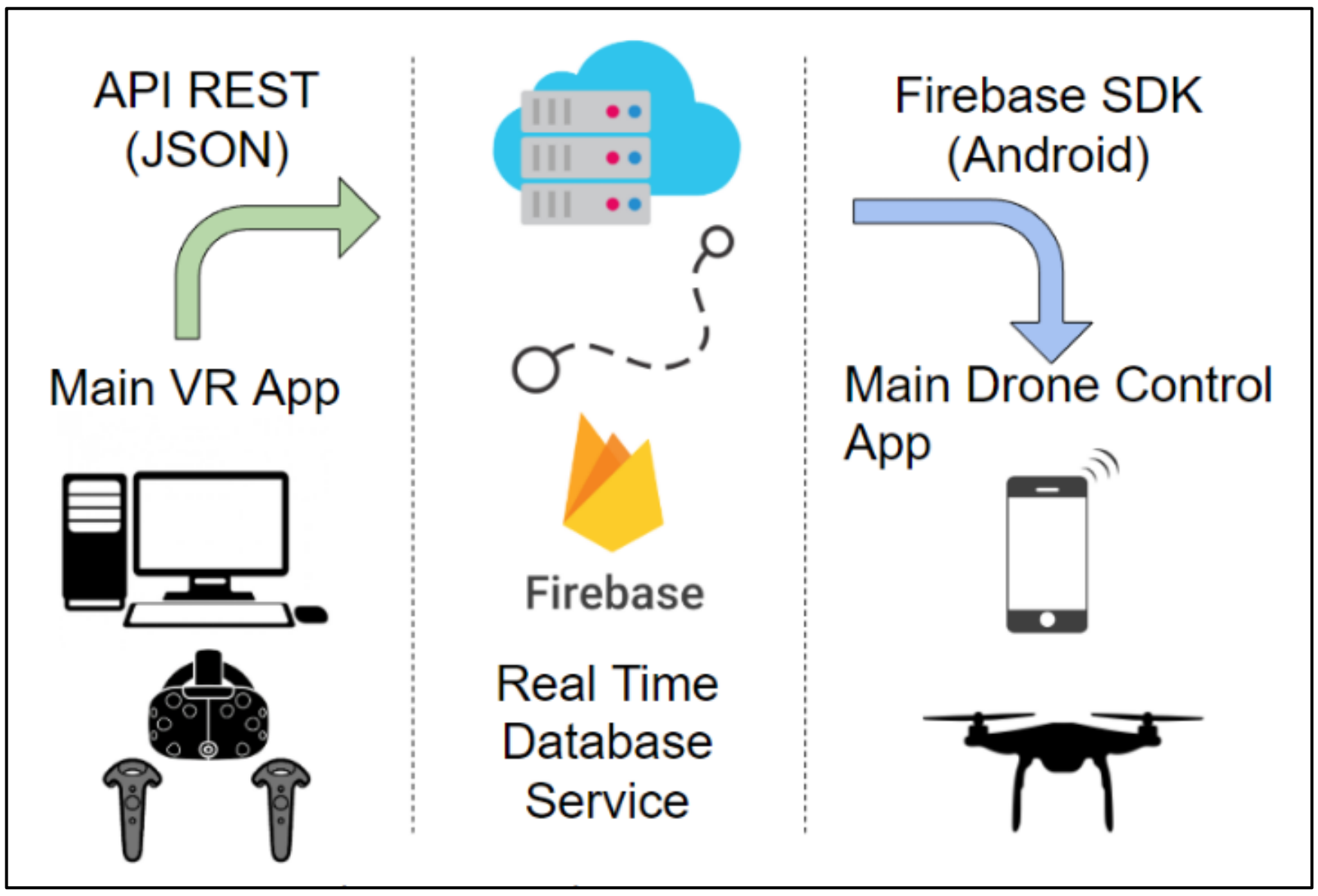

*Figura 6. Esquema de comunicación de una solución para la construcción y el seguimiento de rutas tridimensionales para drones [1]*

A pesar de que los trabajos anteriores reportaron un desempeño satisfactorio, y con el fin de poder extender la solución, identificamos varios puntos de mejora a explorar los cuales son la motivación del proyecto actual.

Tanto en el trabajo de Sánchez como en el de Munera las pruebas de usabilidad y precisión se realizaron de forma exitosa en entornos reales con usuarios potenciales, sin embargo, no se profundizo en el uso de esta herramienta en aplicaciones específicas y no se consideran funcionalidades adicionales asociadas a una tarea particular una vez se define un plan de vuelo.

En otras palabras, una vez un usuario diseña una ruta él o ella podría estar interesado en **automatizar** una gran variedad de tareas ahora que sabe que puede programar un recorrido y hacer seguimiento de este, desde la filmación de escenas cinemáticas de un evento o paisaje, la exploración exterior de una edificación en construcción o la vigilancia y monitoreo de grandes extensiones de terreno. El uso de una herramienta de este tipo en los contextos antes mencionados requiere no solo la capacidad de fácil planeación sino también formas de aprovechar uno de los componentes principales de un dron: su cámara.

Otro aspecto que cabe destacar es que dada las motivaciones de los trabajos anteriores y con el fin de hacer más fácil la planeación de rutas, la arquitectura actual está restringida al uso de hardware especializado de Realidad Virtual para su funcionamiento, específicamente se usó el kit HTC vive VR que se ve en la figura 7(recuperada de htc.com). Por lo tanto, es de especial interés agregar un componente adicional que faciliten la planeación de rutas en 2D (cuando la altura entre puntos es constante) y la planeación/simulación de rutas en 3D por parte de usuarios que no posean este hardware.

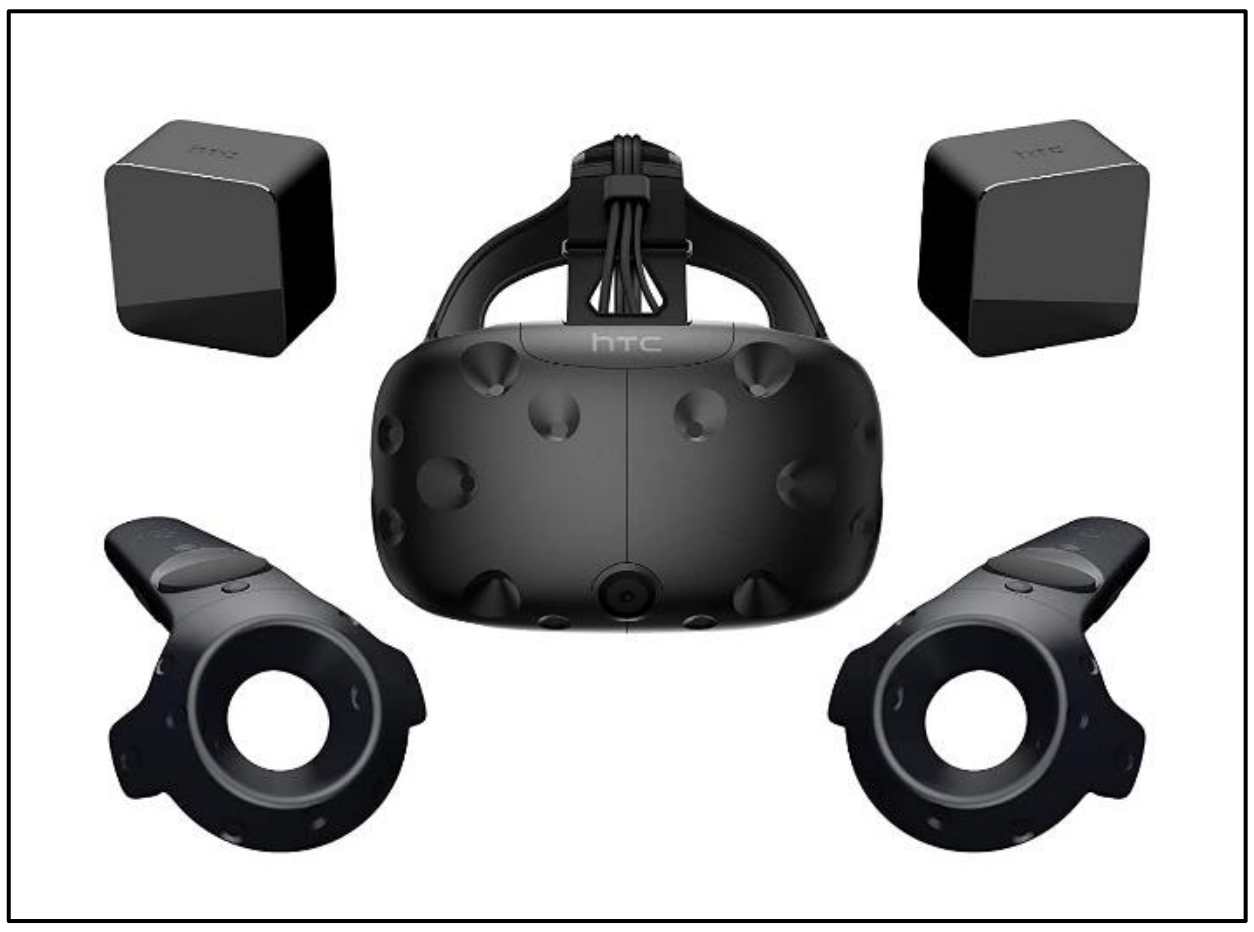

*Figura 7. Kit HTC vive con el que cuenta el laboratorio Colivri*

Finalmente, otra motivación para construir un nuevo componente para creación de rutas se debe al hecho de que la actual solución funciona muy bien si, previo a la etapa de planeación, se ha hecho un trabajo de recrear el espacio donde se diseñara la ruta. Por tanto, es de interés que una de las características de este nuevo componente sea la capacidad de recrear cualquier espacio exterior usando mapas y puntos de referencia en 3D.

# 4. Objetivos del proyecto

Con base en las motivaciones y oportunidades de extensión anteriormente mencionadas definimos los siguientes objetivos para este proyecto

## 4.1 Objetivos generales:

Extender la implementación de una arquitectura preexistente basada en interacción humano-robot para la planeación y seguimiento de rutas de un dron. Esta extensión busca la integración de nuevos y diferentes componentes tecnológicos que soporten la planeación y evaluación de rutas de vuelo de dron y aprovechen las funcionalidades y hardware de la aeronave, específicamente la cámara. Así mismo, estos componentes deben ser fácilmente reemplazables para ajustarse a diferentes contextos y restricciones tecnológicas. Finalmente, se buscará aplicar y evaluar por medio de diferentes métricas y pruebas la precisión y limitaciones de esta arquitectura en un escenario real.

## 4.2 Objetivos específicos:

1. Explorar y familiarizarse con las tecnologías involucradas en el proyecto actual. Estar familiarizado con el entorno de desarrollo móvil Android Studio y el lenguaje de programación Java.
2. Explorar la documentación y referencias del SDK de DJI enfocándose en las librerías que controlan las acciones de la cámara durante el trayecto entre waypoints o cuando el dron está en estado hover sobre waypoints específicos.
3. Extender el esquema de persistencia actual para que se puedan almacenar no solo rutas y waypoints sino también almacenar las diferentes tareas de la cámara que le interesen al usuario. Por otra parte, agregar información de interés relativa a la ruta, como puede ser un campo para una descripción que hable propósito de la ruta y el entorno donde se llevará acabo.
4. Implementar la planeación de tareas de la cámara asociadas a un subconjunto de puntos de una ruta predefinida desde la aplicación móvil existente. Estas tareas involucraran básicamente el aprovechamiento de la cámara del dron durante la ejecución de la ruta. Las tareas que queremos considerar son las siguientes:
    - **Tarea de imágenes panorámicas**: reconstruir una imagen panorámica partir de un conjunto de fotos tomadas sobre un mismo punto mientras el dron está en estado hover (suspensión) rotando sobre su eje;
    - **Tarea de captura por intervalos:** Tomar una foto cada X segundos durante el recorrido del dron entre el punto A y el punto B

- **Tarea de grabación de video:** Iniciar la grabación de video en un punto inicial de la ruta y detener la grabación en un punto final, ambos seleccionados por el usuario.
- **Tarea de captura de fotografías**: tomar fotografía en un punto o serie de puntos sobre la ruta

5. Implementar un componente alternativo para la planeación y simulación de rutas. Se espera que esta aplicación soporte la planeación de rutas y tareas de cámara sin necesidad de hardware de interacción con ambientes de realidad virtual. Adicionalmente, se espera que este componente no solo facilite la planeación en 2D y 3D sino también que proporcione una aproximación de las estructuras en el espacio en donde se ejecutará la ruta y que sirvan como guía para el usuario. Por otra parte, sería ideal que esta herramienta ofrezca opciones de simulación de rutas mediante la animación de un modelo a escala del dron, que este se mueva por la ruta demarcada y tenga indicadores de las tareas de cámara que el dron está ejecutando. Finalmente, este componente también debe integrar la planeación de tareas de cámara definidas anteriormente.

## 4.3 Potencial aplicación de la solución:

Los objetivos anteriormente planteados, además de ser el punto de partida para construir los requerimientos funcionales de la solución, fueron pensados con base a los usos potenciales que se le puede dar en el mundo real y las posibles integraciones que se lograrían con otras tecnologías y plataformas. A continuación, planteamos diversos escenarios en donde la extensión de la solución sería de utilidad:

Una vez se tenga una serie de imágenes panorámicas en diferentes puntos se puede recrear un recorrido virtual interactivo con herramientas como Krpano, que fue usada para construir el recorrido virtual del campus de la universidad de los Andes (Fig. 8 Recuperada de campusinfo.uniandes.edu.co)

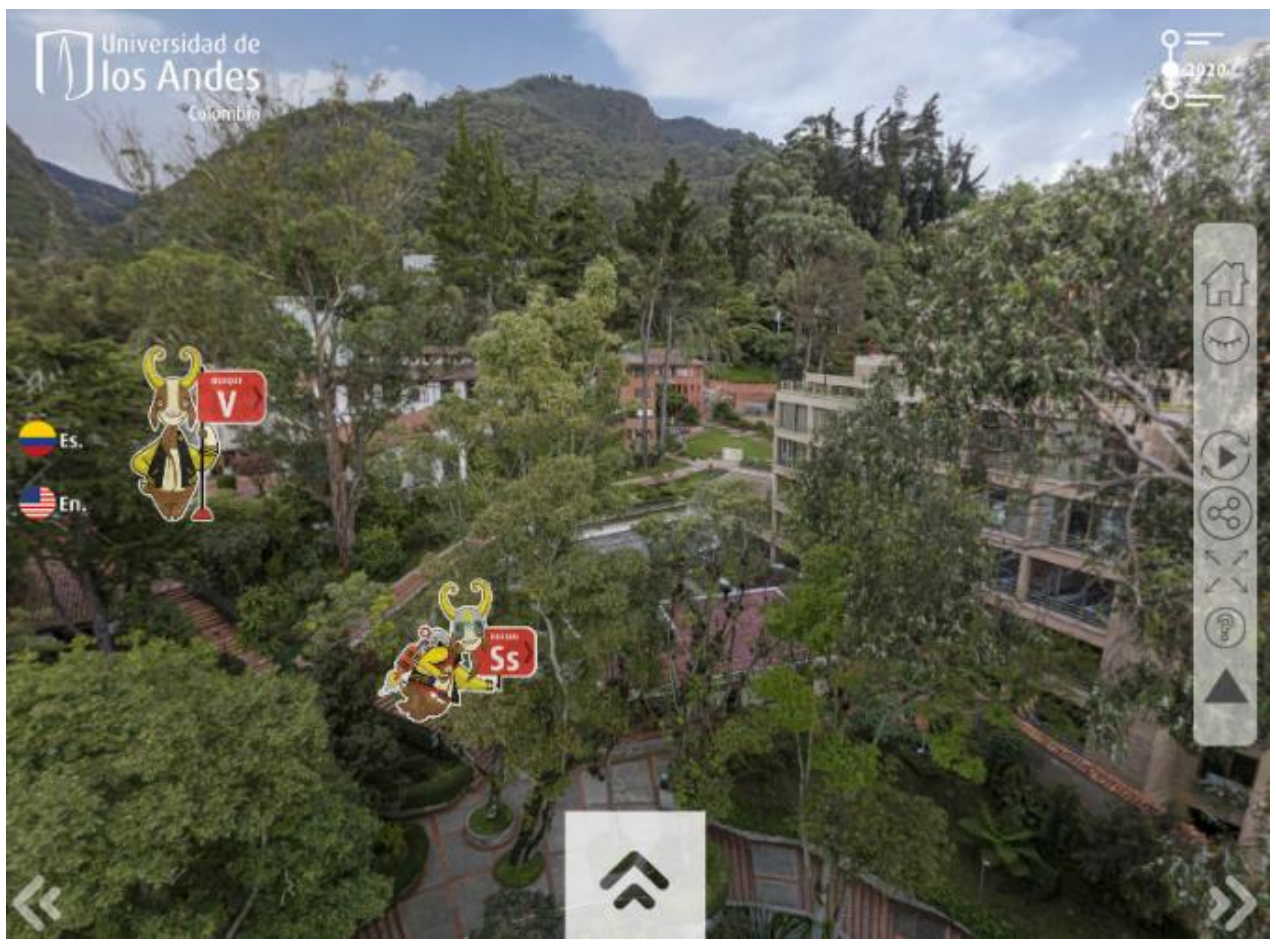

*Figura 8. Interfaz del recorrido virtual del campus Uniandes con Kaprano*

Otro objetivo es generar una solución que le permita al usuario de un dron generar contenido multimedia con potencial de utilidad: recorridos virtuales de un espacio desde una perspectiva aérea o reconstrucción de espacios y estructuras en un ambiente de realidad virtual 3D. Este conjunto de fotos de puntos estratégicos podría ser usado en tareas de reconstrucción de objetos en ambientes de realidad virtual o aumentada (3D object assets) como se muestra en la figura 9 (recuperada de https://unity.com/solutions/photogrammetry).

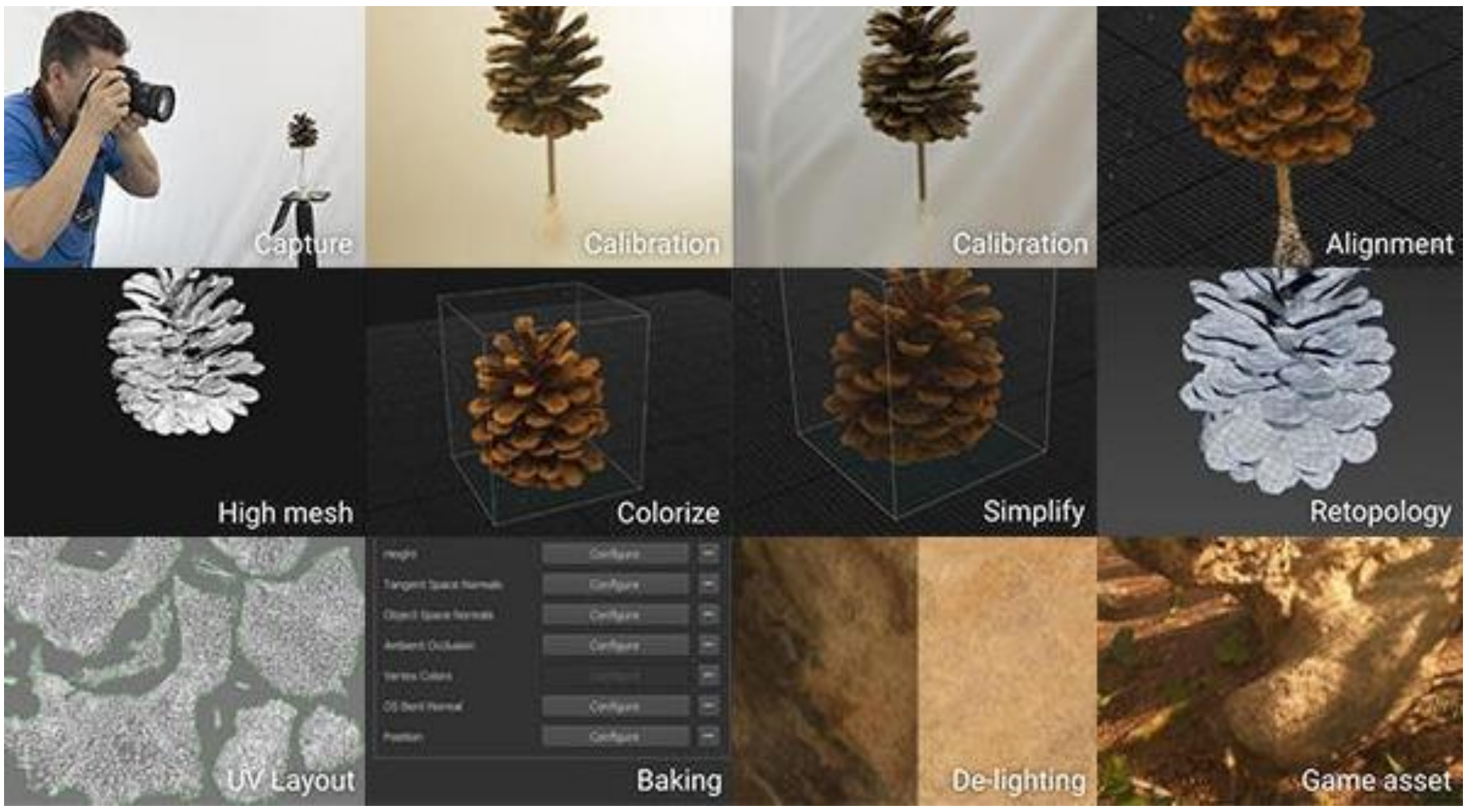

*Figura 9. Flujo de trabajo de una tarea de fotogrametría en el ambiente Unity*

# 5. Propuesta y diseño de solución

## 5.1 Especificaciones

Con el fin de abordar los objetivos planteados anteriormente se definen los siguientes requerimientos que debe cumplir esta extensión de la arquitectura:

### Definición y Edición de Tareas de Cámara

| Requerimiento | RF1: Planear foto en un Waypoint |
|---|---|
| Descripción | Dado un waypoint en una ruta existente, el usuario marca el waypoint para que se tome una fotografía una vez el dron alcance dicho punto |
| Tipo | Funcional |
| Componente asociado | App móvil y App de planeación y Simulación |
| **Requerimiento** | **RF2: Planear inicio de video** |
| Descripción | El usuario planifica la captura de video entre un waypoint inicial y un waypoint final en una ruta existente |
| Tipo | Funcional |
| Componente asociado | App móvil y App de planeación y Simulación |
| **Requerimiento** | **RF3: Planear intervalo de fotos** |
| Descripción | El usuario configura un waypoint para tomar una fotografía cada X segundos entre un waypoint y el waypoint siguiente |
| Tipo | Funcional |
| Componente asociado | App móvil y App de planeación y Simulación |
| **Requerimiento** | **RF4: Planear foto panorámica** |
| Descripción | El usuario configura un waypoint para tomar una fotografía de 360º desde ese punto, la cual se reconstruye a partir de una serie de fotos tomadas a medida que el dron rota sobre su eje vertical |
| Tipo | Funcional |
| Componente asociado | App móvil y App de planeación y Simulación |

## Definición y Edición de Rutas

| Requerimiento | RF5: Ver la lista de rutas planeadas |
|---|---|
| Descripción | El usuario puede ver la lista de las rutas que se han planeado, se muestra el identificador de las rutas, su descripción, el número de puntos a recorrer y los botones de edición o eliminación de cada una |
| Tipo | Funcional |
| Componente asociado | App de planeación y Simulación |
| **Requerimiento** | **RF6: Eliminar una ruta planeada** |
| Descripción | El usuario puede seleccionar eliminar una ruta especifica en la lista, después de confirmar la acción la ruta es borrada de la base de datos |
| Tipo | Funcional |
| Componente asociado | App de planeación y Simulación |
| **Requerimiento** | **RF7: Entrar en modo edición** |
| Descripción | Ya sea que el usuario seleccione crear una ruta o editar una existente, se le dirige a una vista donde puede editar los detalles de la ruta y sus waypoints |
| Tipo | Funcional |
| Componente asociado | App de planeación y Simulación |

| Requerimiento | RF8: Agregar waypoint a la ruta |
|---|---|
| Descripción | En modo edición, ya sea que la ruta ya tenga waypoints existentes o no haya ninguno, el usuario selecciona sobre el mapa y un nuevo waypoint se agrega con una altura por defeco. Una línea guía conecta los waypoints consecutivos |
| Tipo | Funcional |
| Componente asociado | App de planeación y Simulación |
| **Requerimiento** | **RF9: eliminar waypoint de una ruta** |
| Descripción | En modo edición, el usuario abre las configuraciones de un waypoint especifico y selecciona eliminar, después de la confirmación el waypoint seleccionado es eliminado de la ruta |
| Tipo | Funcional |
| Componente asociado | App de planeación y Simulación |
| **Requerimiento** | **RF10: editar altura de un waypoint** |
| Descripción | En modo edición, un usuario selecciona un waypoint y cambia la altura a su disposición, seleccionar guardar preferencias y al altura se actualiza de forma lógica y de manera grafica |
| Tipo | Funcional |
| Componente asociado | App de planeación y Simulación |
| **Requerimiento** | **RF11: editar tarea cámara** |
| Descripción | En modo edición, el usuario selecciona un waypoint sobre la ruta y puede cambiar la tarea de cámara según los requerimientos RF1 – RF4, selecciona guardar preferencias y se aplican los cambios de forma lógica y de forma gráfica con colores e indicadores |
| Tipo | Funcional |
| Componente asociado | App de planeación y Simulación |
| **Requerimiento** | **RF12: agregar descripción a una ruta** |
| Descripción | En modo edición, el usuario puede agregar una cadena de texto en un campo descripción, el cual es opcional |
| Tipo | Funcional |
| Componente asociado | App de planeación y Simulación |
| **Requerimiento** | **RF13: ver detalles de waypoints** |
| Descripción | En modo edición, el usuario puede ver listados los detalles de cada waypoint, como lo es el orden, la latitud y longitud, la altura y la tarea asignada (si tienen alguna) |
| Tipo | Funcional |
| Componente asociado | App de planeación y Simulación |
| **Requerimiento** | **RF14: Guardar/Persistir Ruta** |
| Descripción | En modo edición, cualesquiera que sean los cambios hechos en una ruta, el usuario selecciona la opción de guardar. La ruta queda guardada en la base de datos que comparten los componentes de la arquitectura |
| Tipo | Funcional |
| Componente asociado | App de planeación y Simulación |
| **Requerimiento** | **RF15: Iniciar simulación** |
| Descripción | En modo edición, se hayan guardado en la base de datos o no, los cambios hechos a una ruta y sus waypoints, el usuario selecciona inicial simulación y se da inicio a una animación de un icono de un dron moviéndose por la ruta y se ve indicadores de las tareas planificadas en waypoints o entre ellos |
| Tipo | Funcional |
| Componente asociado | App de planeación y Simulación |

| **Requerimiento** | **RF16: Cambiar inclinación de cámara** |
|---|---|
| Descripción | En modo edición, un usuario puede cambiar la inclinación de la cámara y el zoom de modo que puede moverse entre diversos puntos de vista y ángulos sobre una zona de interés |
| Tipo | Funcional |
| Componente asociado | App de planeación y Simulación |
| **Requerimiento** | **RF17: cambiar a vista satelital** |
| Descripción | En modo edición, un usuario selecciona ver el espacio de forma de que pueda guiarse con fotos satelitales para guiarse en la creación de rutas y tener puntos de referencia más orgánicos y precisos |
| Tipo | Funcional |
| Componente asociado | App de planeación y Simulación |
| **Requerimiento** | **RF18: cambiar a vista de mapa de estructuras 3D** |
| Descripción | En modo edición, el usuario selecciona ver el espacio de forma que pueda ver una representación en 3D de las estructuras arquitectónicas que están modeladas, de esta forma tiene un punto de referencia para las alturas |
| Tipo | Funcional |
| Componente asociado | App de planeación y Simulación |
| **Requerimiento** | **RF19: Ver Live Feed** |
| Descripción | Una vez una ruta es ejecutada desde la app móvil, el usuario selecciona ver live feed para ver en tiempo real lo que está captando la cámara |
| Tipo | Funcional |
| Componente asociado | App móvil |

## 5.2 Exploración y evaluación de tecnologías

Principalmente la definición de los requerimientos son el punto de partida para definir lo que se espera de la solución final, sin embargo, también son de ayuda para poder evaluar la viabilidad del uso de tecnologías y metodologías de desarrollo que faciliten la correcta implementación y éxito de la solución, a continuación, se presenta una evaluación y discusión de las diferentes tecnologías que fueron consideradas para ser usadas en este proyecto.

Con el fin de lograr el objetivo específico de agregar un componente que soporte la planeación y simulación, se evaluaron varias plataformas y tecnologías como una opción para construir una aplicación que soportara mapas 2D/3D y simulación del movimiento de un objeto sobre un escenario 3D.

### 5.2.1 ROS

Primeramente, se evaluó el framework de ROS para dar soporte a este objetivo. ROS es un framework y comunidad en línea que sirve para construir y reutilizar código entre aplicaciones de robótica de código abierto. Este framework cuenta con una gran variedad de paquetes que abarcan todo el espectro de necesidades que un programador de robots podría necesitar, además cuenta con implementaciones ampliamente probadas y constantemente mejoradas de algoritmos de propósito general cuando se trata de robots. Entre sus paquetes de librerías y herramientas están las de visualización y simulación del comportamiento de robots en el espacio 3D. Ros tiene compatibilidad con simuladores como Gazebo y Rviz, los cuales podrían ser usados para visualizar el comportamiento del drone previo a la ejecución de una ruta. En la figura 10 se puede ver el modelo 3D de un drone DJI mavic pro en el simulador Gazebo, como se muestra en la siguiente figura (recuperada de https://dev.px4.io/v1.10_noredirect/en/simulation/gazebo.html).

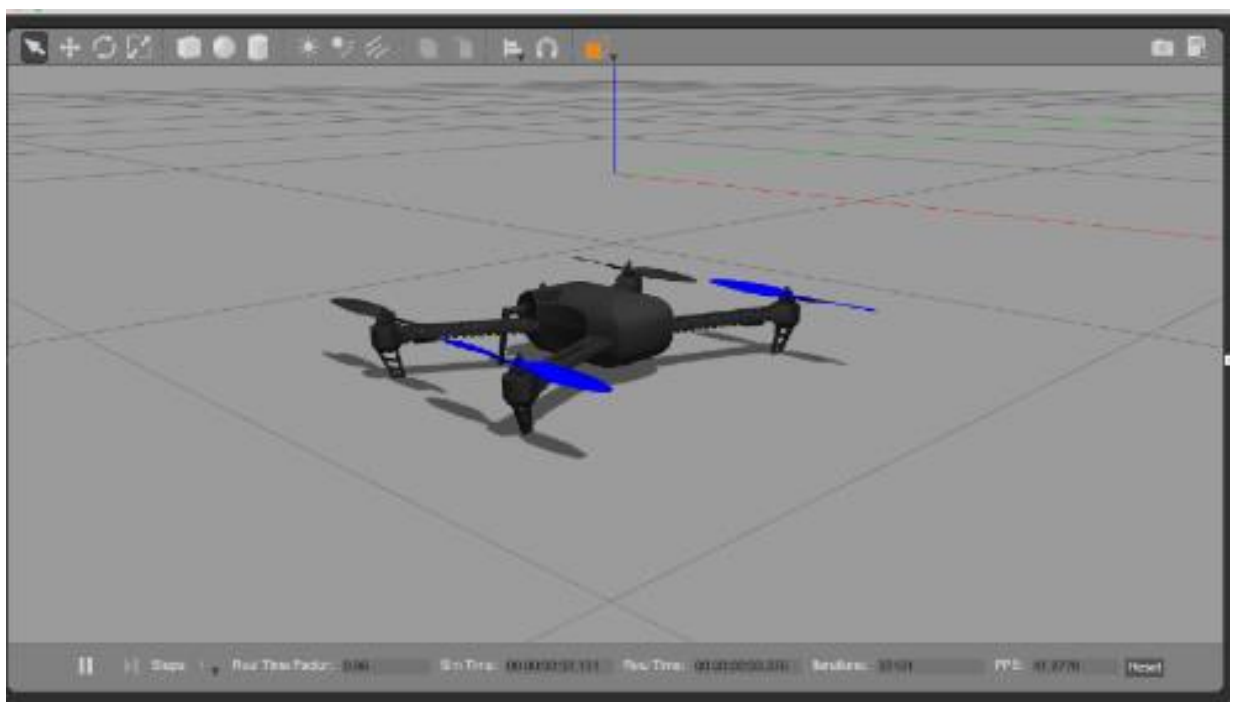

*Figura 10. Simulación de un dron DJI en el espacio 3D usando Gazebo*

A pesar del potencial que ROS tiene para ser usado en contextos que requieren el control de drones, encontramos dos limitaciones que nos llevaron a descartar ROS para su uso en el actual proyecto. En primer lugar, los simuladores compatibles con ROS no soportan una fácil edición de un espacio 3D donde el usuario podría estar interesado en ejecutar una ruta. Al igual que sucede con la mayoría de las aplicaciones que usan Unity o motores de gráficos similares, la representación de un escenario tridimensional, los objetos y estructuras (también llamados assets) tendrían que ser codificados para que el usuario los visualice sobre la escena. Lo anterior significa una restricción en cuanto a la representación de la gran variedad de espacios en los que el usuario podría estar interesado en ejecutar la ruta.

La segunda razón que nos llevó a desistir del uso de este framework es que ya se cuenta con un conjunto de herramientas y librerías (DJI SDK) que nos permitían comunicarnos y programar el comportamiento de este drone en particular (y otros drones DJI que son compatibles con la última versión del DJI Mobile SDK). No obstante, cabe resaltar que ROS sería de gran utilidad si se quisiera extender la arquitectura de solución actual para ser usada con drones de otros fabricantes o drones construidos a partir de componentes electrónicos individuales y que usan microcontroladores y librerías de código abierto para su manejo.

### 5.2.2 Mapbox API

Teniendo en mente que se podría realizar una aplicación web para soportar la planeación y simulación de rutas, se consideró primeramente el uso de Mapbox ya que ofrece un API basado en JavaScript que puede ser usado en aplicaciones web para ver mapas en 3D. En la siguiente figura (recuperada de https://docs.mapbox.com/mapbox-gl-js/example/3d-buildings/) se puede ver un ejemplo de mapa con edificios en 3D.

A pesar de ser un API que se pude usar de forma gratuita, esta opción fue descartada debido a las limitaciones encontradas para la representación de marcadores con altura relativa al suelo y uso de assets 3D o iconos dinámicos que se graficaran sobre el mismo.

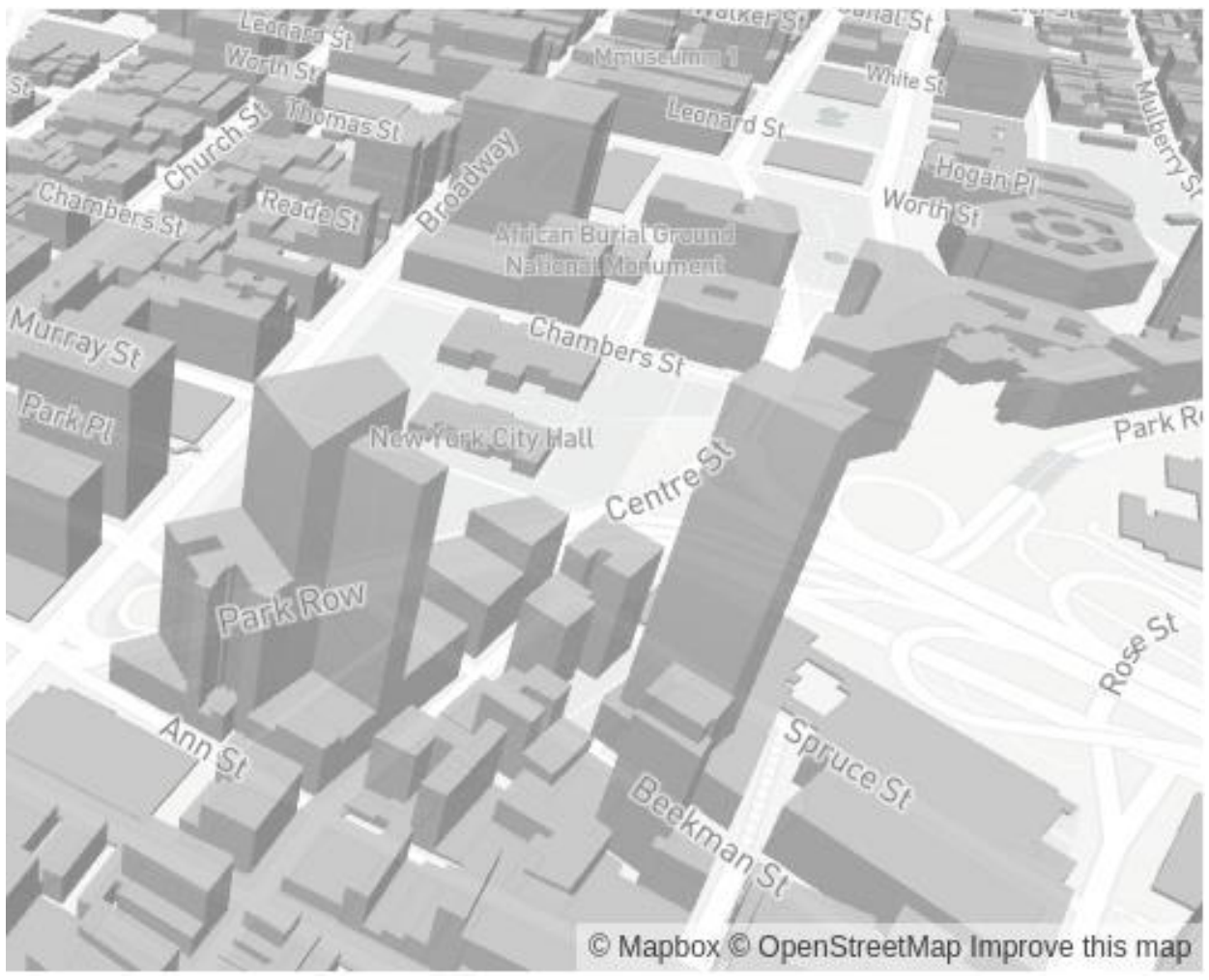

*Figura 11. Mapas 3D de Mapbox*

### 5.2.3 Google Maps JavaScript API

Finalmente se tomó la decisión de usar las librerías de mapas que ofrece Google maps para aplicaciones web. Las razones principales para su elección radican en el hecho de que esta API está muy bien documentada, cuenta también con soporte para estructuras 3D mediante estilos personalizados y más recientemente en el Google I/O 2021 conference se hizo el reléase de nuevas características que involucran el uso de WebGL para mejorar la experiencia e interacción en estos mapas 3D (WebGL powered maps). Estas nuevas características serian de utilidad en esta solución pues se pueden usar marcadores que respondan a la altura de un waypoint, como se muestra en la figura 12 (recuperada de recuperado de https://geo-devrel-io2021-oobe.web.app), en cuanto a simulación se podría usar fácilmente un asset o icono de un dron para que sea el objeto que se mueva sobre una ruta mediante una animación, tal como se muestra en las figuras 13 y 14 recuperadas de la misma fuente antes citada.

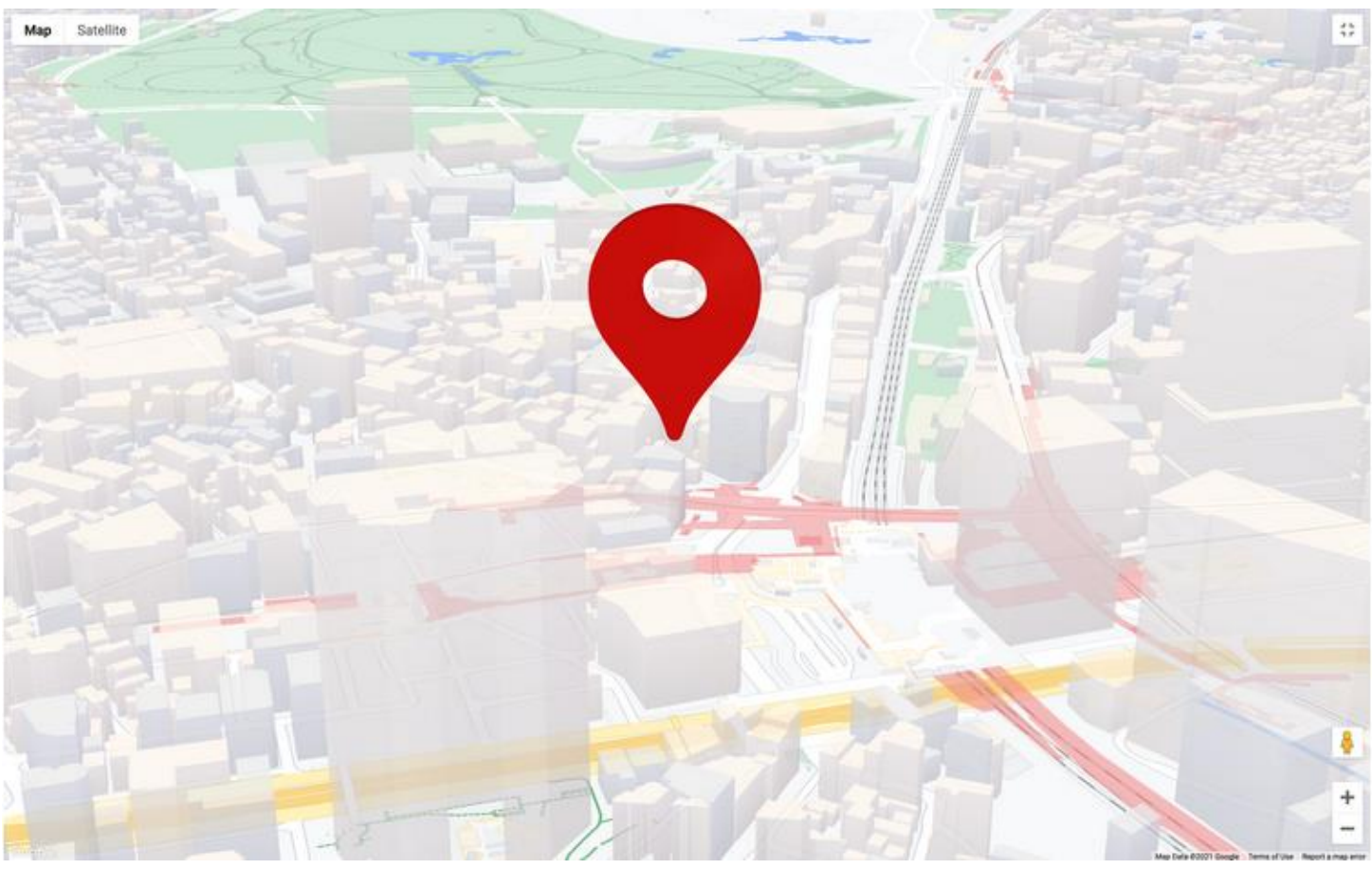

*Figura 12. Creación de un marcador con tamaño y altura relativa al espacio en Google Map*

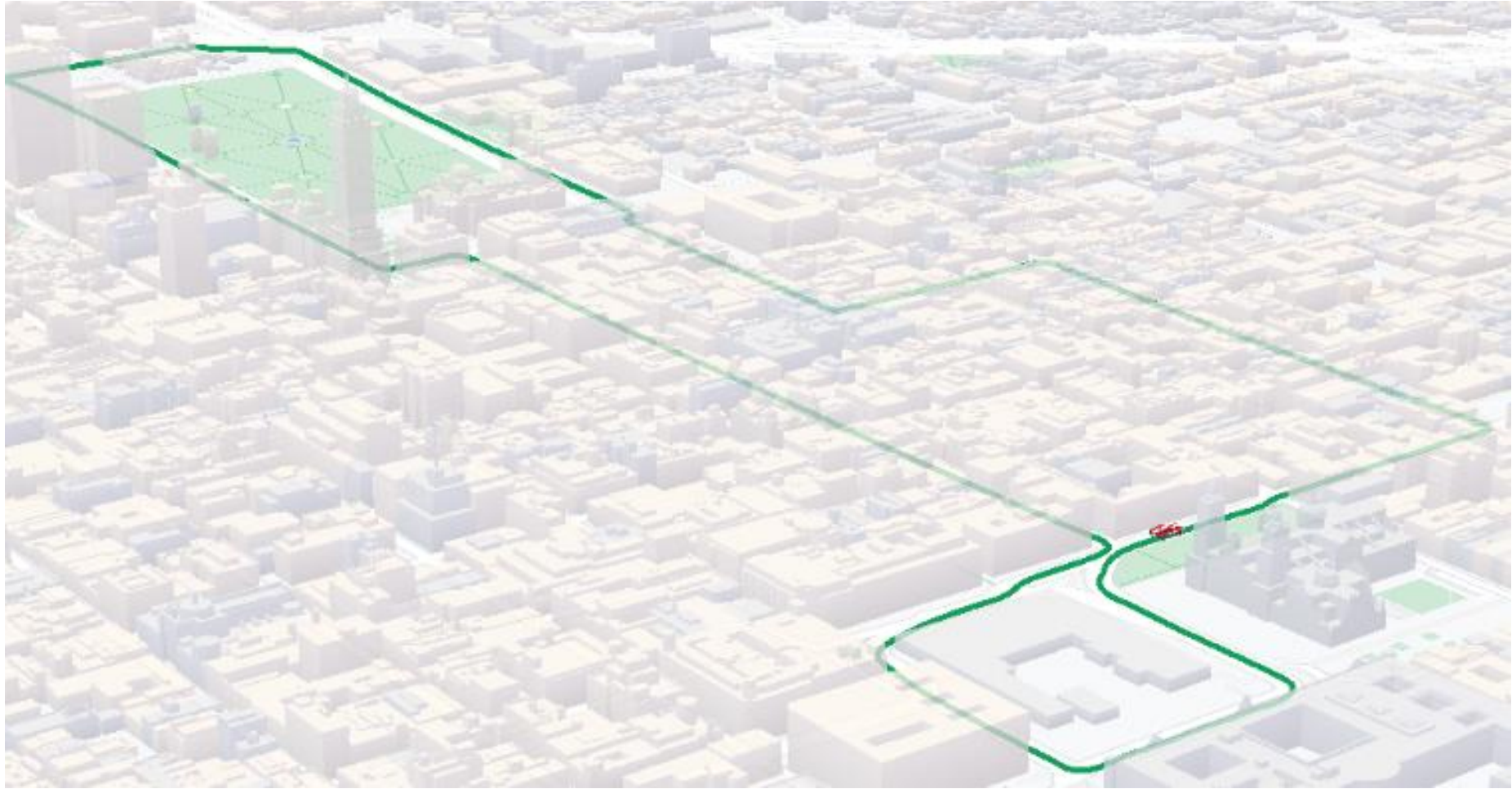

*Figura 13. Creación de una animación del recorrido de un path mediante un asset 3D*

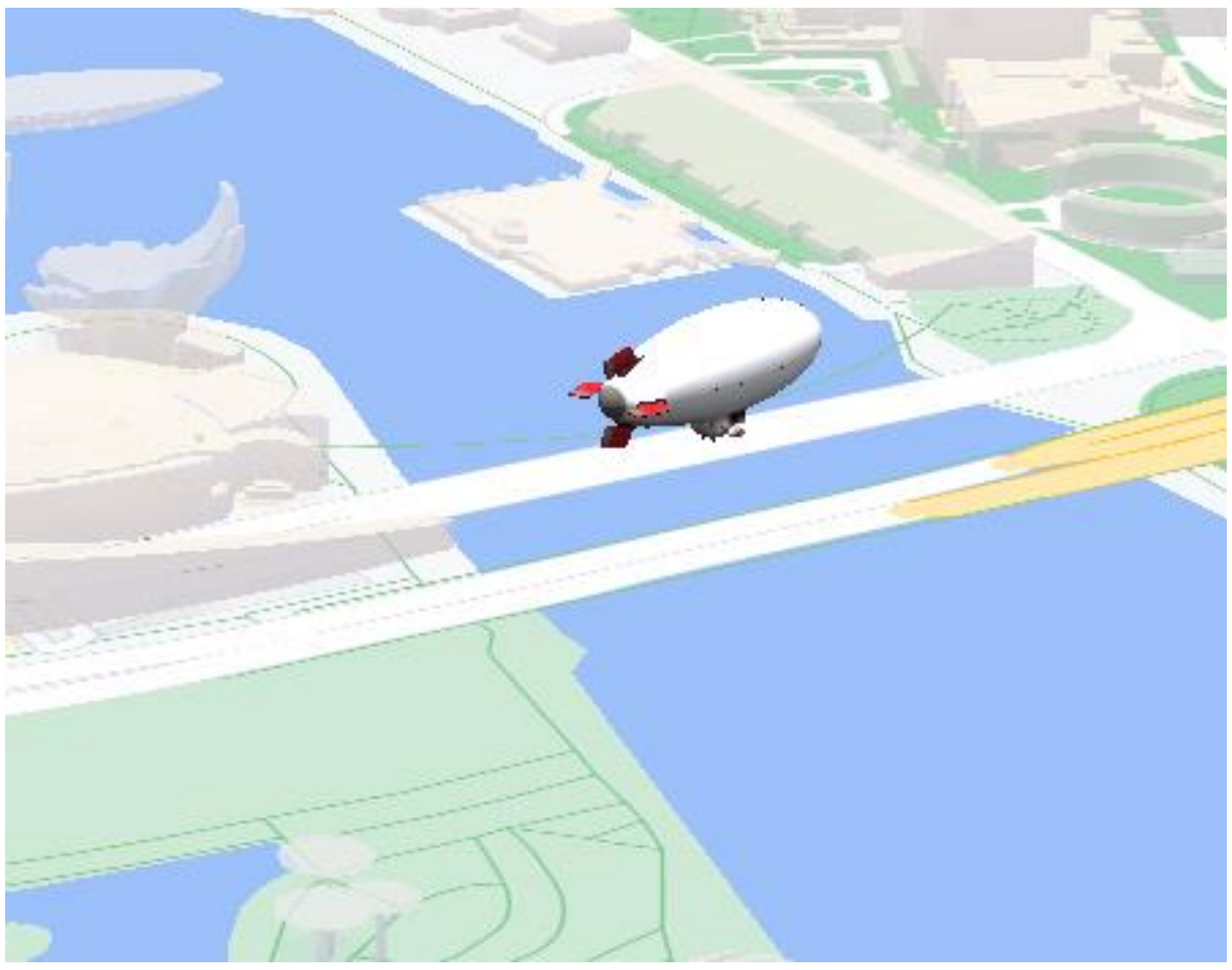

*Figura 14. Animación de un objeto volador teniendo en cuenta altura*

### 5.2.4 Angular

Después de verificar la viabilidad del API de mapas de Google para JavaScript el siguiente paso era buscar un framework que facilitara la creación de una aplicación web que manejara datos dinámicos y que estos se actualizarán en los elementos HTML de la página a medida que se interactúa con los mismos y se modifica la información relacionada a rutas y waypoints de manera local y una vez son persistidos en una base de datos. Para este proyecto se usó la versión estable 12.1.4.

Entre otras opciones estaban los frameworks de Django y React, pero decidimos usar el framework Angular debido a que los desarrolladores de Google ya han implementado un componente que se integra muy bien con las librerías y paquetes necesarios para mostrar mapas dinámicos, además que están en constante mejora y cuentan con una documentación muy completa.

Puede encontrar información de este componente en el repositorio oficial de angular: Google Maps Angular Component [10].

### 5.2.5 Angular Material

Con el fin de dar una sensación al usuario de que está usando una solución con componentes integrados, en términos estéticos se decidido usar la librería de iconos y elementos HTML de Material Design para Angular [11], con el fin de que haya, una equivalencia, en términos de diseño de interfaz, entre la aplicación móvil existente y la aplicación web de planeación y simulación.

## 6. Propuesta de arquitectura de solución

La arquitectura que se propone se extiende de las desarrolladas por Sánchez [1] y Múnera [2], la cual se encuentra representada en la siguiente imagen

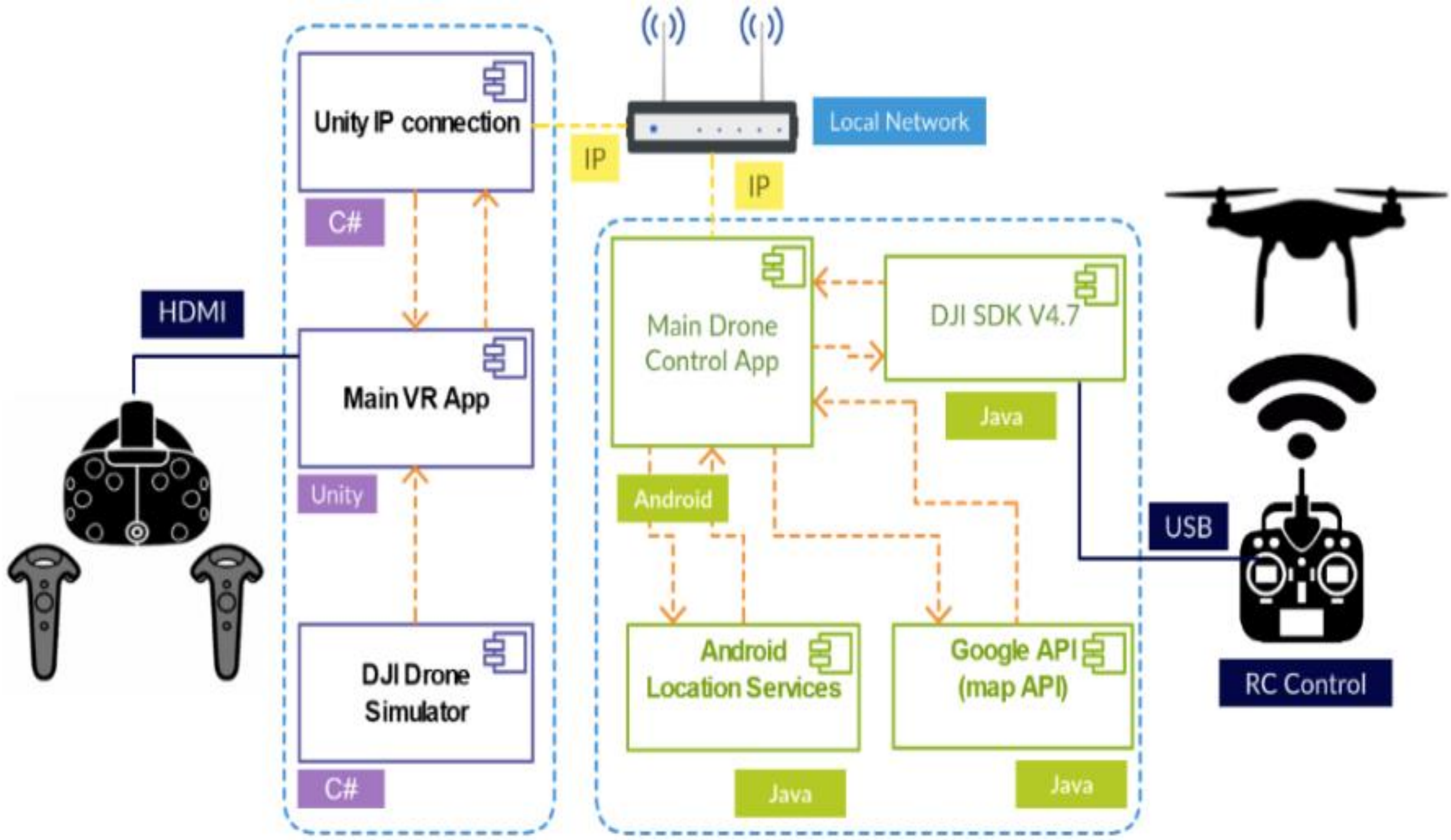

*Figura 15. Arquitectura de solución propuesta por Sánchez y reutilizada por Múnera*

La arquitectura anterior consta de un módulo de creación de rutas en un entorno de Realidad Virtual y un módulo de operación de rutas en un entorno móvil (Android). El módulo de creación consta de una aplicación de escritorio desarrollada en Unity (Main VR App), la cual permite la creación de rutas de vuelo en un ambiente 3D virtual. Una vez se crea la ruta, esta es almacenada en una base de datos en Firebase, donde se almacena la altura, latitud y longitud de cada punto de la ruta. Por otro lado, el módulo de operación consiste en una aplicación móvil desarrollada en Java para sistemas operativos Android. Esta app descarga las rutas de la base de datos y configura el comportamiento del dron durante el vuelo. Finalmente, carga la ruta de vuelo configurada al dron y permite iniciar la operación, configurar parámetros adicionales y hacer seguimiento a su posición durante la trayectoria.

A partir de esta arquitectura, se propone implementar las siguientes extensiones:

- Ampliar el módulo de planeación de rutas: Permitir al usuario planear rutas 2D y 3D sin la necesidad de hacer uso del entorno de Realidad Virtual. Así mismo, permitir al usuario simular el vuelo del dron sobre las rutas planeadas y agregar tareas de cámara.

- Ampliar el módulo de operación: Hacer uso de las funcionalidades que ofrece la cámara integrada al dron, como hacer capturas de imágenes y videos. Adicionalmente, realizar transformaciones sobre estos outputs para obtener información relevante sobre las rutas.

Con base en esto, se propone la siguiente arquitectura para el proyecto actual

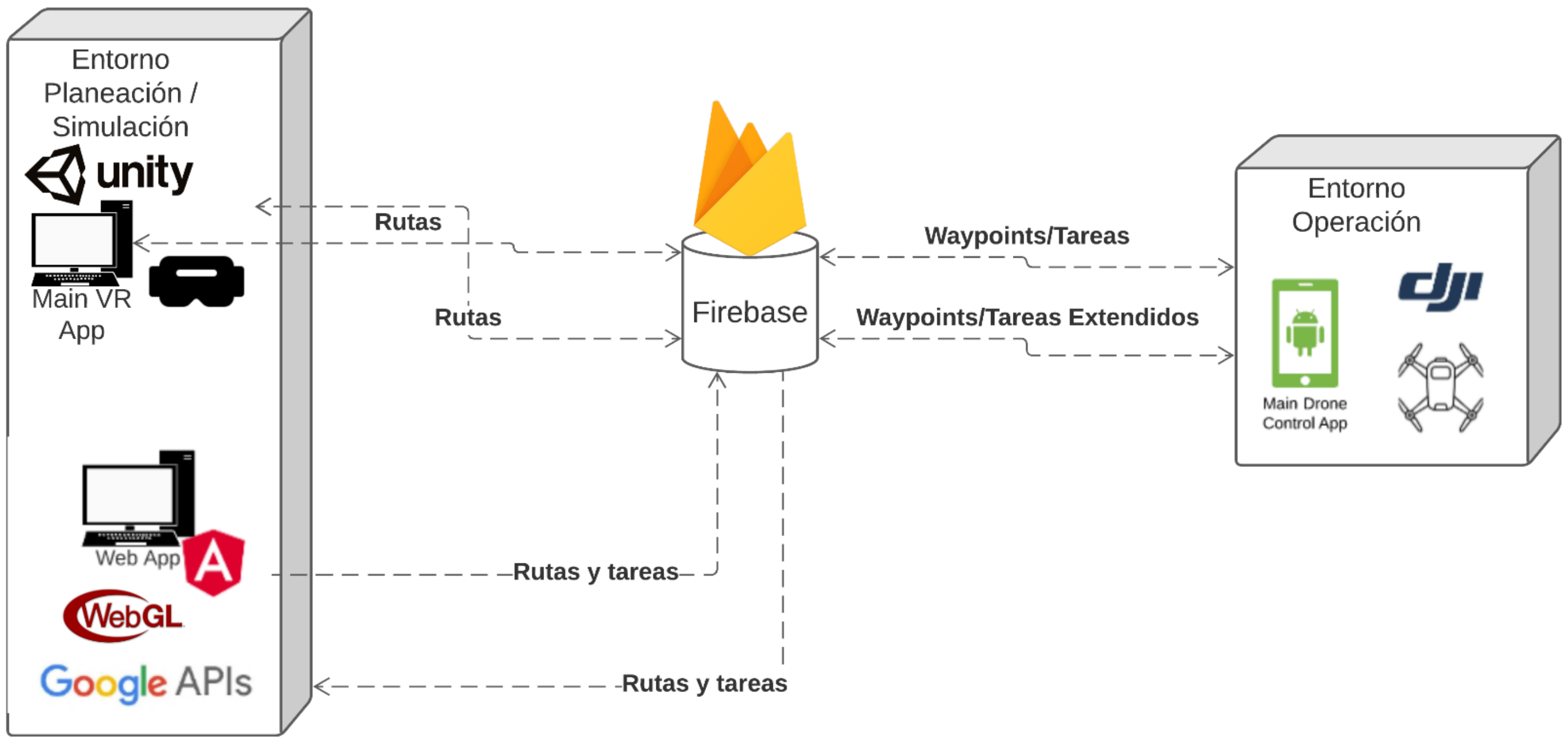

*Figura 16. Arquitectura de solución propuesta. Un módulo de planeación/simulación, un módulo de operación, y uno de comunicación*

Esta arquitectura mantiene la separación de módulos previos y extiende sobre estos mismos. Para extender el módulo de planeación, se propone implementar un aplicativo web desarrollado en Angular que permite la planeación de rutas, tanto 2D como 3D, gracias a la integración de tecnologías como Angular, Google Maps JavaScript API y WebGL. Así mismo, permite la simulación de rutas planeadas recuperándolas de la base de datos en Firebase.

Para extender el módulo de operación, se propone desarrollar nuevas funcionalidades sobre la app de Android, las cuales permitan aprovechar al máximo las capacidades de la cámara. Para esto, se propone desarrollar nuevas actividades y vistas que soporten estas funcionalidades. Así mismo, se planea integrar la app con librerías y APIs de OpenCV que permitan realizar el procesamiento de imágenes y video, para realizar tareas de "stitching" de imágenes (para panorámicas).

## 7. Desarrollo de la solución

A continuación, se hará una descripción del desarrollo realizado a partir de las especificaciones y requerimientos definidos anteriormente. Primeramente, se presentarán los diagramas de componentes de la app web y la extensión de la app móvil. Por otra parte, se explicará el diagrama de clases que corresponde a la extensión del esquema de persistencia. Finalmente, se hará un recorrido por cada requerimiento para explicar detalles particulares de la implementación y el resultado final.

### 7.1 Diagrama de componentes de la aplicación web

A continuación, se muestra una figura (Fig. 17) que ilustra cómo se comunican los diferentes componentes de la aplicación web:

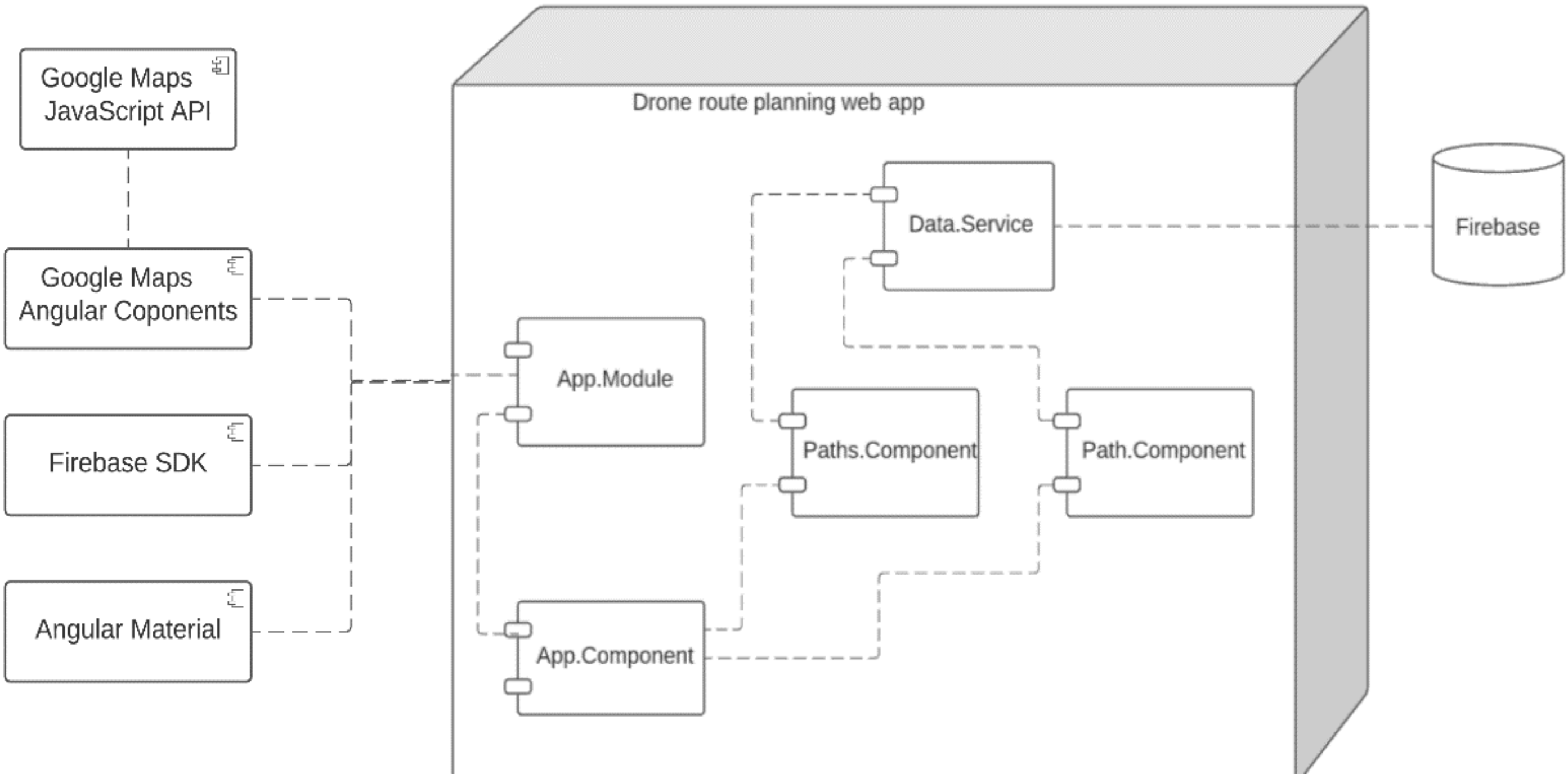

*Figura 17. Diagrama de componentes de la aplicación web Drone Route Planning Web App*

En primer lugar, se tiene el módulo principal App.Module el cual se encarga de importar las librerías y paquetes externos para que el resto de los componentes puedan hacer uso de estos. Entre las librerías externas está el componente de mapas de Google oficial para Angular, la dependencia de este paquete a la librería de Google Maps JavaScript escrita en TypeScript, el SDK de firebase para el intercambio y tareas CRUD de bases de datos alojadas en esta plataforma y finalmente el paquete de componentes de Angular Material que contiene reglas de diseño de elementos HTML definidas en CSS.

El siguiente componente en la lógica de la aplicación es el App.Component y se encarga encapsular todos los elementos HTML que se creen en otros componentes y también se encarga del re-direccionamiento entre páginas.

El componente Data.Service, haciendo uso del SDK de Firebase, es el encargado de definir los métodos de CRUD a la base de datos y servir como interfaz para que el resto de los componentes puedan realizar operaciones sobre la base de datos

Por otra parte, el componente Paths.Component es el que representa la página principal de la app que la que se listan y muestran detalles básicos de las rutas alojadas en la base de datos. Finalmente, El componente Path.Component es el que representa la página del modo de edición de la aplicación, en la que se mostraran de forma detallada los atributos y elementos de una ruta y se despliegan los controles para manipularla.

## 7.2 Diagrama de componentes de la aplicación móvil Main Control App

Para el componente de operación, se extendió el funcionamiento de la aplicación Main Drone Control App. Para esto, se agregaron nuevas actividades que soportan las funcionalidades planeadas. Estas nuevas actividades se muestran en el siguiente diagrama:

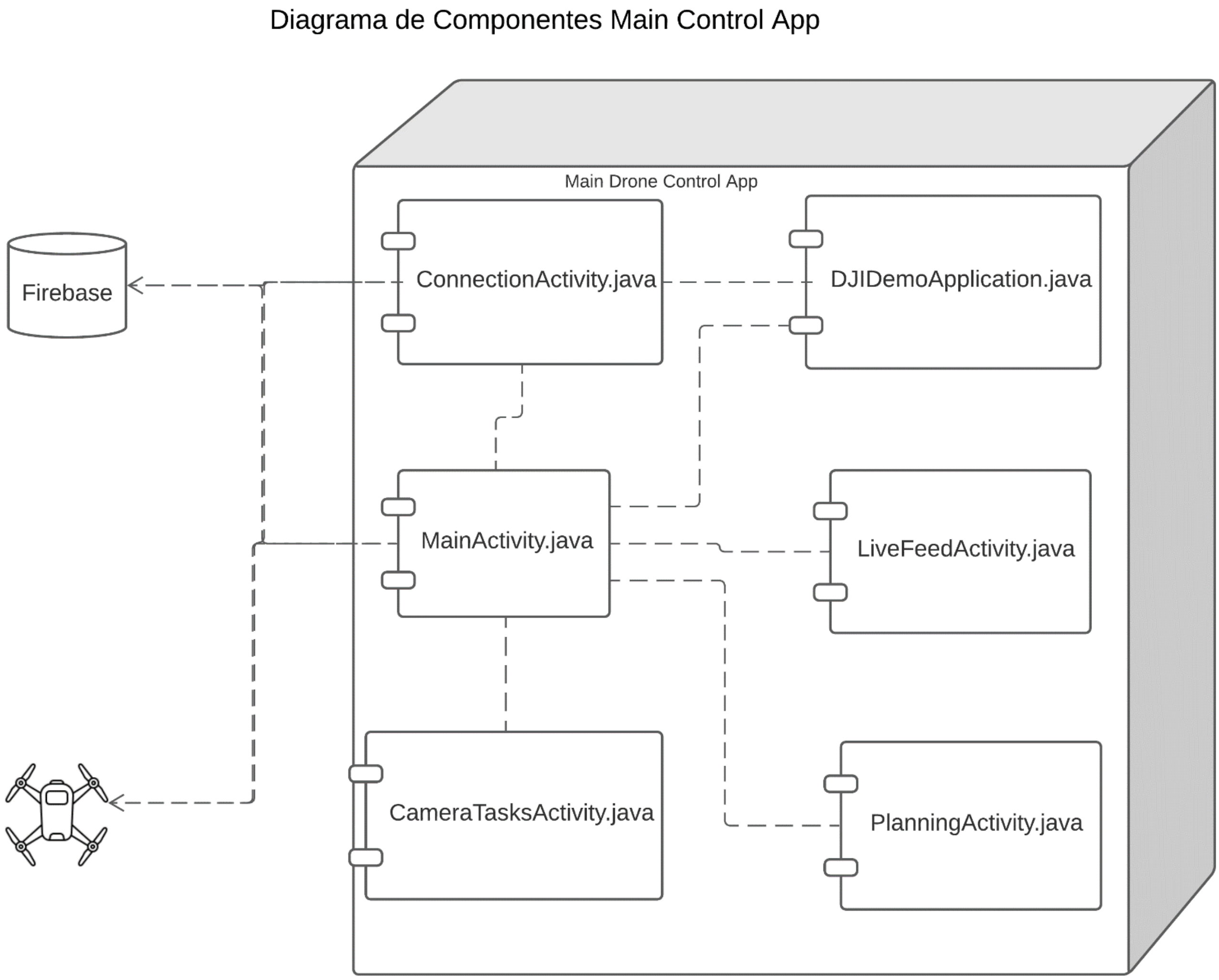

*Figura 18. Diagrama de componentes de la aplicación móvil Main Control App*

La aplicación móvil original cuenta con dos vistas principales: ConnectionActivity, la cual realiza la conexión con el dron y con la base de datos, y MainActivity, la cual se encarga de visualizar y configurar las rutas de vuelo, realizar su carga al dron y ejecutarlas. Adicionalmente, esta última actividad permite observar información acerca de las rutas cargadas y las rutas ejecutadas por el dron.

La primera actividad se comunica con el control del dron al cual el celular está conectado y realiza la conexión con este. Posteriormente, el usuario puede indicar el identificador de la ruta que desee cargar.

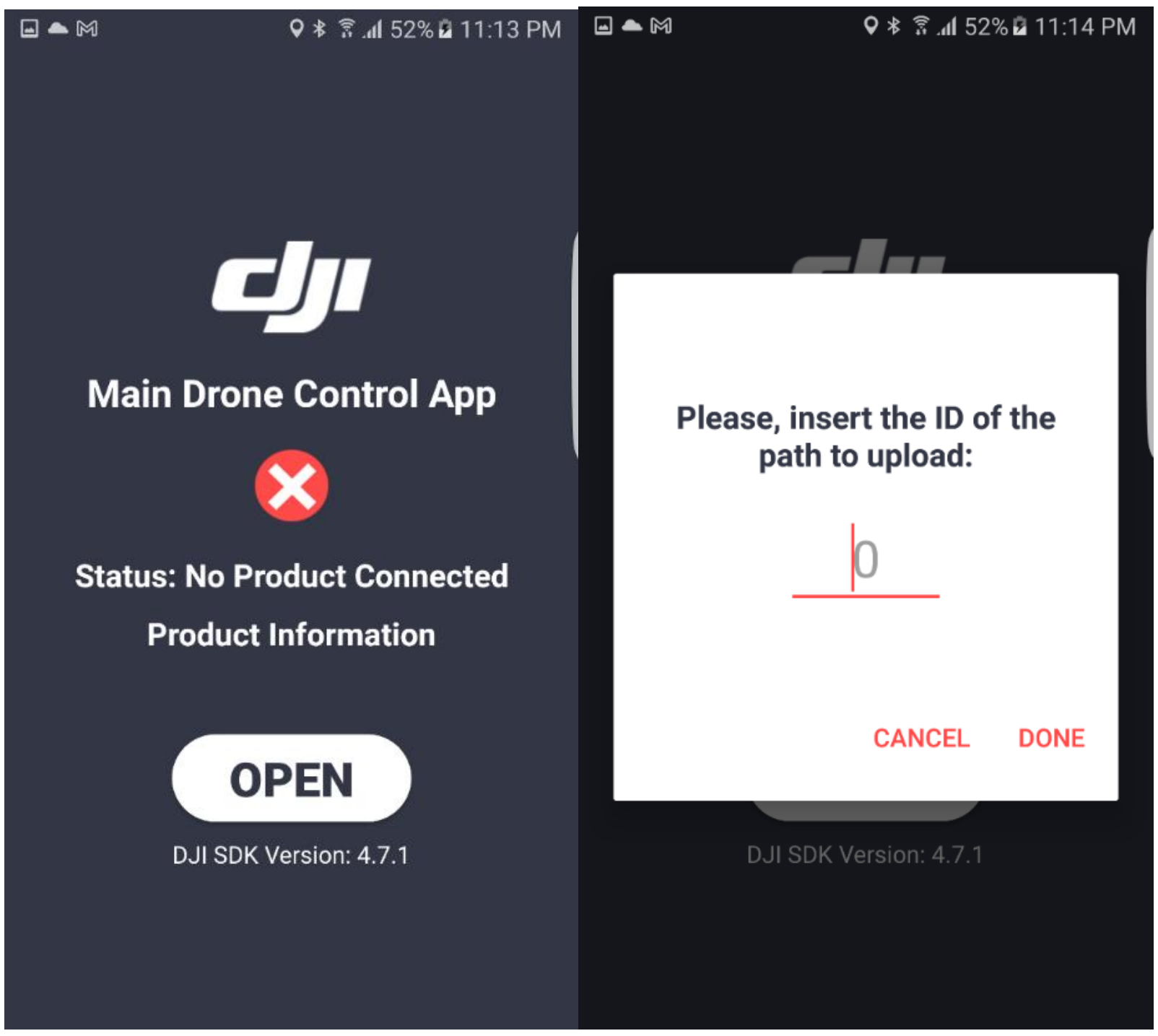

*Figura 19. Vistas de la actividad Connection*

Una vez ingresado el identificador, la actividad principal es ejecutada y se despliega un mapa con los puntos de la ruta cargada desde la base de datos. En esta vista (Fig. 20), el usuario puede navegar por el mapa e interactuar con la ruta de vuelo. Así mismo, puede configurar los parámetros de vuelo e indicarle al dron que inicie o detenga la ejecución de la ruta.

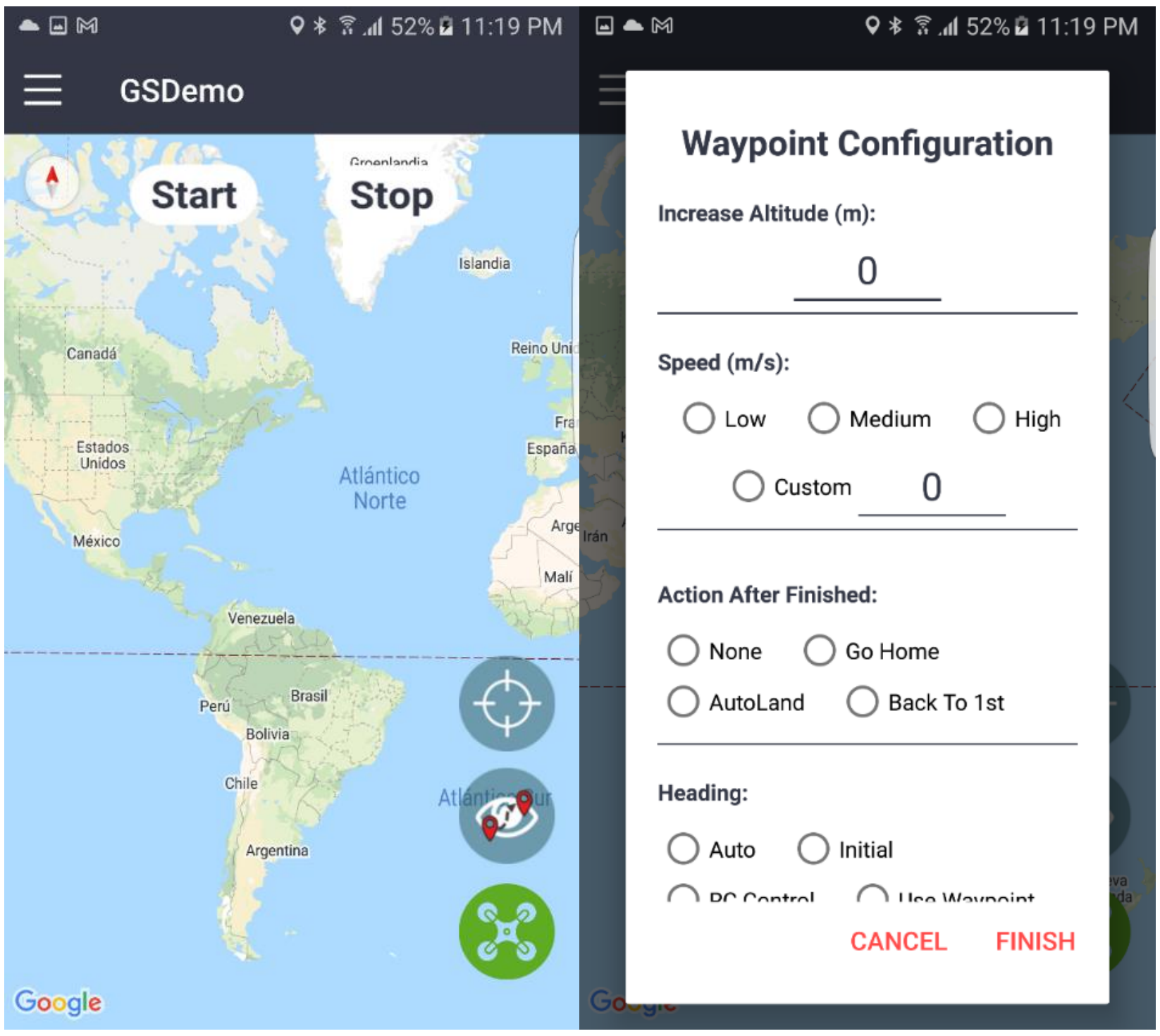

*Figura 20. Vistas de visualización y configuración de rutas de la actividad principal*

De igual manera, el usuario puede consultar información acerca de la ruta cargada desde la base de datos, así como de la ejecución de la ruta realizada por el dron (Fig. 21).

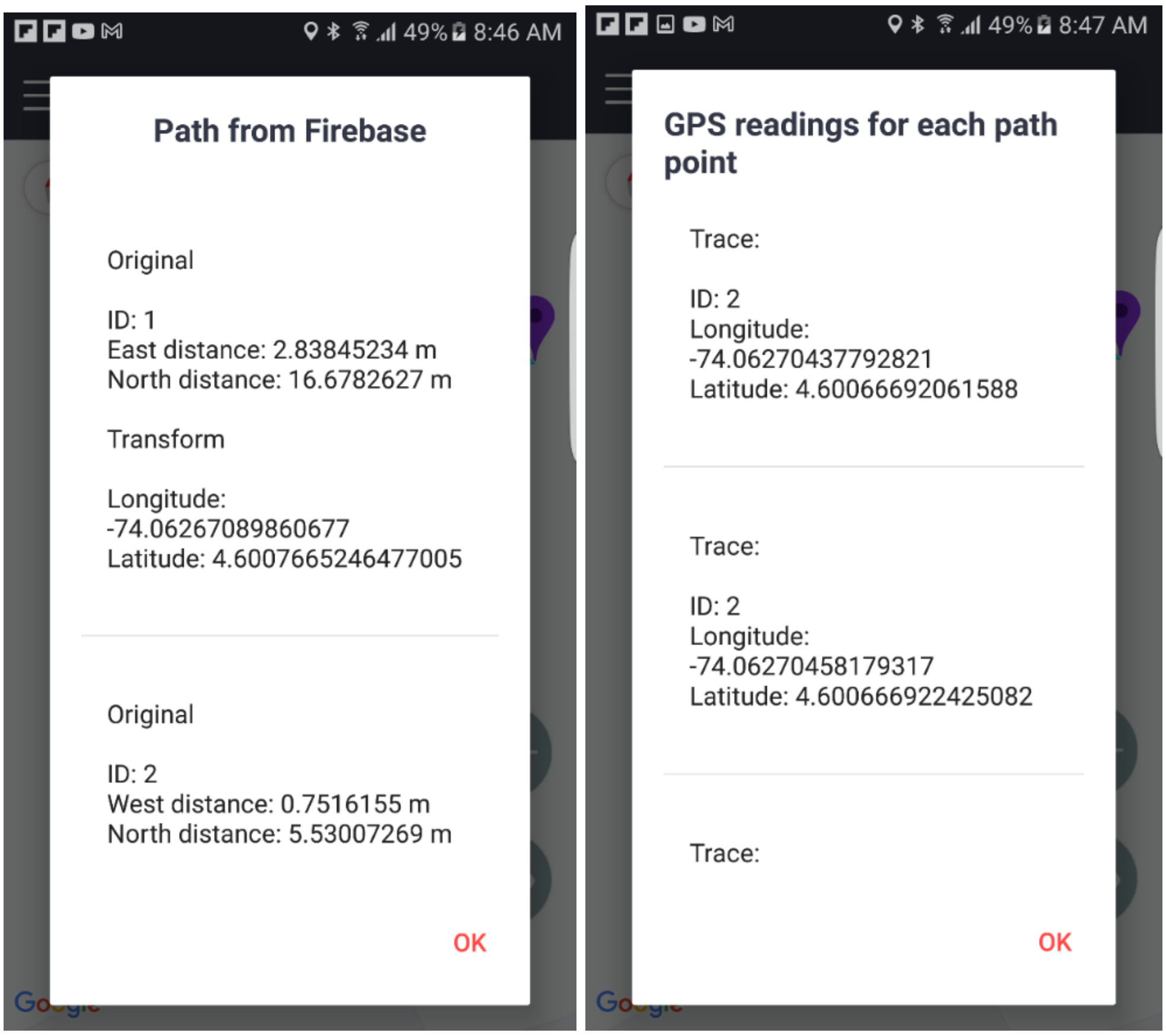

*Figura 21. Vistas de información de la actividad principal*

Para el desarrollo de la arquitectura diseñada, se implementaron tres nuevas actividades que enriquecen el funcionamiento de la app, las cuales se describen a continuación:

**CameraTasksActivity**: Esta actividad se encarga de realizar la configuración de las tareas asociadas a la cámara. Para esto, se despliega un mapa con los puntos asociados a la ruta, en el cual, al seleccionar un punto, se le puede asignar una tarea.

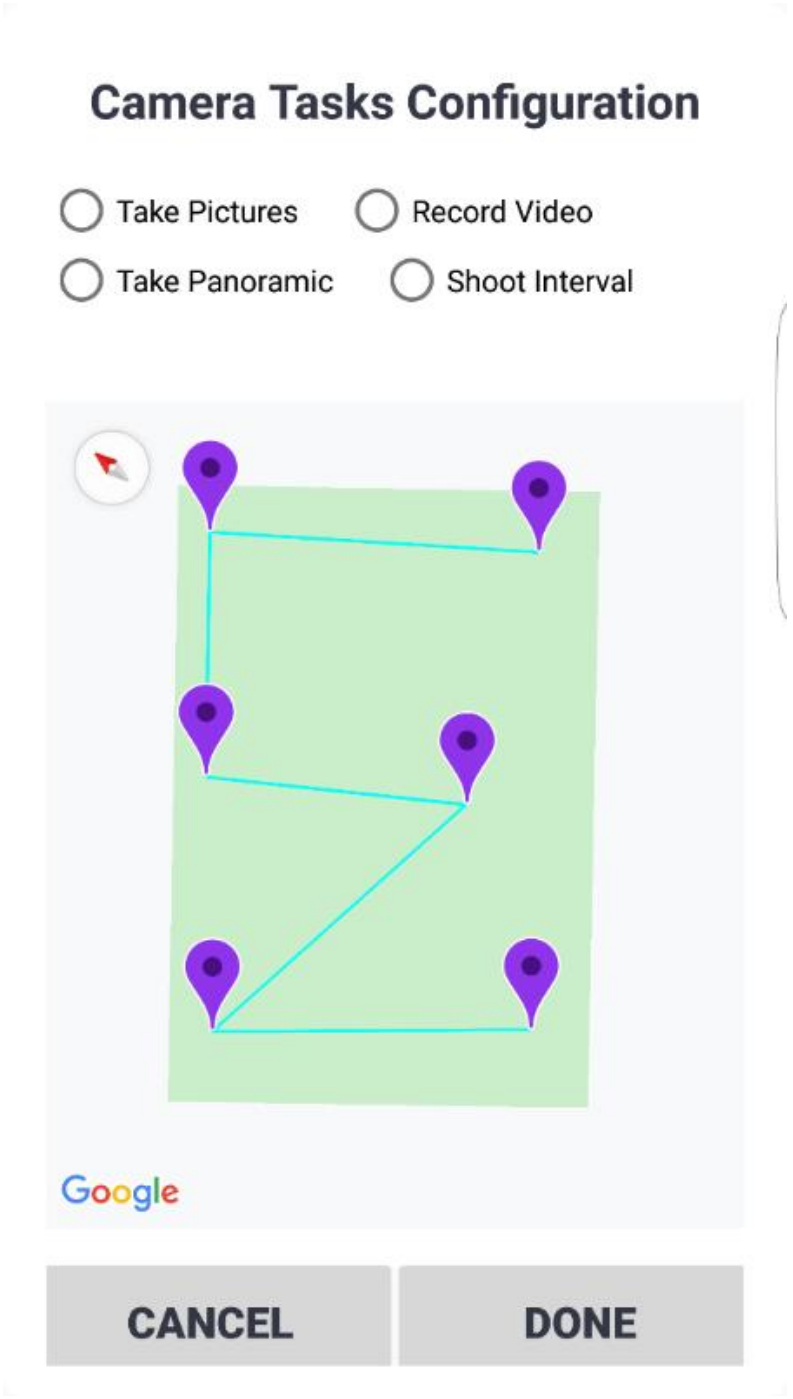

*Figura 22. Vista de la actividad CameraTasks*

Como se puede apreciar en la anterior vista (Fig. 22), el usuario tiene la posibilidad de seleccionar las tareas que desee ejecutar y los puntos sobre los cuales se realizarán. En la siguiente figura (Fig. 23) se muestran los puntos a los que el usuario ha asignado tareas:

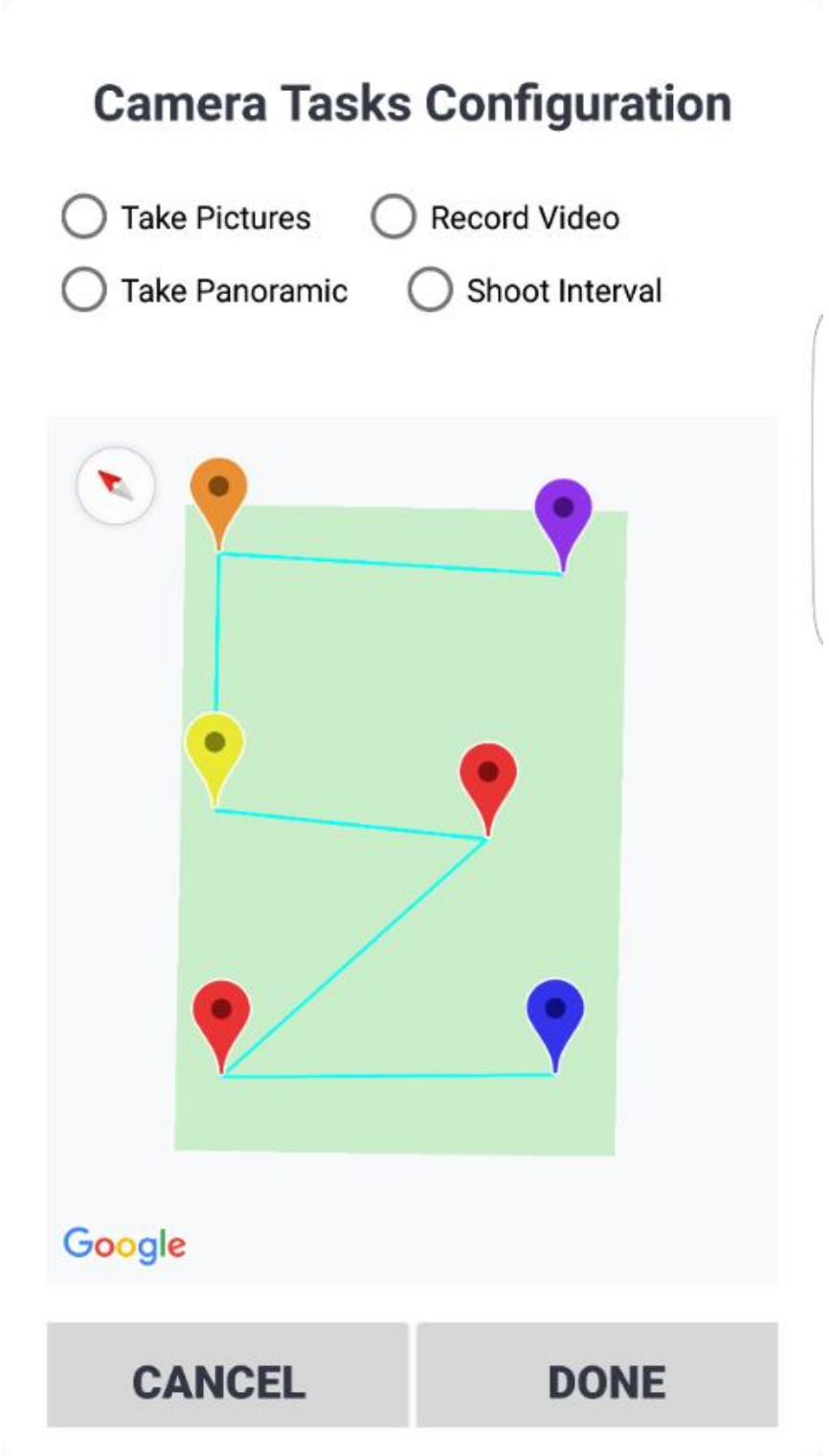

*Figura 23. Vista de la actividad CameraTasks con tareas seleccionadas*

Por convención, para la aplicación móvil, se definió un esquema de colores asociados a los marcadores de mapa para cada tarea de cámara. Este esquema se muestra a continuación (Fig. 24):

| None | Take Picture | Record Video | Take Panoramic | Shoot Interval |
|---|---|---|---|---|
| | | | | |

*Figura 24. Esquema de colores asociado a tareas de cámara*

Posteriormente, cuando el usuario ha definido las tareas, se puede oprimir el botón "Done" para finalizar la configuración de tareas. La selección de tareas es enviada devuelta a la actividad principal MainActivity, la cual actualiza la lista de WayPoints y agrega las tareas asociadas a cada uno. Así mismo, los marcadores de ruta del mapa de esta actividad se actualizan, como se puede observar en la siguiente imagen (Fig. 25):

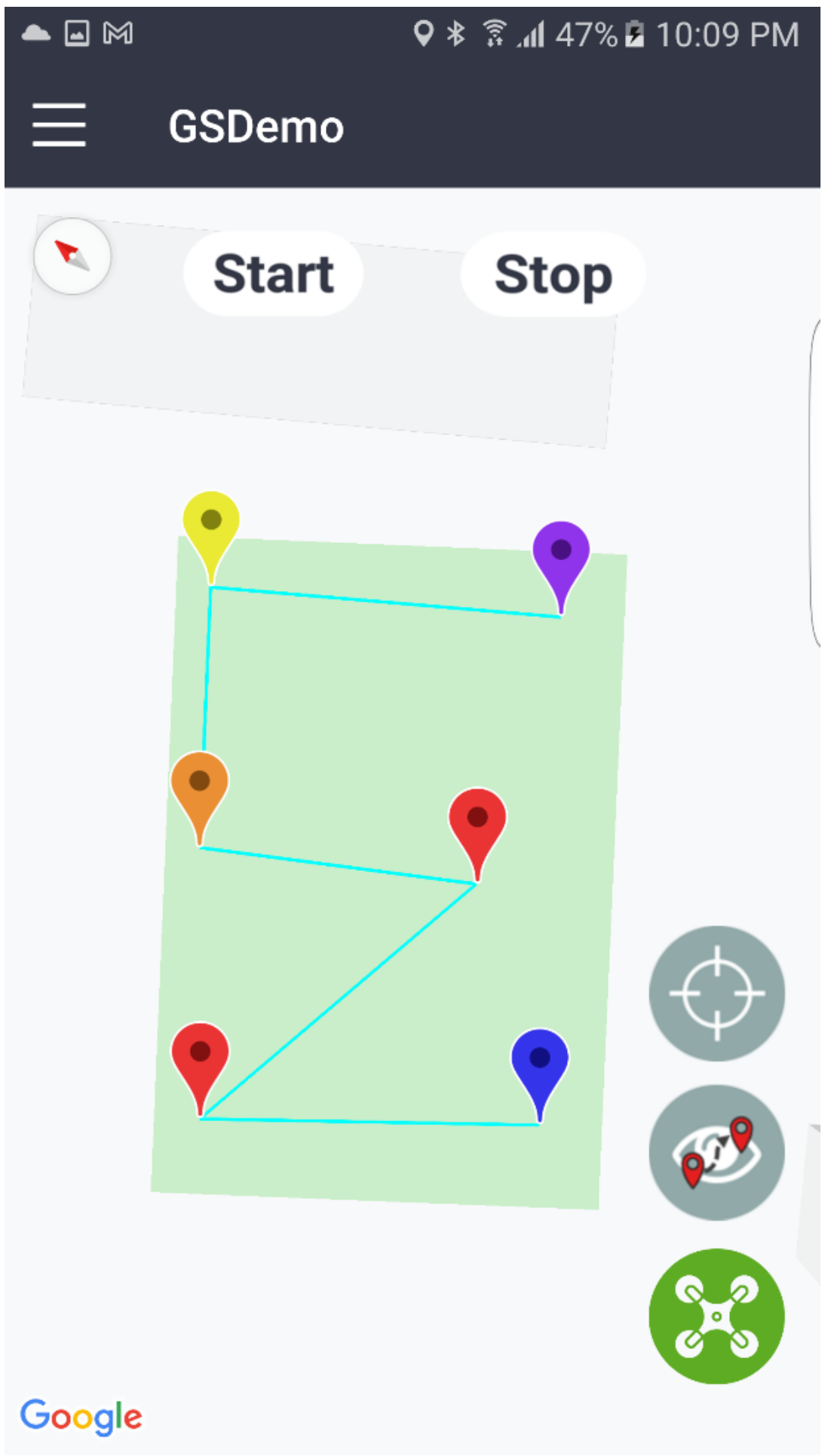

*Figura 25. Vista de la actividad principal con tareas de cámara actualizadas*

Finalmente, el usuario puede terminar de configurar los parámetros de vuelo e iniciar la ruta. Si se configuraron tareas de cámara, estas serán ejecutadas por el dron en los WayPoints seleccionados por el usuario. Al finalizar la ejecución de la ruta, el usuario puede descargar al celular los archivos multimedia generados por las tareas de cámara. Así mismo, estos archivos se encuentran en la memoria extraíble del dron, y pueden ser accedidos por medio de un ordenador conectado al dron o extrayendo la memoria.

**LiveFeedActivity**: Esta actividad permite observar lo que la cámara del dron captura en tiempo real. Así mismo, si no se han configurado tareas de cámara en la ruta cargada, permite al usuario capturar imágenes y video por medio de la cámara del dron.

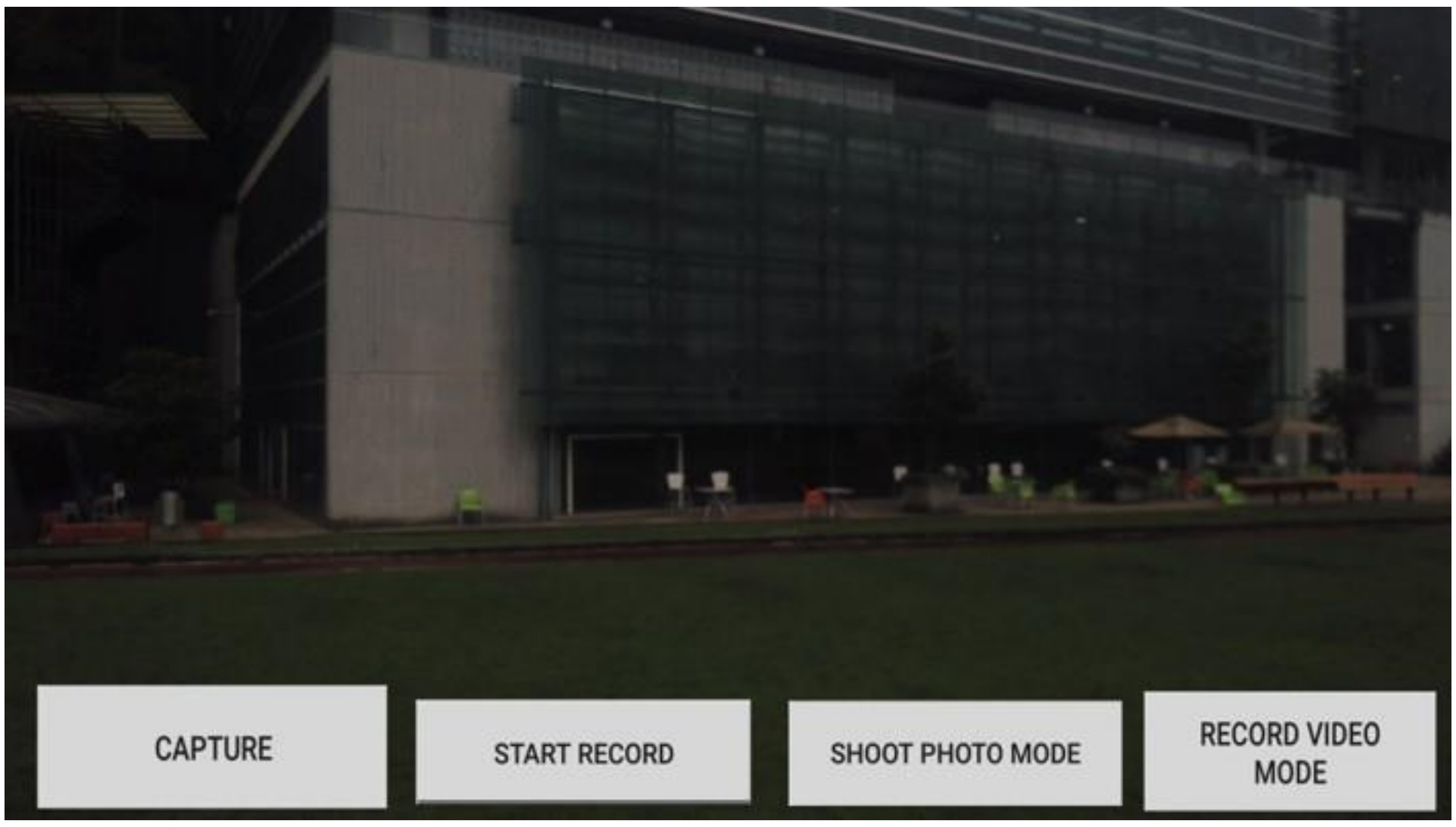

*Figura 26. Vista de la actividad LiveFeed*

Como se puede observar en la imagen (Fig. 26), además de la transmisión en vivo de la cámara del dron, se le puede indicar al dron que capture imágenes y video. Sin embargo, esto únicamente se encuentra habilitado si no se han configurado tareas de cámara en la ruta, con el fin de que estas se ejecuten sin problema. En caso de que haya tareas configuradas, la barra de botones desaparece y únicamente se muestra la transmisión de la cámara del dron.

**PlanningActivity**: Esta actividad permite al usuario crear una ruta de vuelo desde la aplicación móvil. Una vez iniciada la actividad, el usuario podrá seleccionar puntos en el mapa sobre los cuales se trazará la ruta. La vista de esta actividad se puede apreciar en la siguiente imagen (Fig. 27):

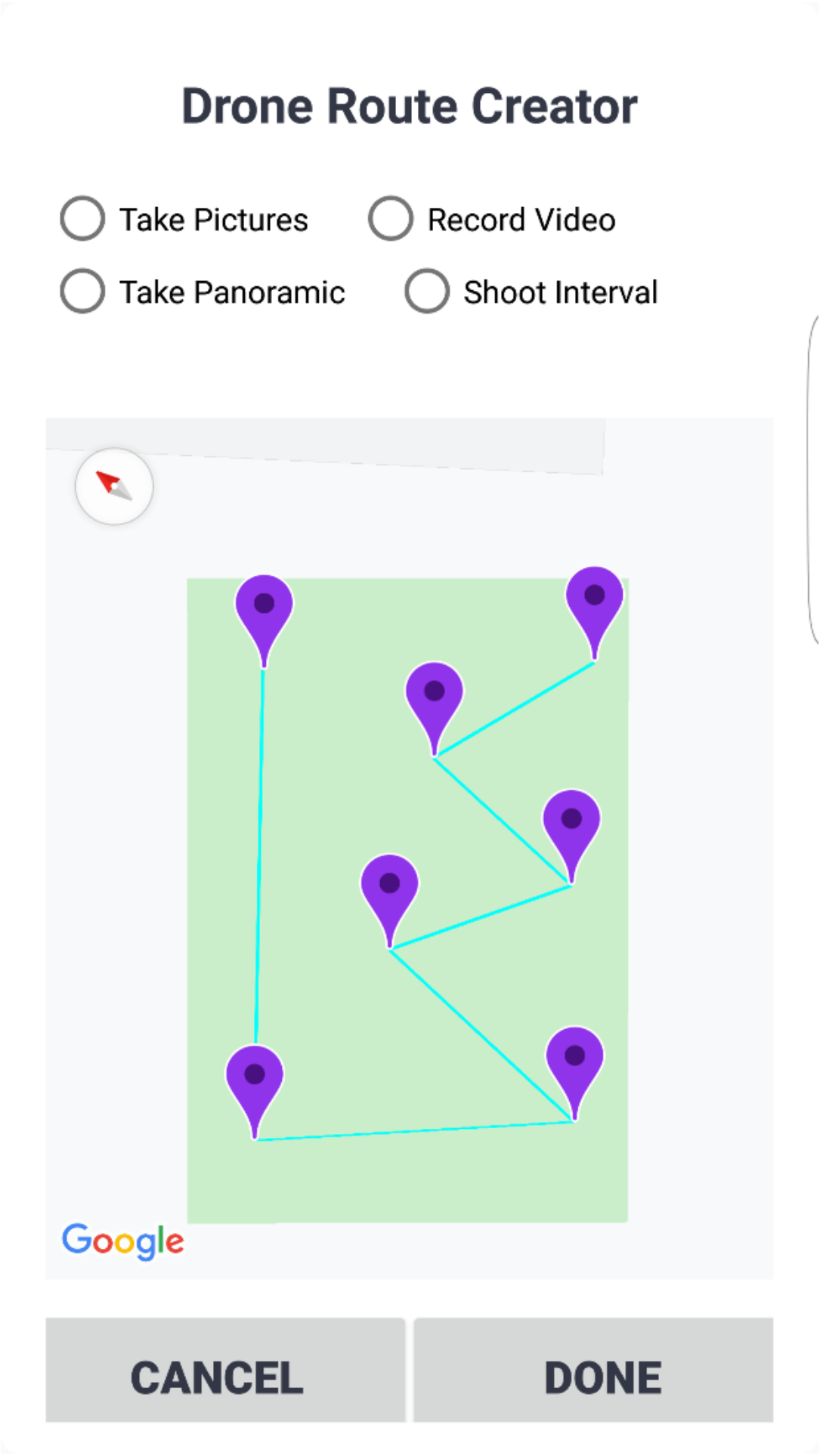

*Figura 27. Vista de la actividad Planning con marcadores creados por él usuario.*

Así mismo, el usuario puede configurar tareas de cámara para cada marcador que se genere. Se puede modificar marcadores existentes o agregar nuevos con una tarea ya definida. A continuación, se muestra una ruta creada con marcadores con tareas asignadas (Fig. 28)

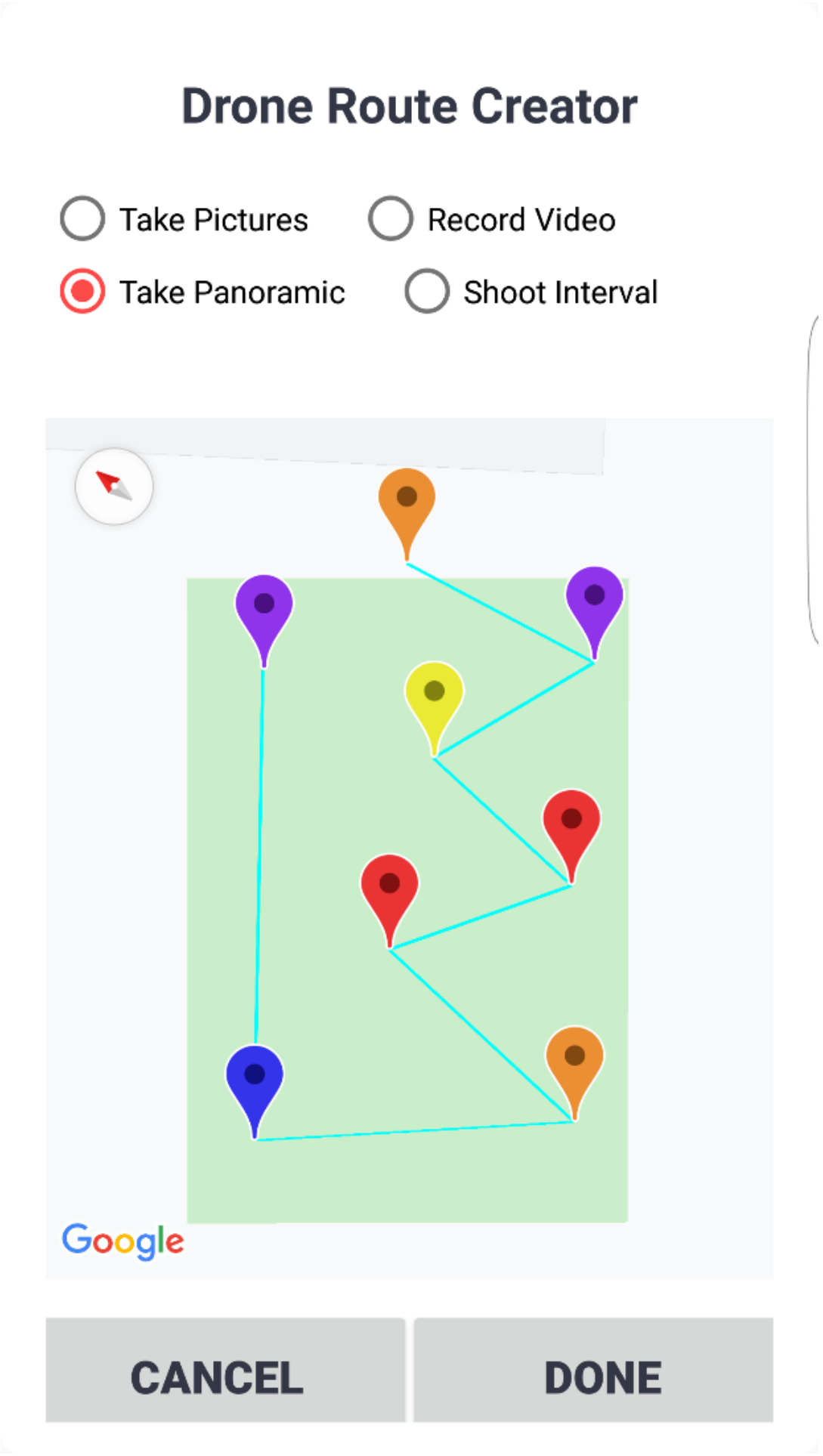

*Figura 28. . Vista de la actividad planning con marcadores con tareas asignadas*

Finalmente, cuando el usuario termine la creación, la información será enviada de regreso a la actividad principal y se actualizará el mapa con los marcadores creados por el usuario. En este punto, la ruta será persistida en la base de datos y el usuario podrá continuar con la configuración y ejecución de la ruta de vuelo.

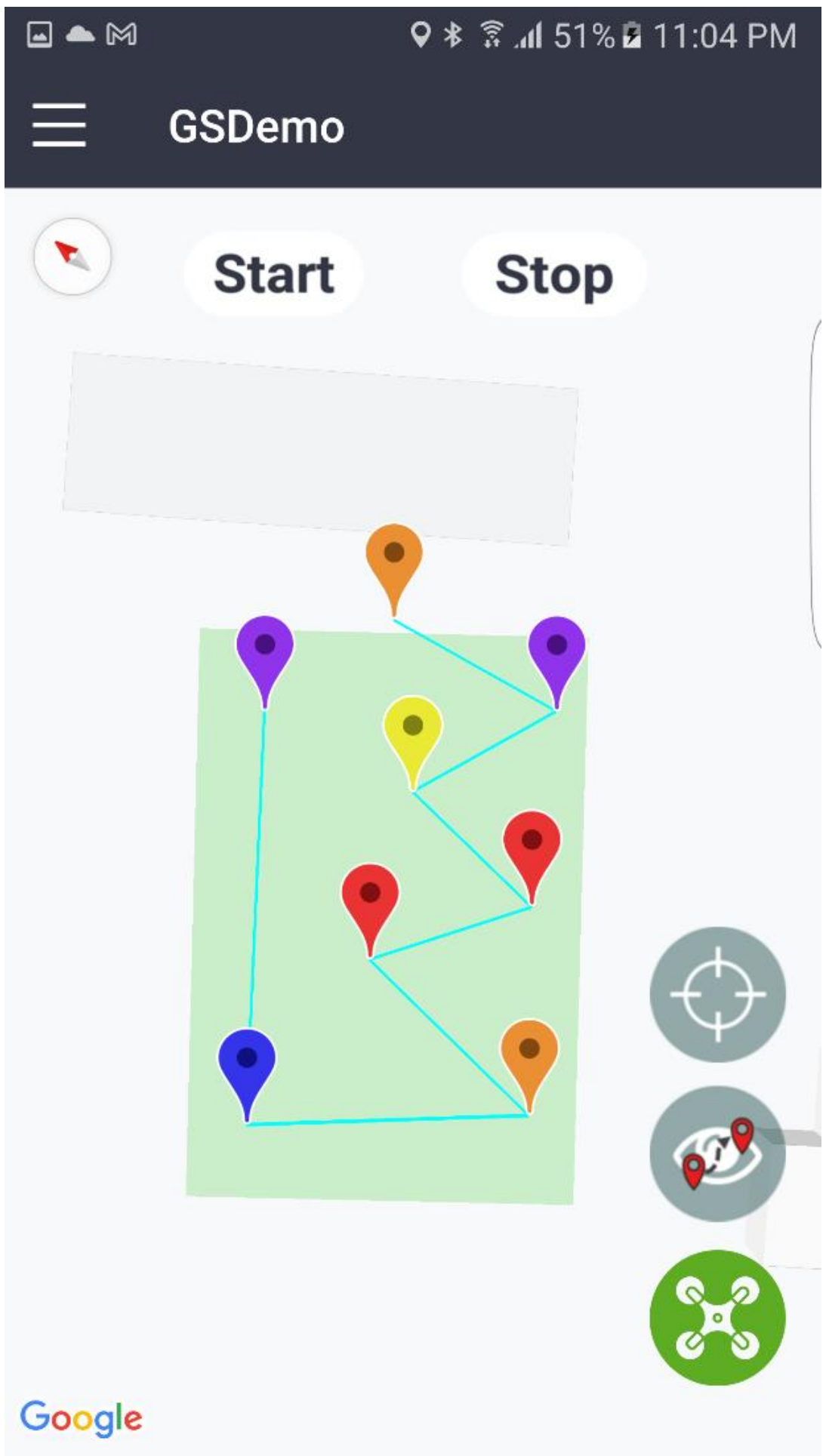

*Figura 29. Vista de la actividad principal con la ruta creada por él usuario*

Estas actividades y vistas implementadas pueden ser accedidas por el usuario por medio del menú con el que cuenta la aplicación. Aquí, el usuario puede elegir qué actividad ejecutar para complementar la configuración de la ruta. En la siguiente imagen se encuentran las opciones del menú, resaltando las implementadas en este proyecto:

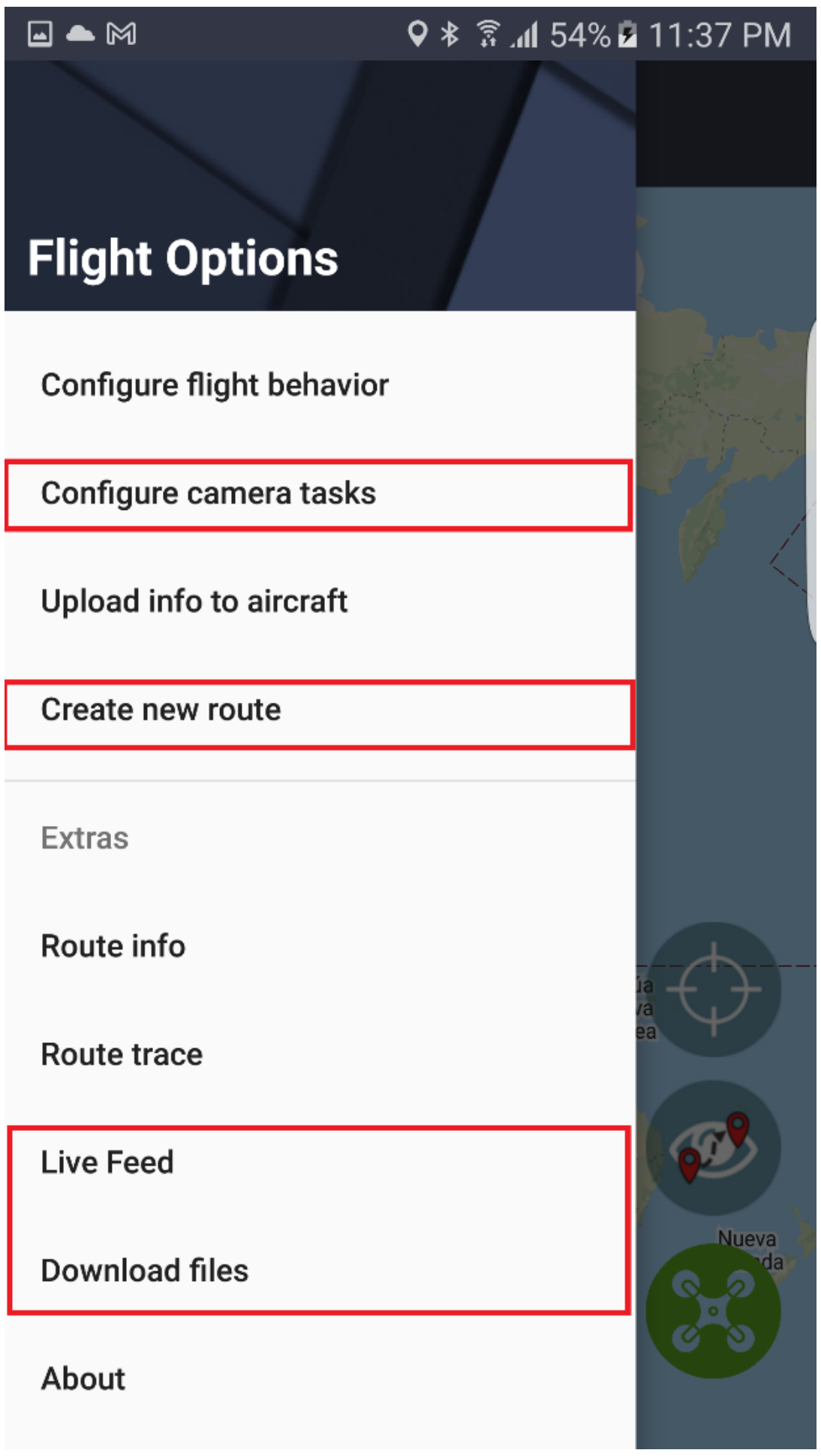

*Figura 30. Menú de opciones de la aplicación*

## 7.3 Extensión del esquema de persistencia

El esquema actual de persistencia solo se está considerando objetos de tipo Ruta los cuales están compuestos por una serie de waypoints. A continuación, se muestra la representación de una ruta de prueba con id 1, la cual tiene 3 waypoints.

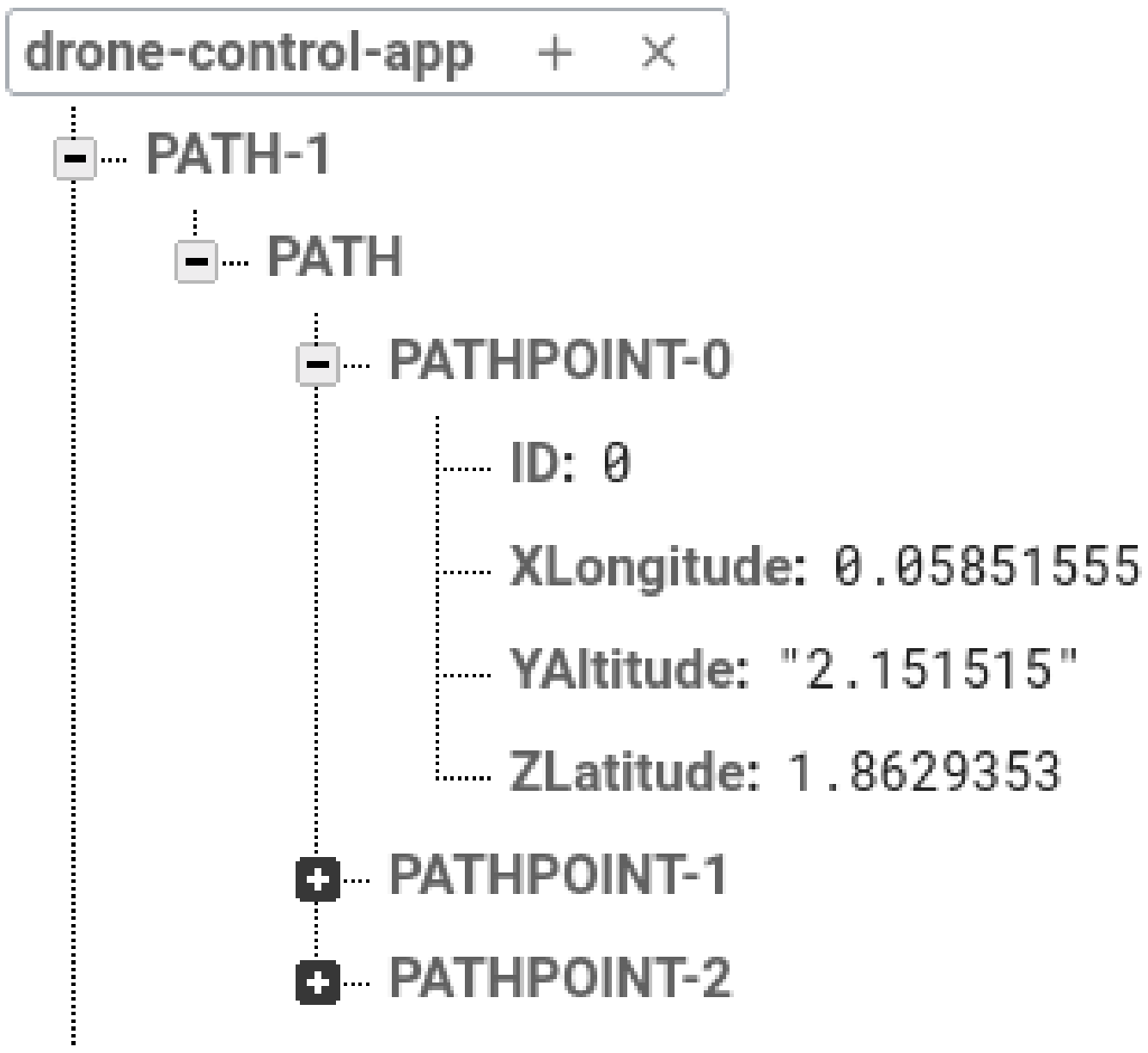

*Figura 31. Representación de una ruta con el esquema de la arquitectura previa*

Dado que la base de datos es de tipo no relacional, una ruta se representa como un objeto, que tiene como nombre PATH-id donde id es el identificador único de la ruta, este documento u objeto tiene un documento hijo llamado PATH que representa la lista de puntos y, por tanto, tiene como documentos hijos a objetos PATHPOINT-order, donde order representa el id del waypoint y también el orden de ejecución, siendo order=0 el waypoint inicial. En el último nivel de esta representación estarían ubicados los atributos de un documento PATHPOINT los cuales son el id del waypoint, la longitud en grados denotada como XLongitude, la latitud en grados denotada como ZLatitude y finalmente YAltitude que representa la altura en metros. Aunque el esquema actual es suficiente debemos extenderlo con el fin de incluir una descripción de una ruta, y tareas de cámara asociadas a waypoints. A continuación, se muestra un diagrama de clases que busca representar esta lógica:

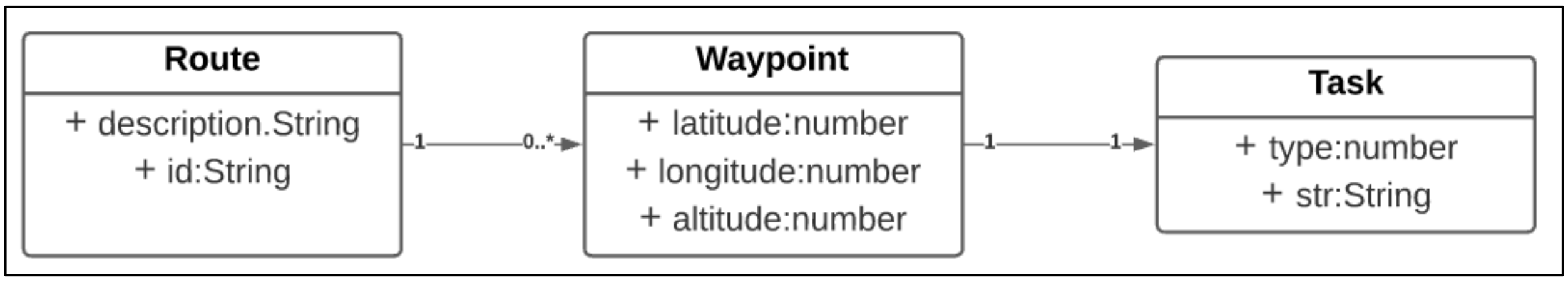

*Figura 32. Esquema extendido de la base de datos para soportar persistencia de tareas de cámara asociadas a puntos en la ruta*

En la implementación se tomó la decisión de que la representación de un objeto PathPoint, además de tener los atributos mencionado en el esquema anterior, se incluyeran los atributos “task” e “instruction” que son equivalentes a los atributos de la clase Tas, respectivamente, del diagrama de clases anterior. La definición de la clase PathPoint para la arquitectura extendida quedaría de la siguiente manera:

```
export class PathPoint {
    constructor(public ID: number,
                public ZLatitude: number,
                public XLongitude: number,
                public YAltitude: number = 0,
                public task: string = '0',
                public instruction: string ='' )
```

*Figura 33. Definición de la clase PathPoint para la extensión de la arquitectura*

Como se puede ver, una tarea está representada con un código numérico, siendo el código por defecto task=”0” que representa que un waypoint cuando es creado no tiene una tarea de cámara asociada.

Con el fin de que los valores posibles para el atributo de tarea sean representativos en la lógica y en la interfaz de usuario, se definió la siguiente interfaz para que todo posible atributo que se almacene en el campo de tarea no solo tenga el código respectivo, pero también un nombre o descripción de la tarea:

```
interface Task {
  value: string;
  viewValue: string;
}
```

*Figura 34. Interfaz para valores de una tarea de cámara*

En el componente Path.Component se definió el siguiente arreglo de valores posibles que el usuario puede elegir para ser asociados a un punto:

```
//Opciones de tareas para los diferentes pathpoints
//value es lo que se guarda como atrubuto
//viewValue es lo que se muestra en pantalla y representa el valor
tasks: Task[] = [
  {value: '0', viewValue: 'do nothing'},
  {value: '1', viewValue: 'Take Picture'},
  {value: '2', viewValue: 'Start video'},
  {value: '3', viewValue: 'Start interval'},
  {value: '4', viewValue: 'Take Panorama Picture'}
];
```

*Figura 35. Definición de las posibles tareas de una cámara, sus códigos y significado*

Además de que una tarea tenga un código que la representa y una cadena de texto que es explica la tarea, también se define tiene un color asociado para que no solo haya una guía semántica sino también una guía visual para la asignación de tareas. Como se ve en la siguiente figura, una tarea con código “0” (“do nothing”) tiene asociado un color negro, una tarea con código “1” (“take picture”) tiene asociado el color azul, una tarea con código “2” (“start video”) tiene asociado el color rojo. una tarea con código “3” (“start Interval”) tiene asociado el color verde y finalmente una tarea con código “4” (“take panorama picture”) tiene asociado el color amarillo. El atributo “svgColor” es una variable que será pasada al icono del marcador de un waypoint.

```
//configurar color segun tarea
let svgColor = "#000000";//negro para waypoints sin tareas
if(currentPoint.task == '1') svgColor = "blue";
if(currentPoint.task == '2') svgColor = "red";
if(currentPoint.task == '3') svgColor = "green";
if(currentPoint.task == '4') svgColor = "yellow";
```

*Figura 36. Asociación de colores a las tareas de en la lógica*

Por otra parte, para responder al requerimiento de que el usuario pueda agregar una descripción en la ruta, se agregó un atributo opcional a la representación lógica de una ruta, aparte de su lista de waypoints. La siguiente figura muestra esta extensión:

```
/**
 * Clase que representa la planeacion de una ruta
 */
export class Path {
    constructor(public description?: string,
                public PATH: PathPoint[]=[])
    {}
```

*Figura 37. Definición de la clase Path para representar una ruta en la arquitectura extendida*

Finalmente se muestra cómo se representan las rutas en la base de datos con esta extensión del esquema. La ruta de prueba con id 97 tiene 5 waypoints y una descripción de la misma, se puede ver que una vez el dron llegue al waypoint inicial y a la altura definida, está programado para tomar una fotografía (task = “1”):

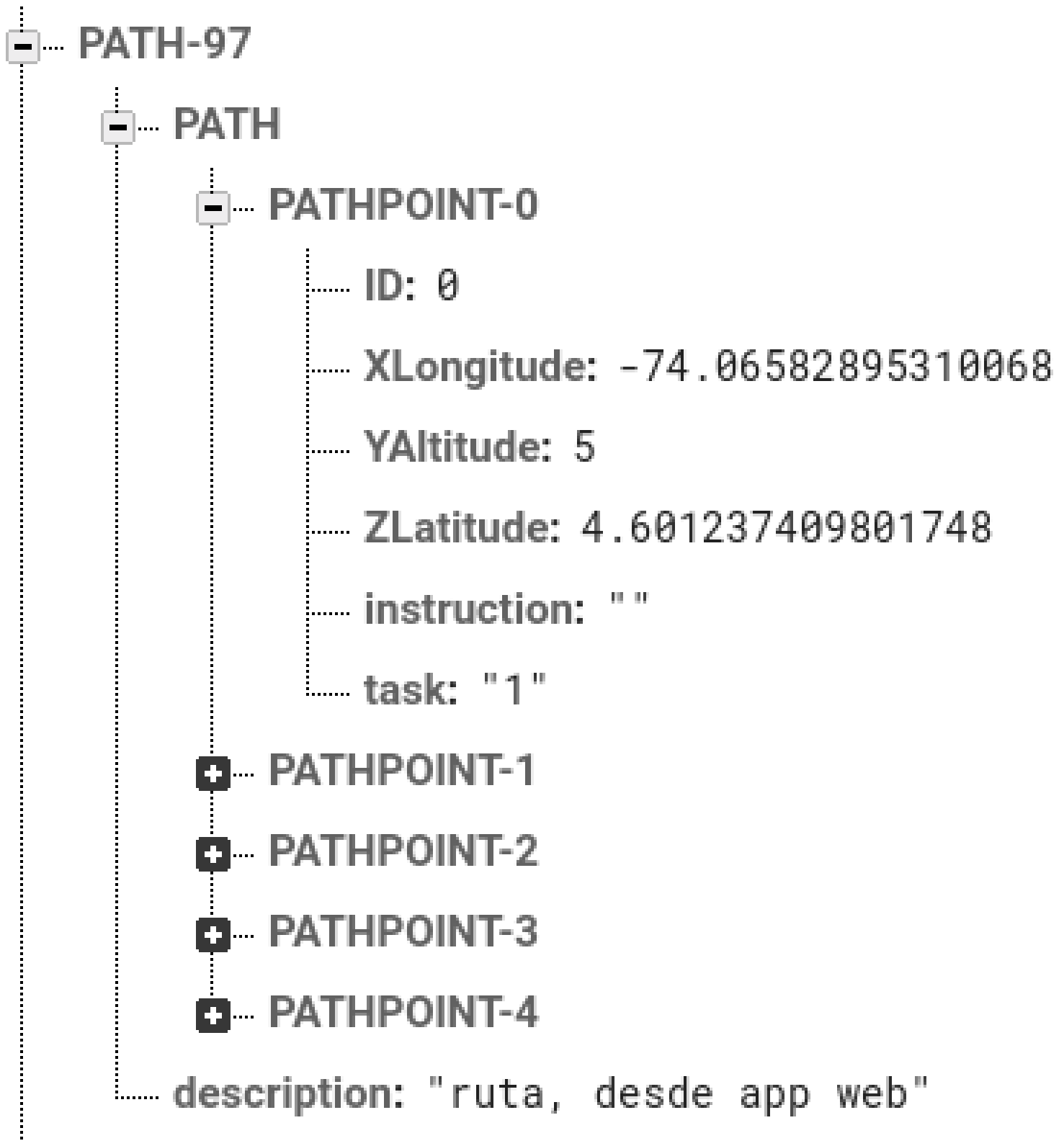

*Figura 38. Representación de una ruta con el esquema de la arquitectura extendido*

La sección anterior completaría la descripción de lo que respecta a la extensión de la persistencia, ahora se hará una descripción de los detalles de implementación de cada uno de los requerimientos y se mostrara el resultado que el usuario verá en la interfaz gráfica del respectivo componente.

## 7.4 Implementación de requerimientos

### RF1: Planear captura de foto

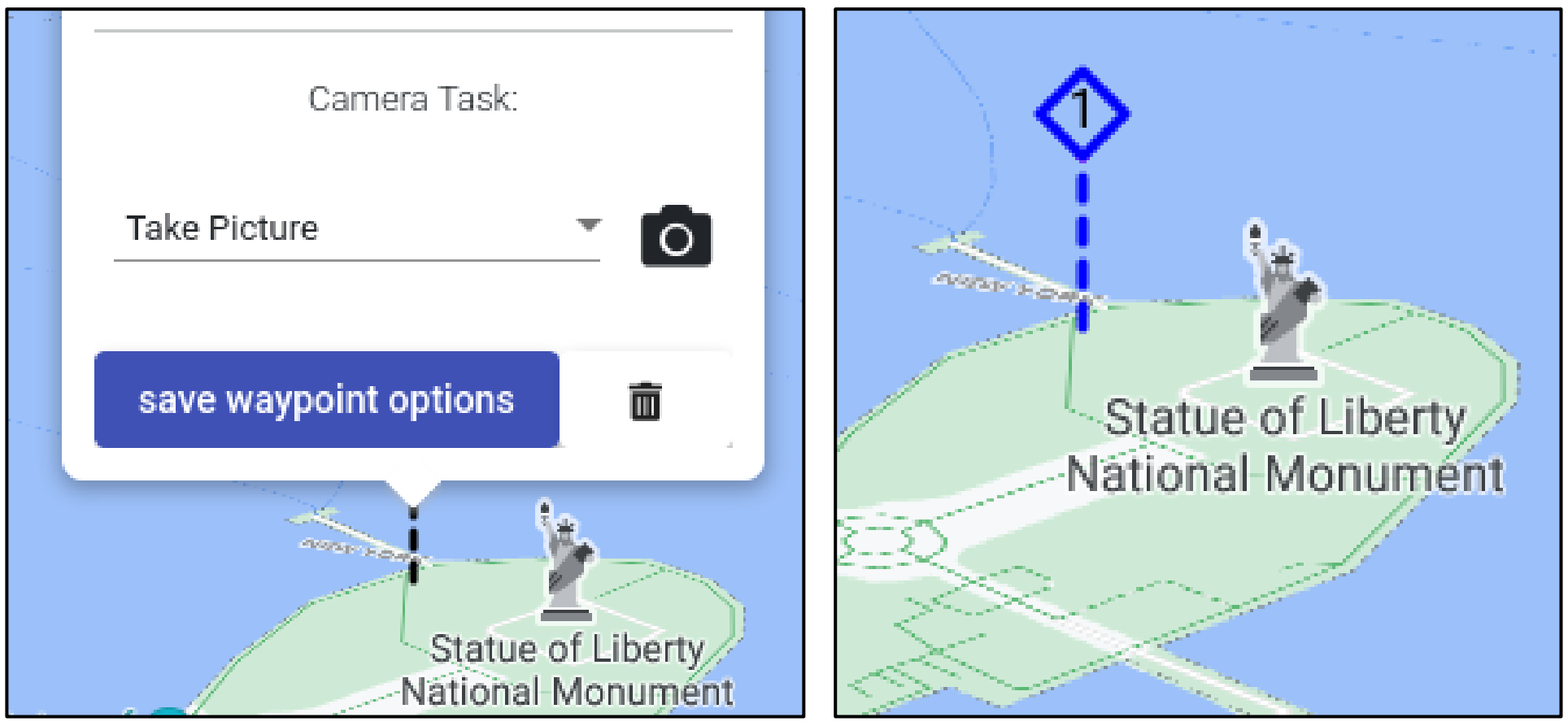

*Figura 39. RF1 Asignar tarea de foto a un waypoint (Entorno Web)*

Como se puede observar en la figura 39, el usuario puede ingresar a las opciones de un waypoint seleccionar “take picture”. Luego presionar “save waypoint options”, se guarda localmente la configuración de la tarea y gráficamente el icono elevado el waypoint cambia a color azul, que significa tomar fotografía en ese punto.

Por medio de la aplicación móvil, el usuario puede configurar la tarea de capturar foto como se observa en la figura 40. Para esto, el usuario debe seleccionar primero la opción “Take Pictures” en la parte superior de la vista. Posteriormente, debe seleccionar a que marcadores desea asignar la tarea y presionarlos. Estos marcadores cambiarán su color a azul. Por último, debe presionar el botón “Done” y la tarea será cargada al dron.

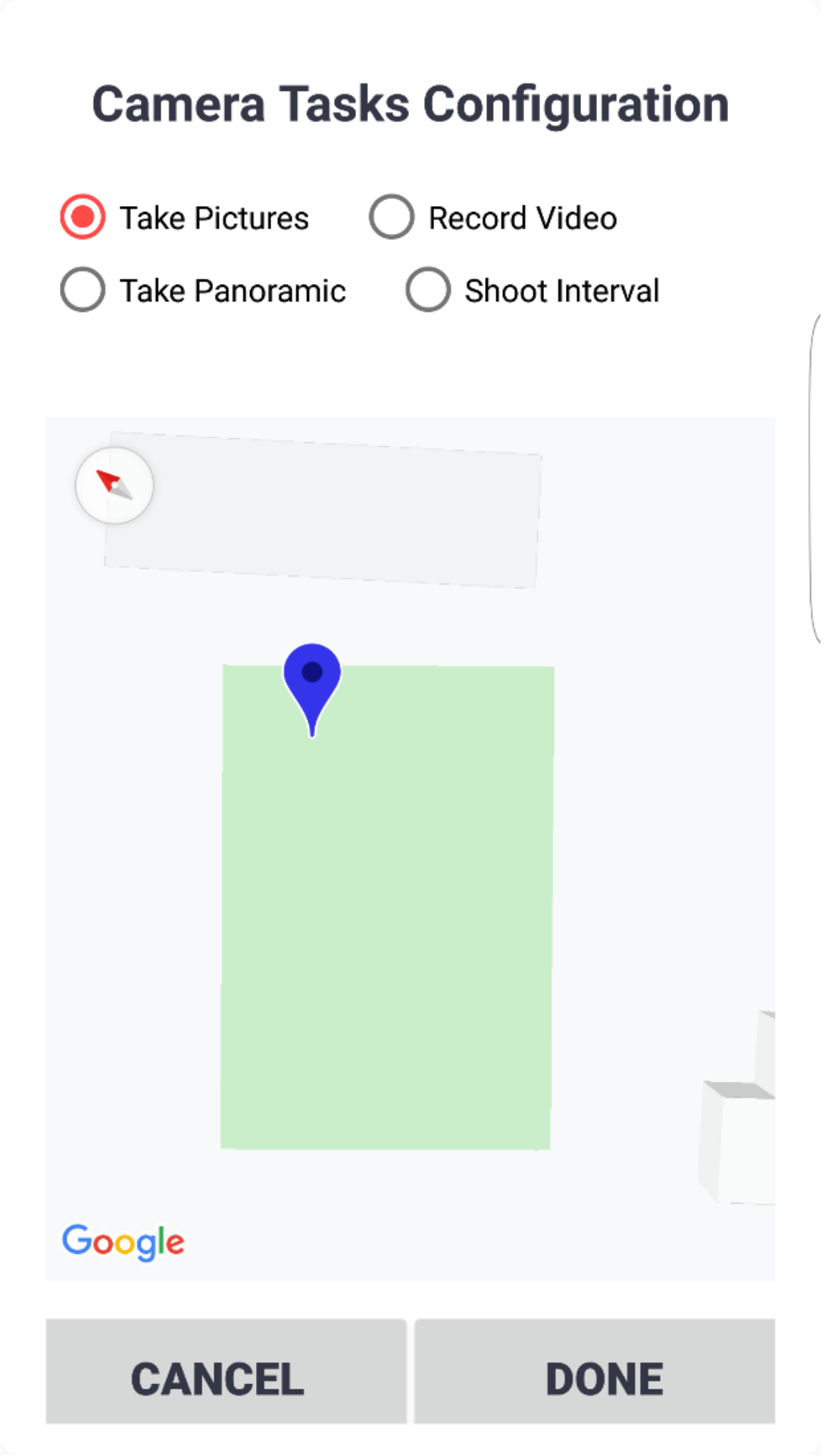

*Figura 40. RF1 Asignar tarea de foto a un waypoint (Entorno móvil)*

## RF2: Planear inicio de video

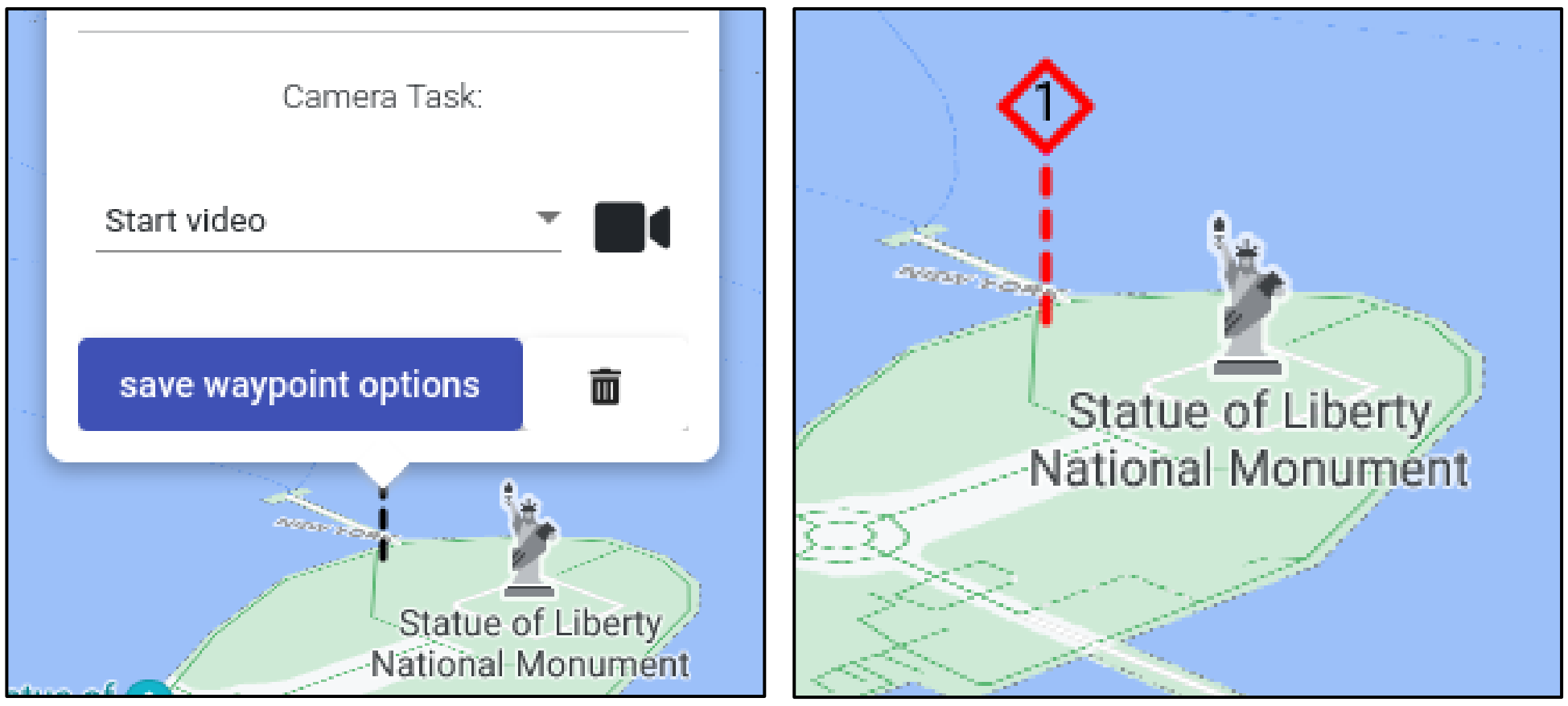

*Figura 41. RF2 Asignar tarea de inicio de video a un waypoint (Entorno Web)*

Como se puede observar en la figura 41, el usuario puede ingresar a las opciones de un waypoint seleccionar start video. Luego presionar “save waypoint options”, se guarda localmente la configuración de la tarea y gráficamente el icono elevado el waypoint cambia a color rojo, que significa iniciar captura de video fotografía en ese punto.

Por medio de la aplicación móvil, el usuario puede configurar la tarea de inicio de video como se observa en la figura 42. Para esto, el usuario debe seleccionar primero la opción “Record Video” en la parte superior de la vista. Posteriormente, debe seleccionar a que marcadores desea asignar la tarea y presionarlos. Estos marcadores cambiarán su color a rojo. Por último, debe presionar el botón “Done” y la tarea será cargada al dron.

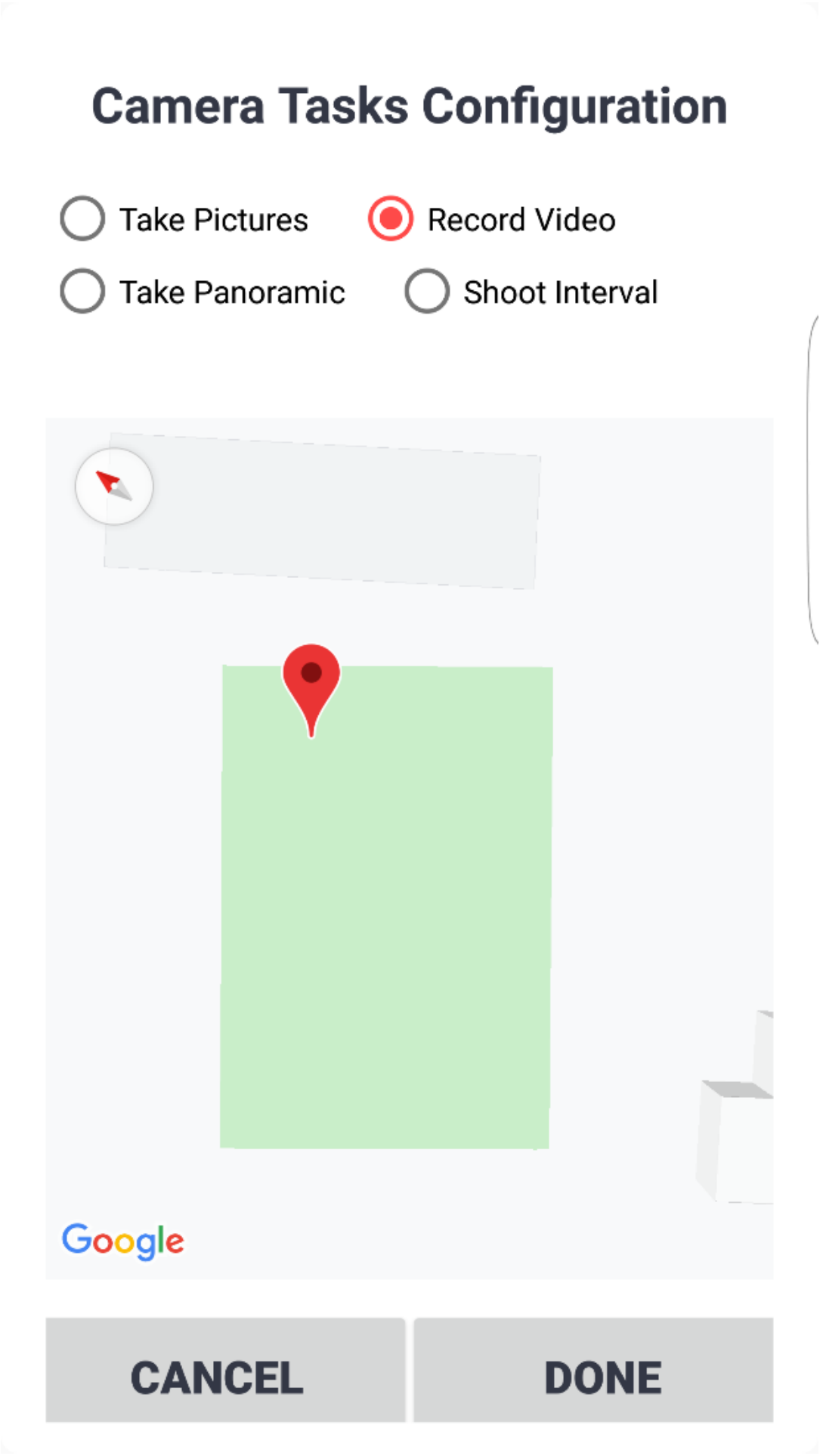

*Figura 42. RF2 Asignar tarea de inicio de video a un waypoint (Entorno móvil)*

## RF3: Planear captura de fotos en intervalo

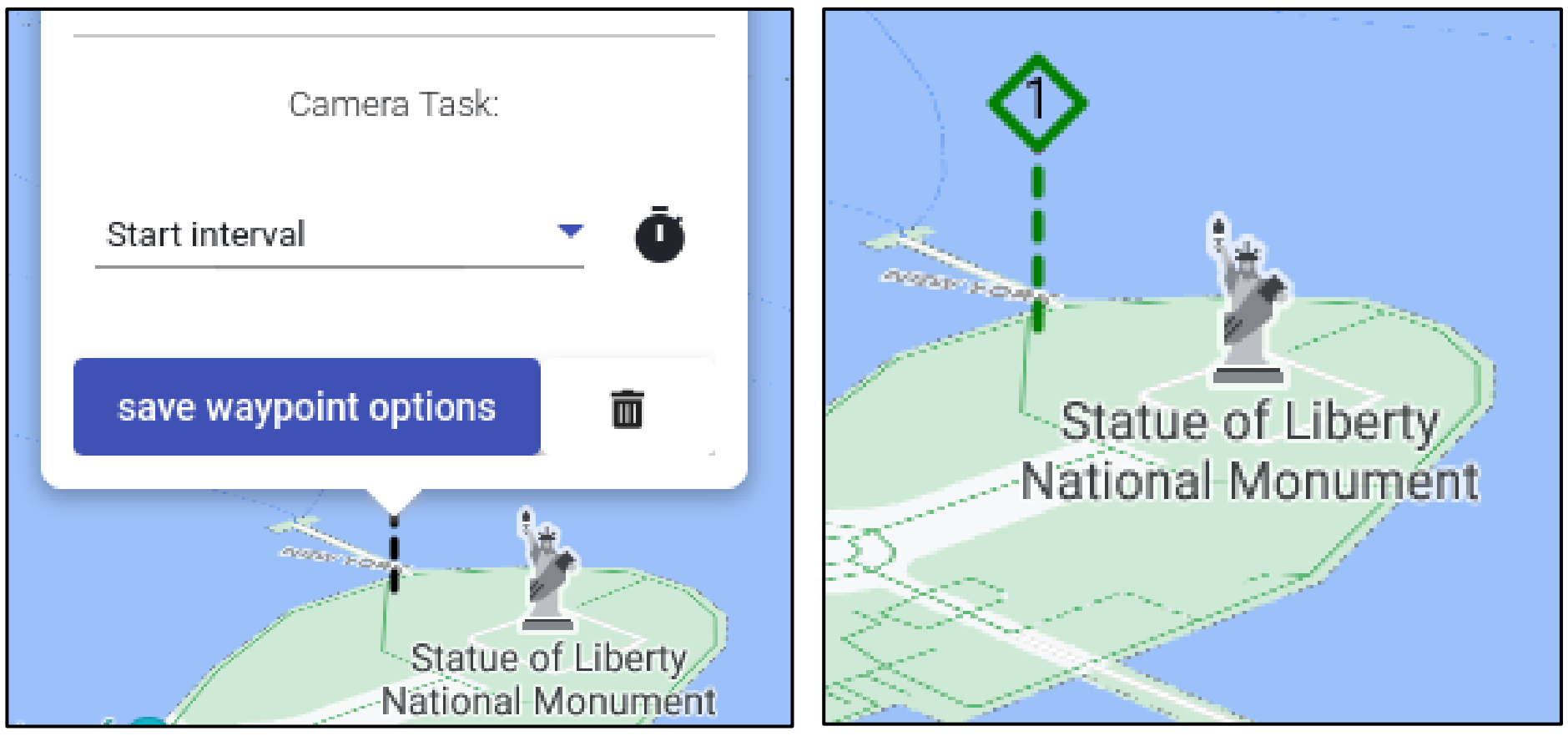

*Figura 43. RF3 Asignar tarea de inicio de captura de fotos en intervalo a un waypoint (Entorno Web).*

Como se puede observar en la figura 43, el usuario puede ingresar a las opciones de un waypoint seleccionar “start Interval”. Luego de presionar “save waypoint options”, se guarda localmente la configuración de la tarea y gráficamente el icono elevado el waypoint cambia a color verde, que significa iniciar captura de fotografía por intervalo desde ese punto.

Por medio de la aplicación móvil, el usuario puede configurar la tarea de captura de fotos en intervalo como se observa en la figura 44. Para esto, el usuario debe seleccionar primero la opción “Shoot Interval” en la parte superior de la vista. Posteriormente, debe seleccionar a que marcadores desea asignar la tarea y presionarlos. Estos marcadores cambiarán su color a anaranjado. Por último, debe presionar el botón “Done” y la tarea será cargada al dron.

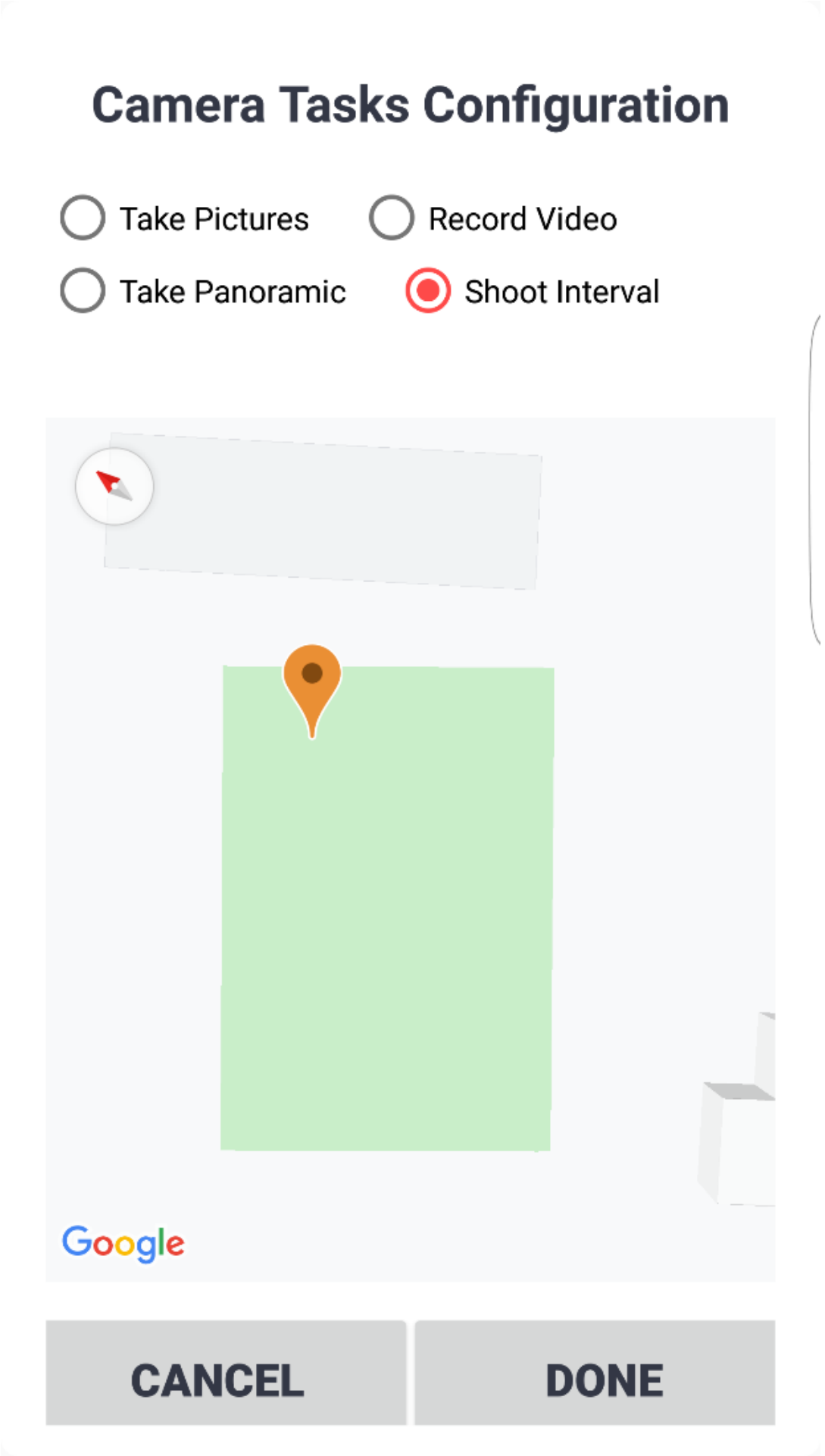

*Figura 44. RF3 Asignar tarea de inicio de captura de fotos en intervalo a un waypoint (Entorno móvil)*

## RF4: Planear captura de foto panorámica

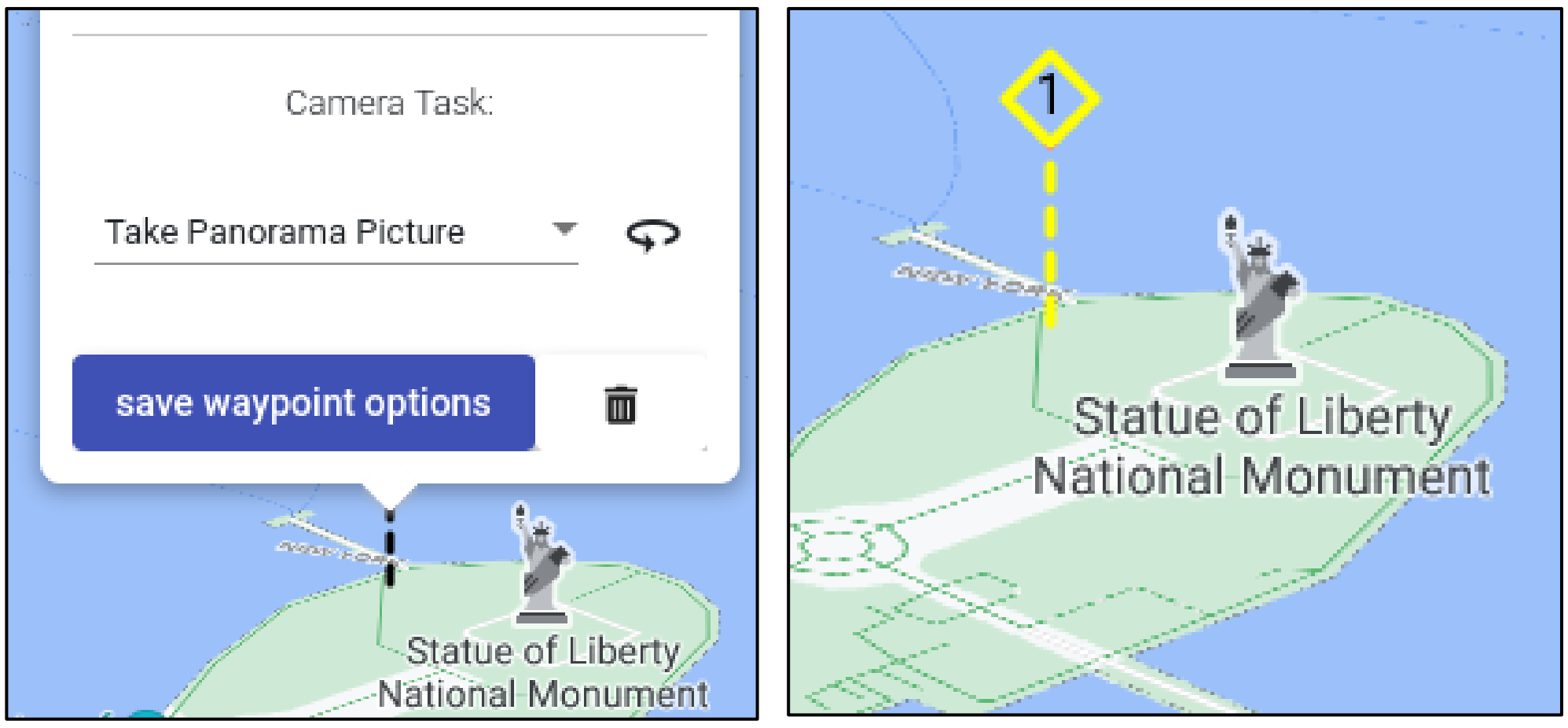

*Figura 45. RF4 Asignar tarea de captura de foto panorámica a un waypoint (Entorno Web)*

Como se puede observar en la figura 45, el usuario puede ingresar a las opciones de un waypoint seleccionar “take panorama picture”. Luego de presionar “save waypoint options”, se guarda localmente la configuración de la tarea y gráficamente el icono elevado el waypoint cambia a color amarillo, que significa tomar foto de 360 grados en ese punto.

Por medio de la aplicación móvil, el usuario puede configurar la tarea de captura de foto panorámica como se observa en la figura 46. Para esto, el usuario debe seleccionar primero la opción “Take Panorama” en la parte superior de la vista. Posteriormente, debe seleccionar a que marcadores desea asignar la tarea y presionarlos. Estos marcadores cambiarán su color a amarillo. Por último, debe presionar el botón “Done” y la tarea será cargada al dron.

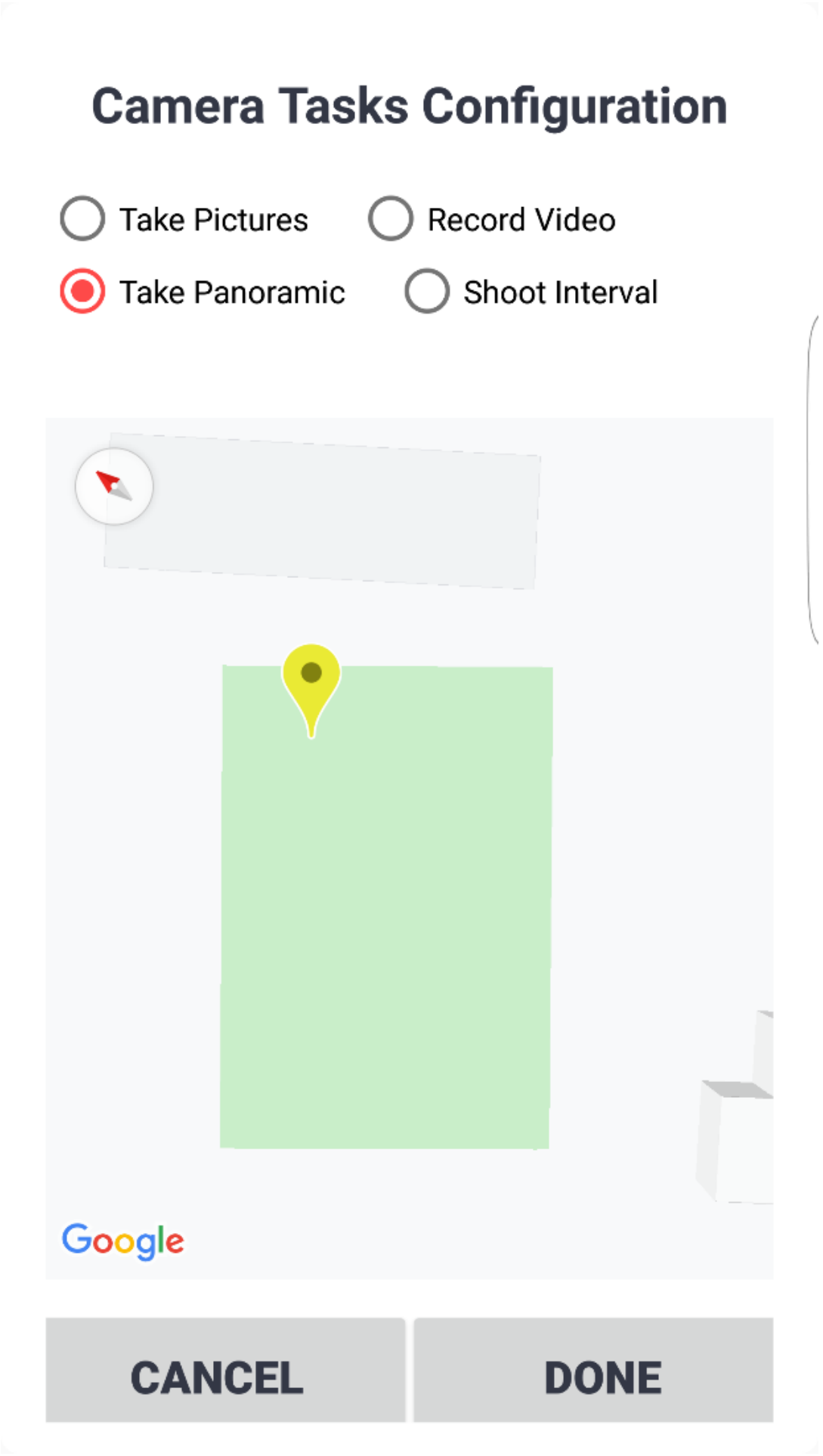

*Figura 46. RF4 Asignar tarea de captura de foto panorámica a un waypoint (Entorno móvil)*

## RF5: Ver la lista de rutas planeadas

Este requerimiento describe la pantalla principal una vez el usuario ingresa a la aplicación. Se listan todas las rutas alojadas en la base de datos, se muestran detalles básicos y unos controles.

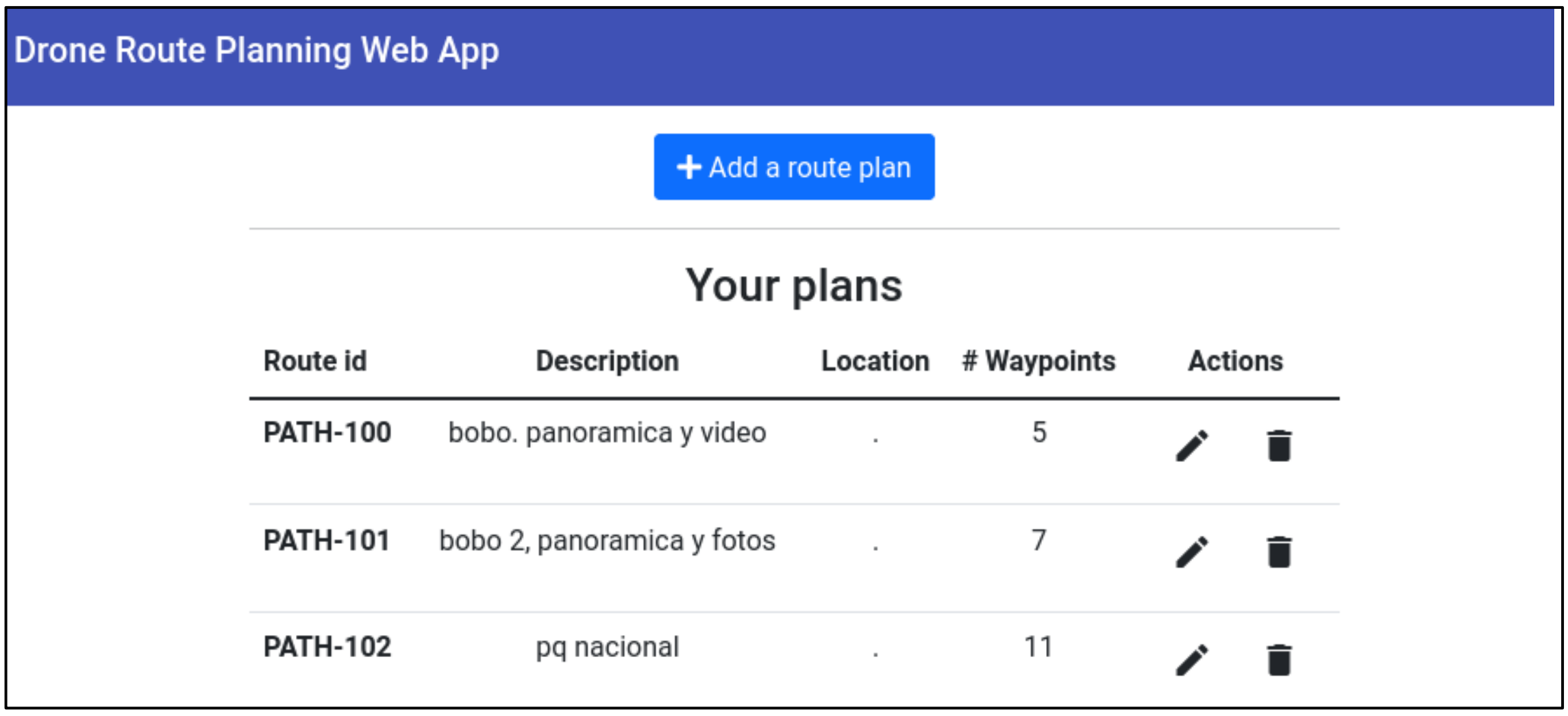

*Figura 47. RF5 ver lista de rutas planeadas*

## RF6: Eliminar una ruta planeada

Cuando el usuario está en la pantalla de lista de rutas, y selecciona el icono de basura de una ruta específica, aparece un mensaje para confirmar la acción, una vez es confirmada la ruta es eliminada de la base de datos:

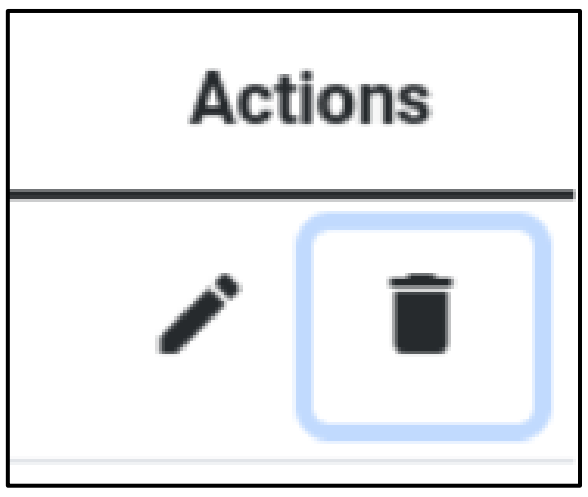

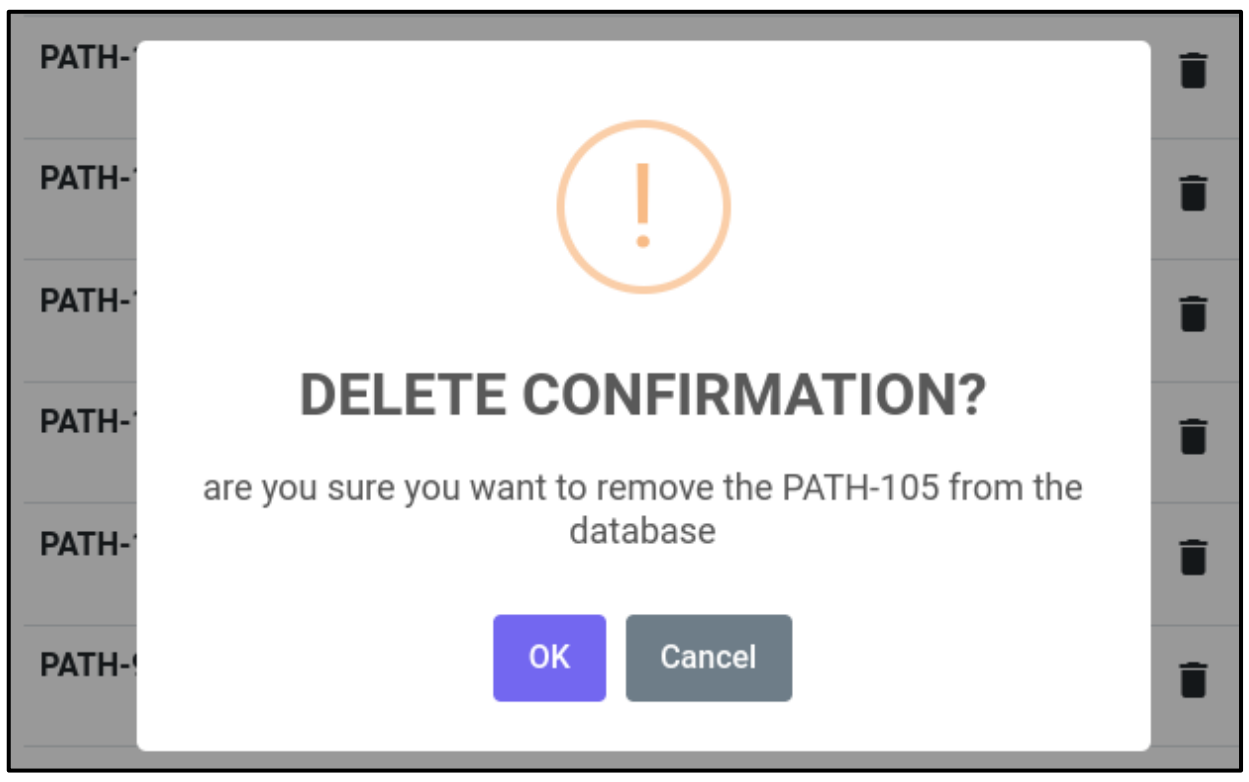

*Figura 48. RF6 eliminar una ruta de la lista*

## RF7: Entrar en modo edición

Hay dos formas de entrar en el modo edición desde la pantalla de lista de rutas. La primera es cuando el usuario selecciona el botón “Add a route plan”. Se redirige entonces al modo edición con una ruta vacía con un mapa general de Bogotá:

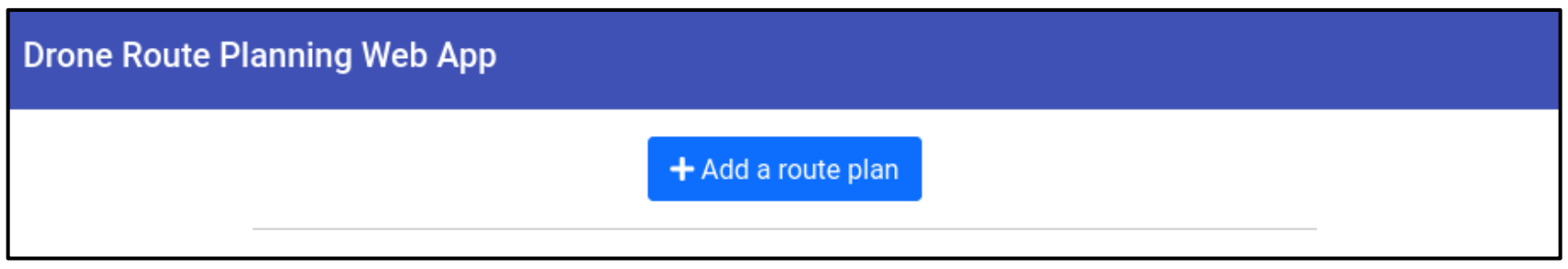

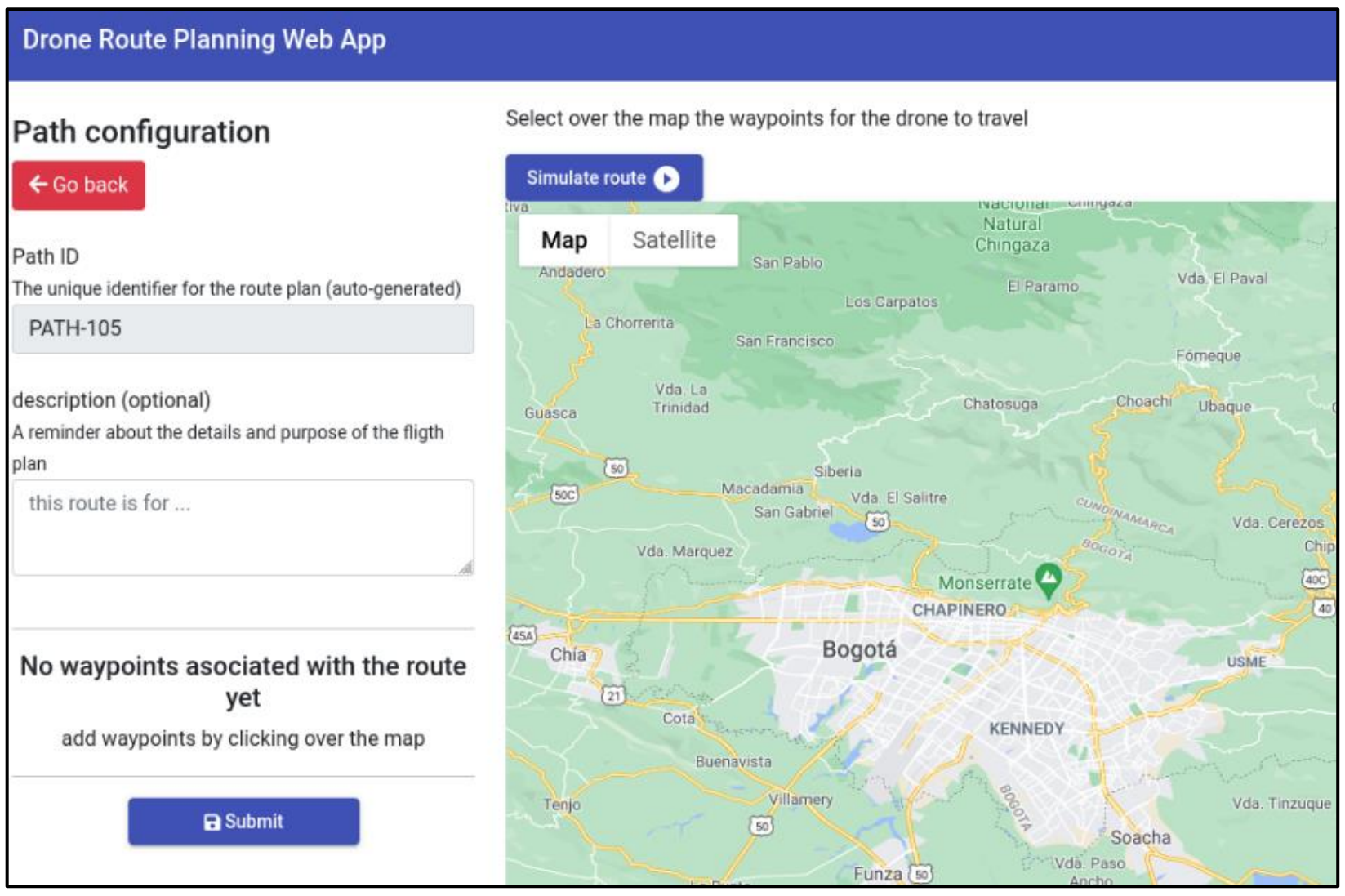

Figura 49. RF7 modo edición ruta vacía

La segunda forma es seleccionando el botón de lápiz para editar una ruta existente, en este caso se abre la pantalla de edición y el mapa se centra y su zoom se ajusta a los puntos de la ruta:

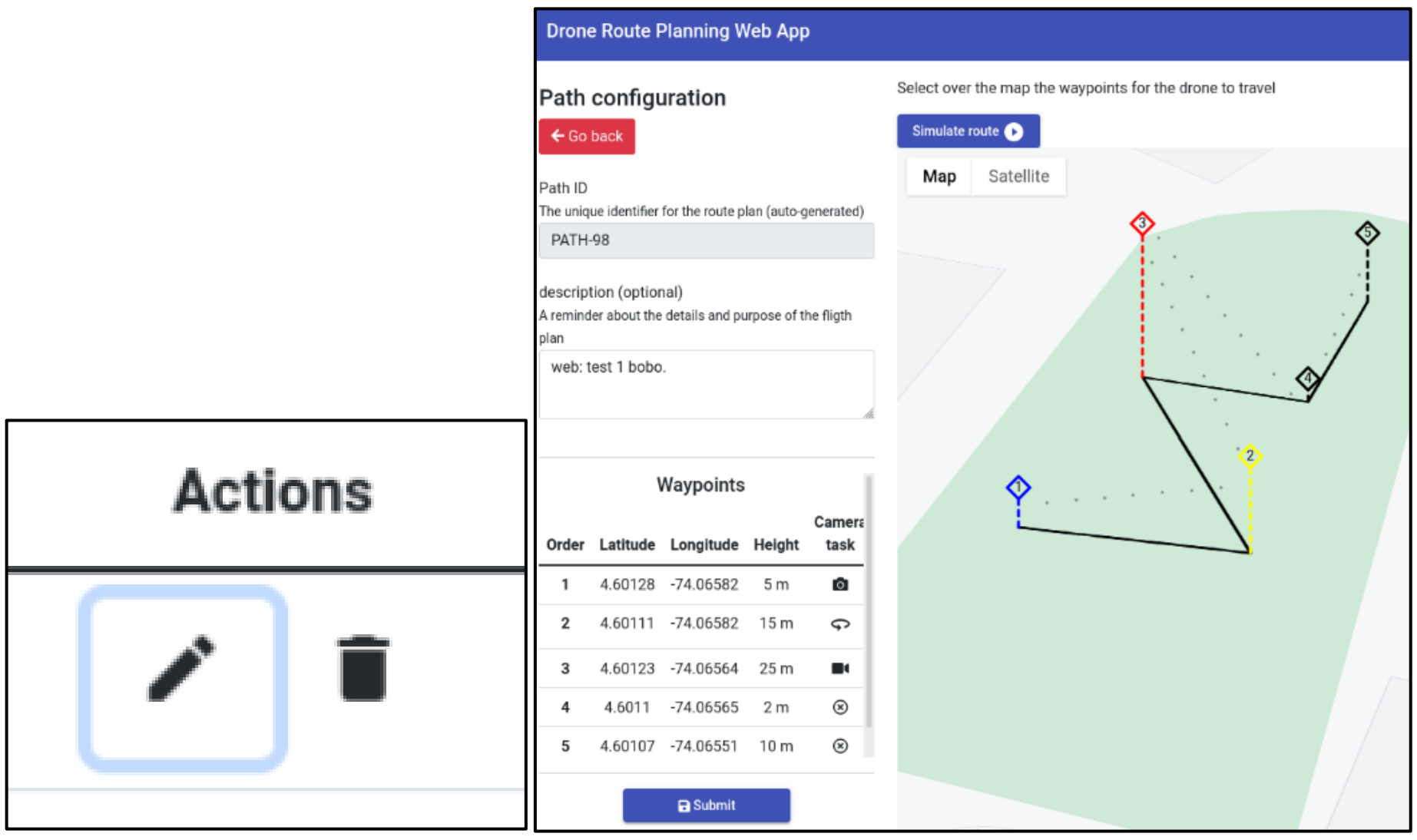

Figura 50. RF7 modo edición ruta existente

## RF8: Agregar waypoint a la ruta

Para este requerimiento tiene dos escenarios. El primero se ilustra en la figura 51 (izquierda), cuando la ruta no tiene waypoints previos y el usuario selecciona un punto en el mapa para agregar uno nuevo. El icono del waypoint se agrega con la altura por defecto de 10m. El segundo caso es cuando ya existe al menos un waypoint en la ruta y el usuario selecciona sobre el mapa para agregar uno nuevo, esto se ilustra en la figura 51 (derecha), el nuevo waypoint se crea, una línea continua al nivel el suelo conecta los puntos y una línea punteada elevada conecta los waypoints según sus alturas respectivas.

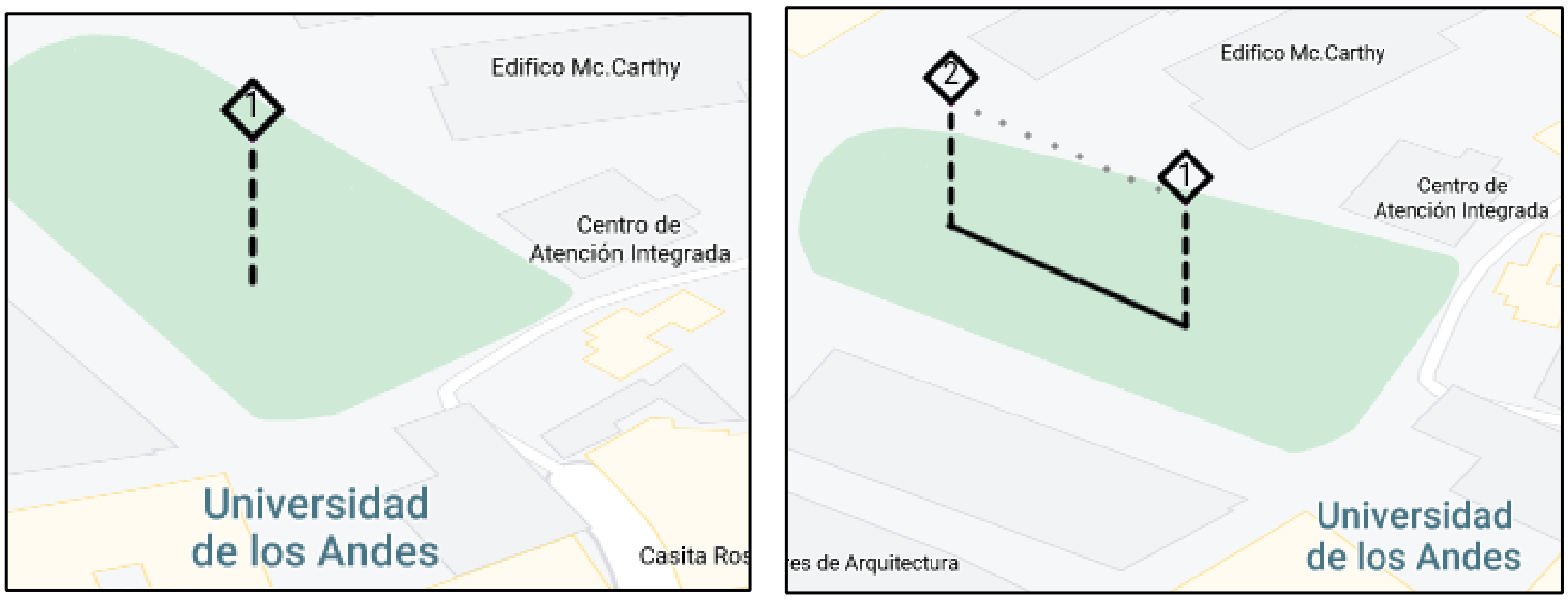

*Figura 51. RF8 agregar nuevo waypoint*

## RF9: Eliminar waypoint

Una vez un usuario abre las preferencias de un waypoint puede seleccionar el icono de papelera y el waypoint será eliminado:

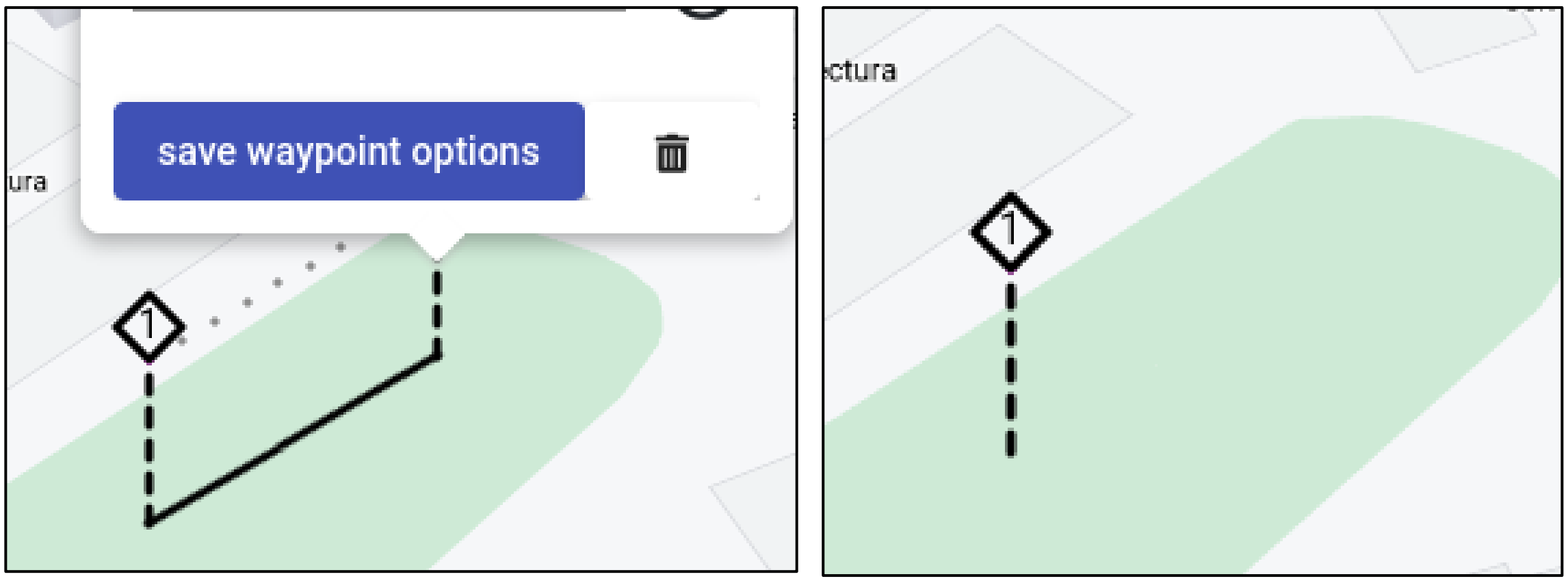

*Figura 52. RF9 eliminar waypoint*

## RF10: Editar altura de un waypoint

Para editar la altura de un waypoint el usuario hace clic sobre este y puede usar el “slider” para cambiar la altura o ingresar el valor de la altura al input de tipo numérico. Ambas opciones son equivalentes. Las siguientes figuras ilustran el resultado cuando se baja o sube la altura de un waypoint:

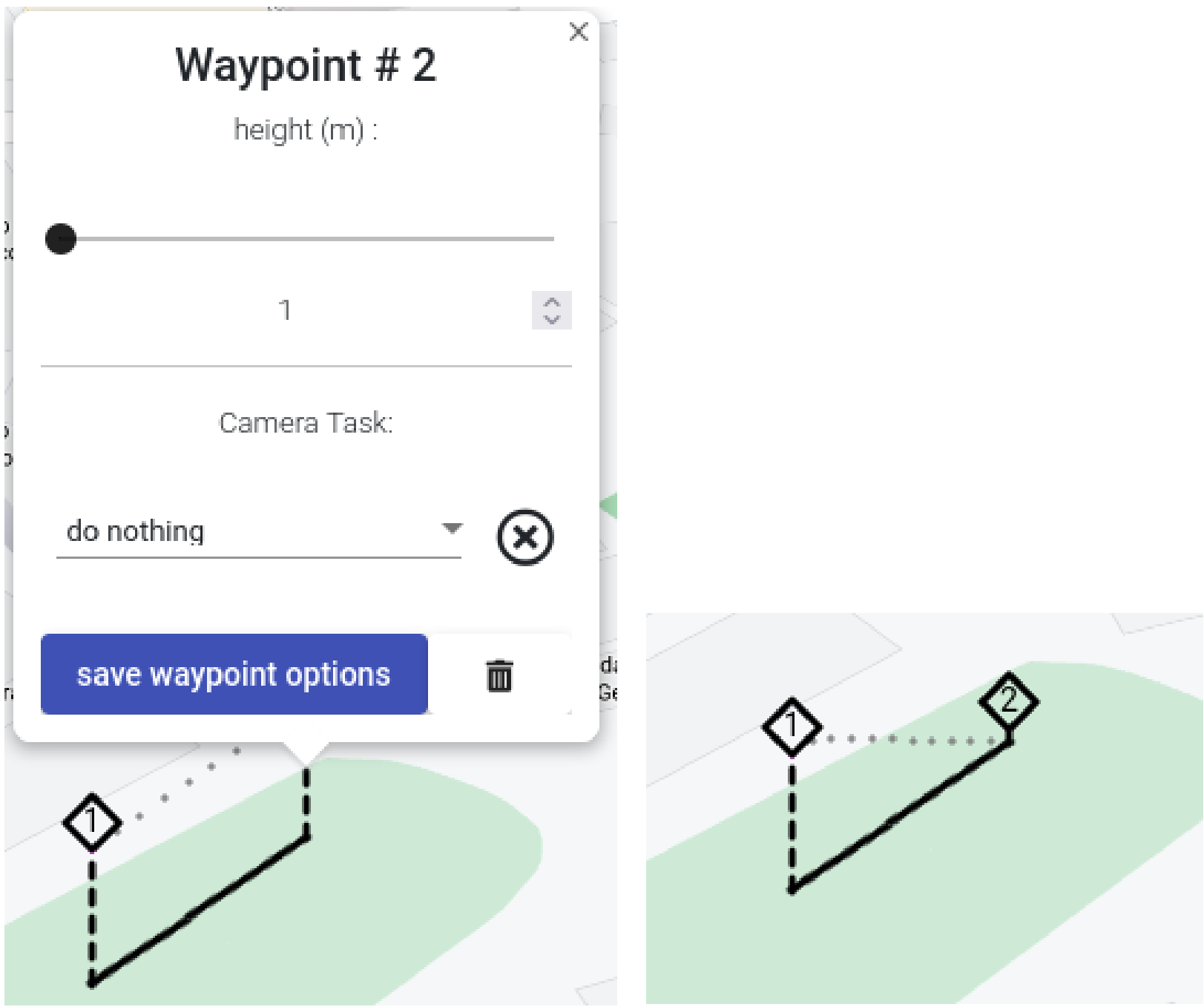

*Figura 53. RF10 disminuir altura*

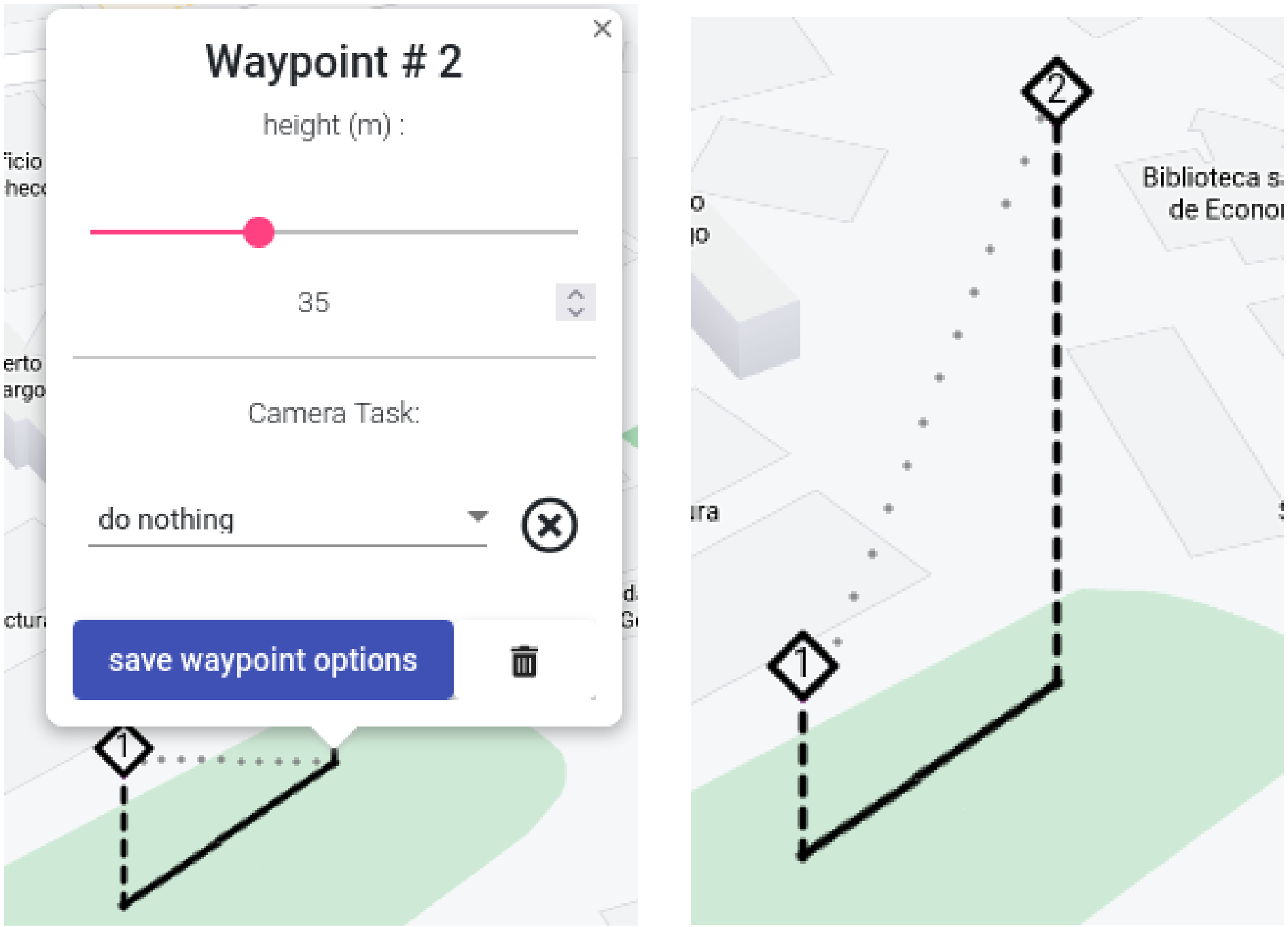

*Figura 54. RF10 aumentar altura*

## RF11: Editar tarea cámara

Este requerimiento tiene que ver con la implementación de un selector para las tareas de cámara, la asignación de cada tarea se describe en la implementación de los RF1 al RF2:

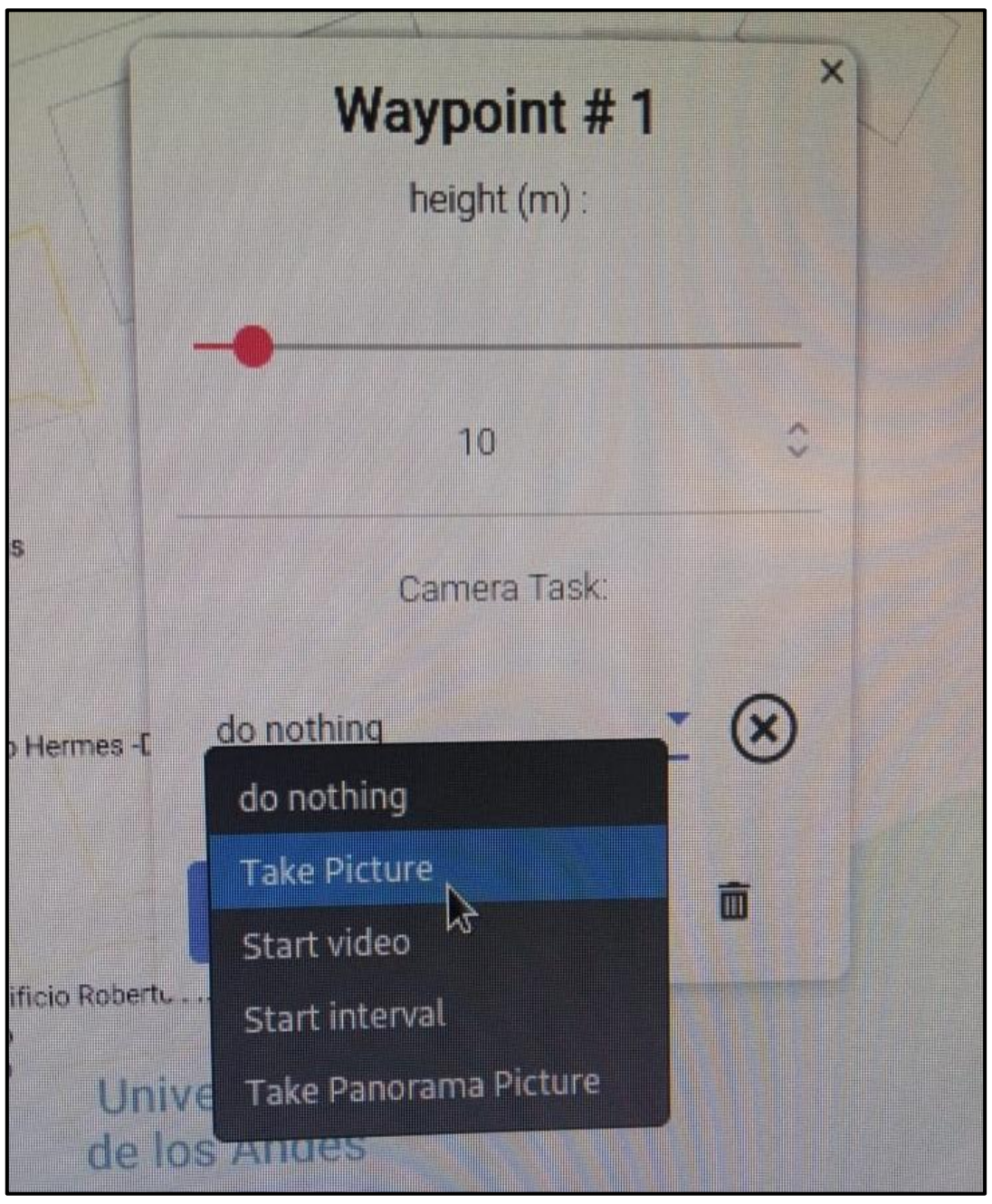

*Figura 55 . RF11 seleccionar tareas de cámara*

## RF12: Agregar descripción de una ruta

En el “sidebar” de la pantalla del modo de edición el usuario puede agregar o no una cadena de texto como descripción:

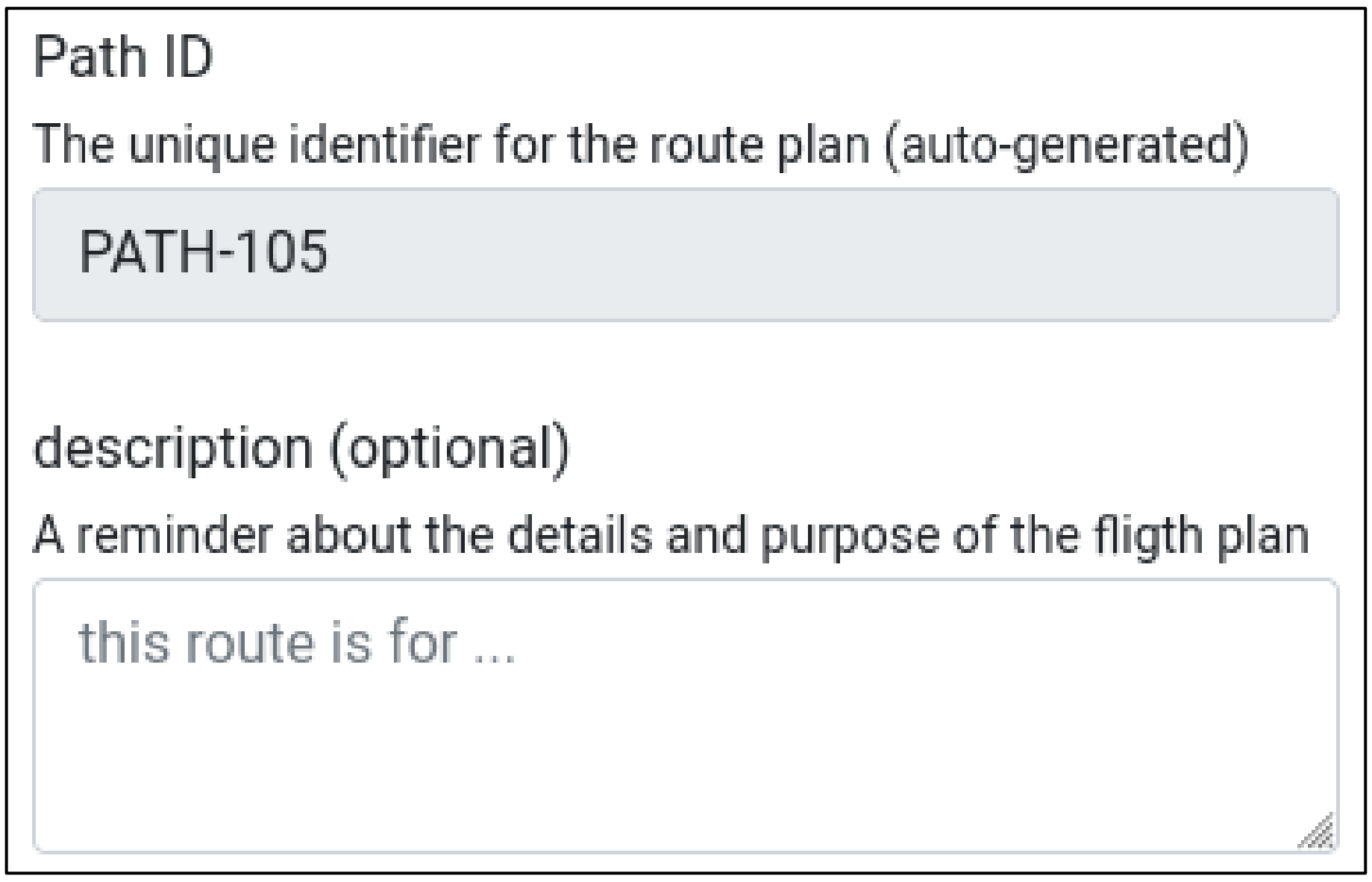

*Figura 56. RF12 agregar descripción a la ruta*

## RF13: Ver detalles de waypoints

En el modo edición, en el “sidebar” el usuario puede ver a lista de waypoints, con sus latitudes, alturas, tareas y orden de ejecución específicos:

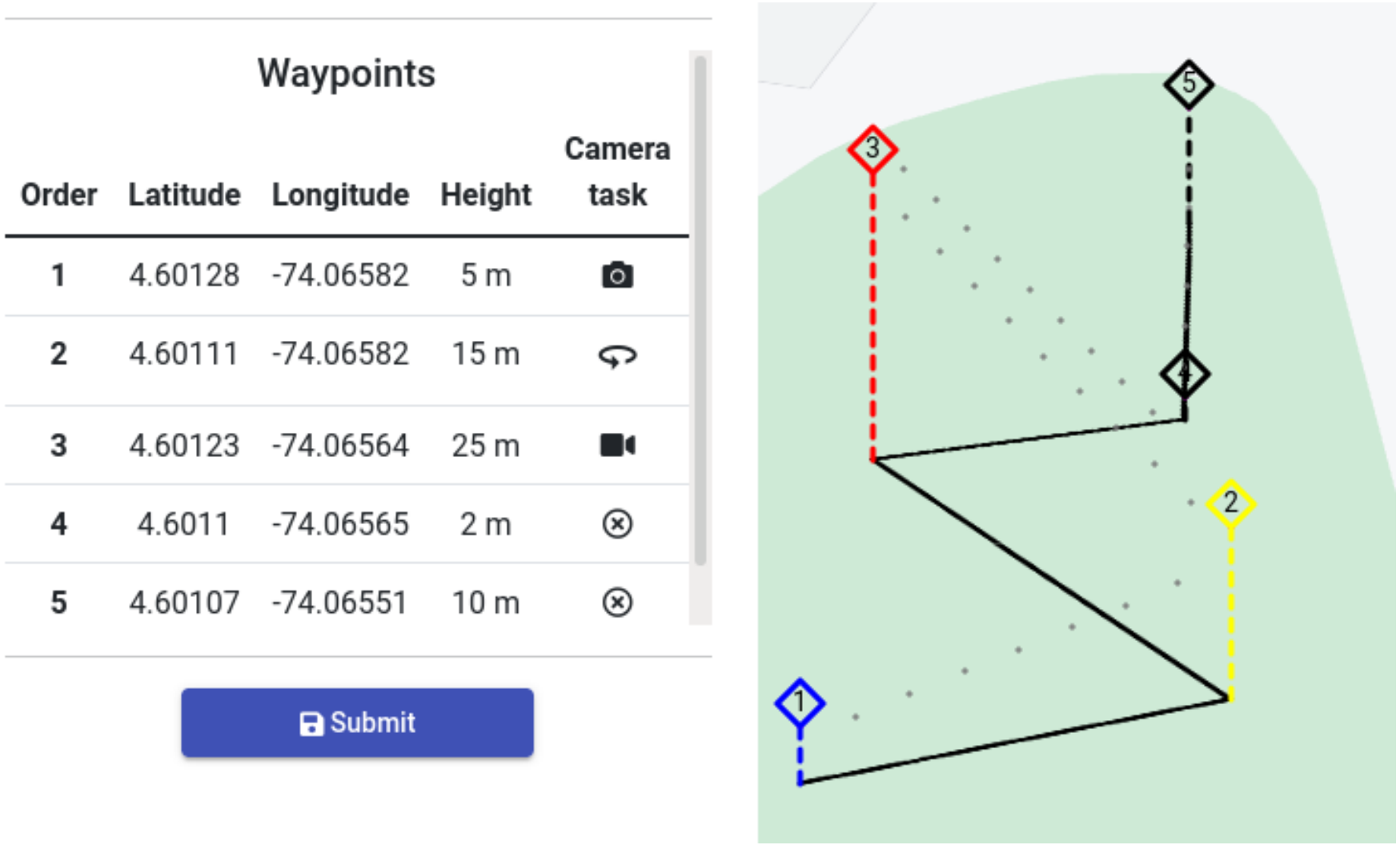

| Order | Latitude | Longitude | Height | Camera task |
|---|---|---|---|---|
| 1 | 4.60128 | -74.06582 | 5 m | |
| 2 | 4.60111 | -74.06582 | 15 m | |
| 3 | 4.60123 | -74.06564 | 25 m | |
| 4 | 4.6011 | -74.06565 | 2 m | |
| 5 | 4.60107 | -74.06551 | 10 m | |

*Figura 57. RF13 ver detalles waypoints*

## RF14: Guardar/persistir ruta

También en el “sidebar” el usuario selecciona “submit” y recibe una confirmación de la operación:

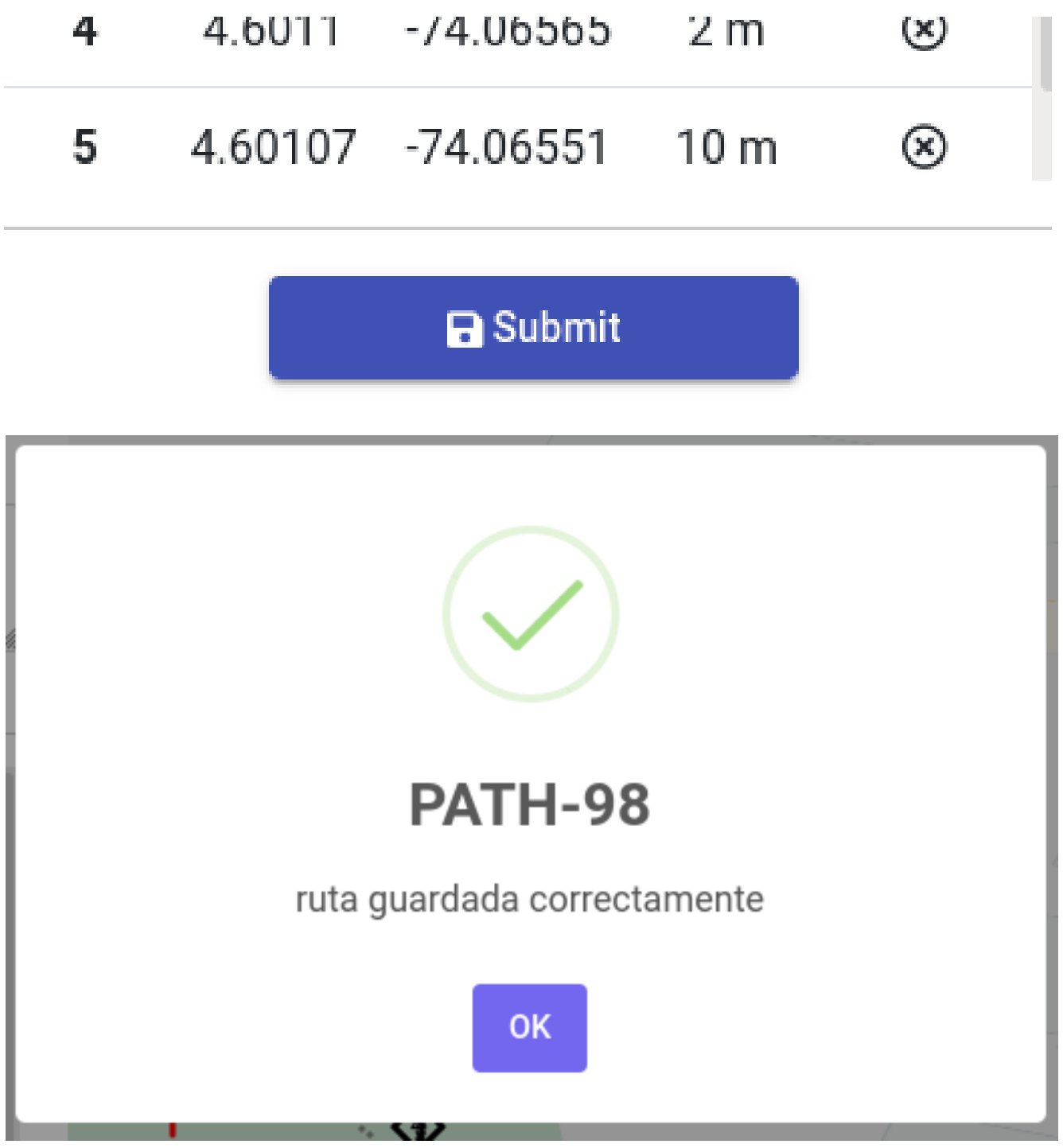

*Figura 58. RF14 guardar ruta y confirmación*

## RF15: Iniciar simulación

Una vez el usuario selecciona “Simulate route”, una animación del icono de un dron se empieza a mover por la guía elevada, debajo del botón de simulación hay un texto que indica la tarea actual y el color del dron indica la tarea (Fig. 59). En el último waypoint sale un aviso de terminación (Fig. 60):

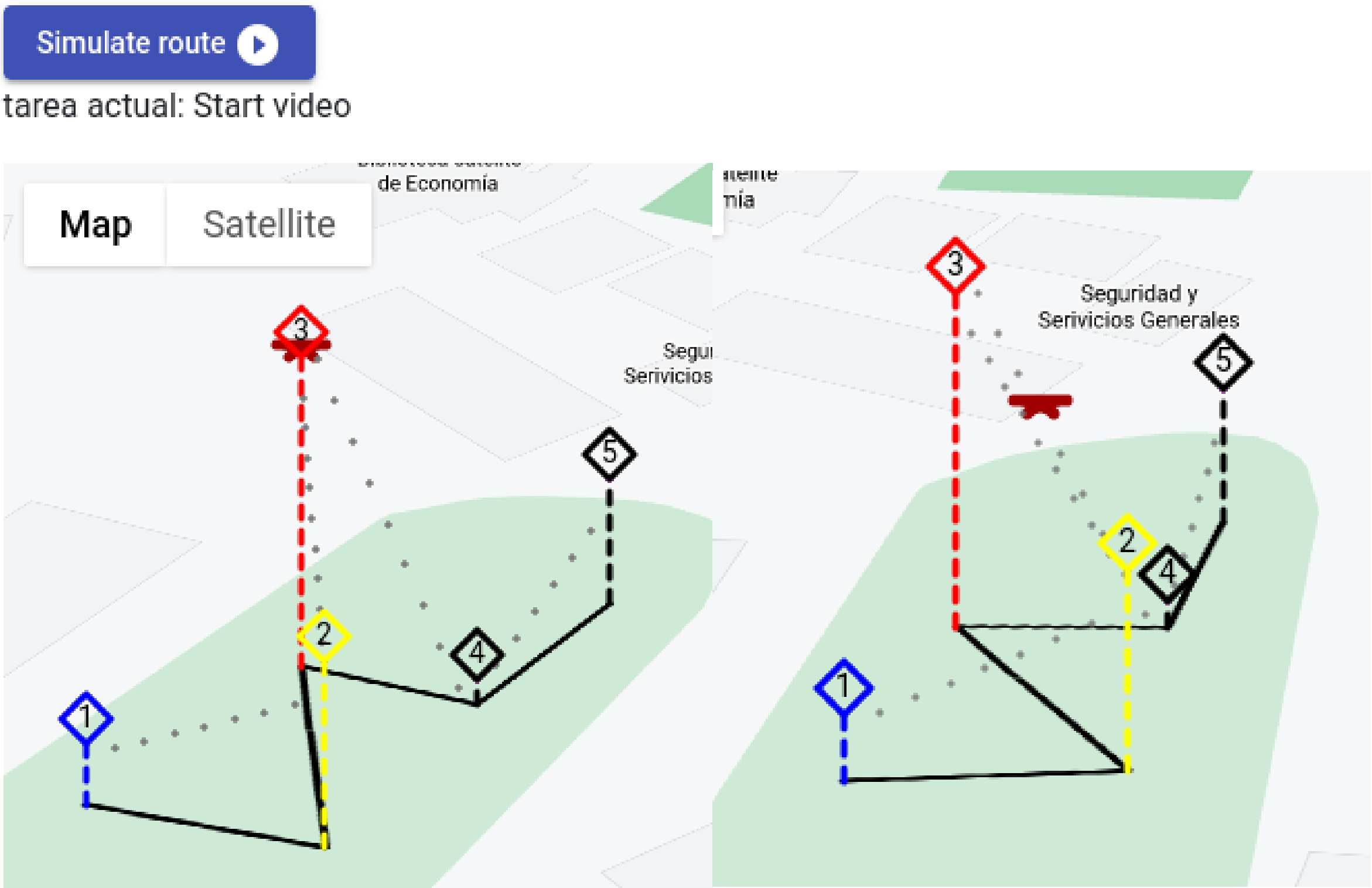

*Figura 59. RF15 simulación en progreso*

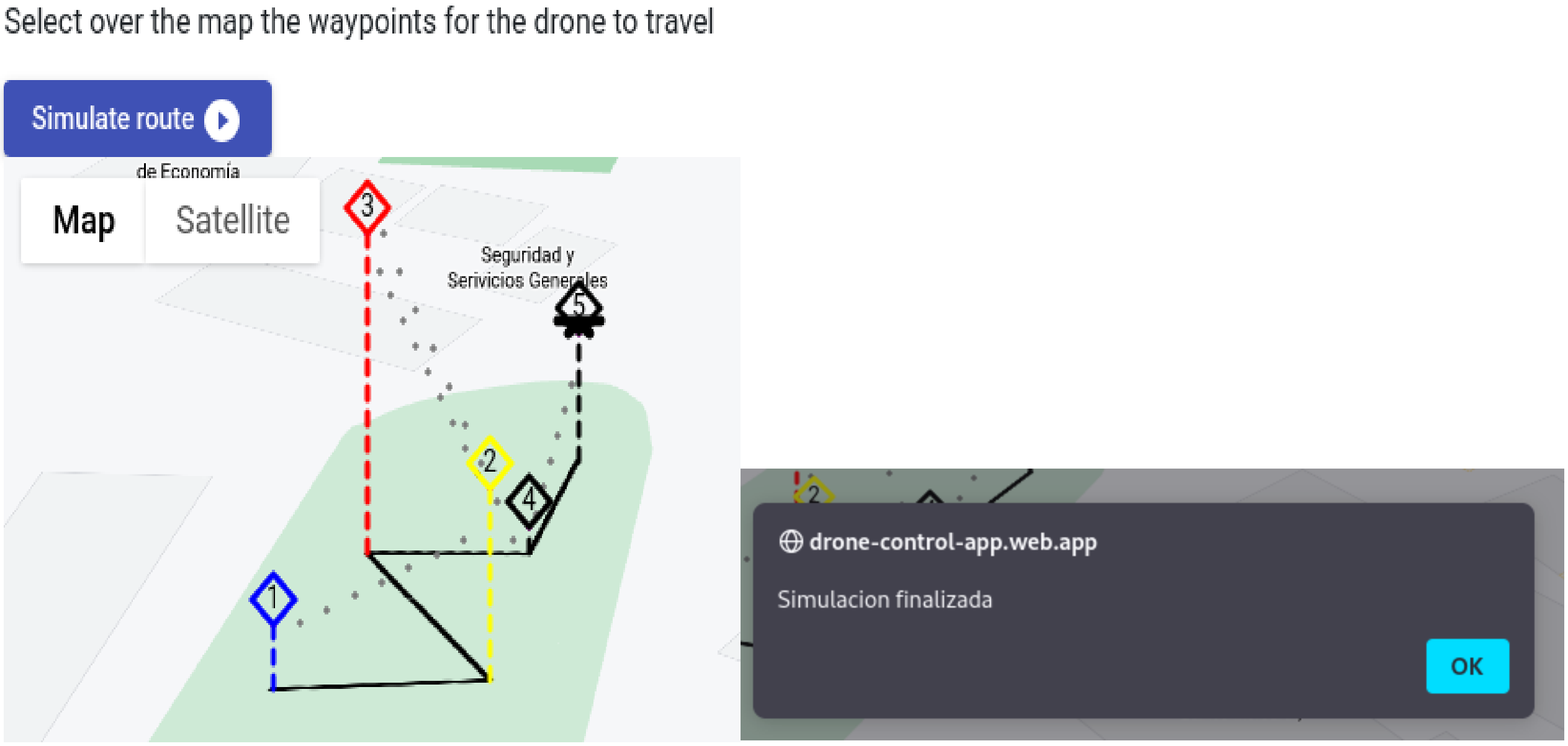

*Figura 60. RF15 Ultimo waypoint de la simulación y aviso de finalización*

## R16: Cambiar inclinación cámara

Esta funcionalidad ya viene implementada por defecto en el componente de mapas, siempre y cuando se esté usando un mapa con un estilo e id personalizado en la plataforma de Google Cloud. Dicho estilo debe tener habilitadas las opciones de mapas vectorizados e inclinación. Al hacer llamada al API, seguido de la key, se debe especificar v=beta ya que los mapas con tecnología 3D de webGL aún siguen en etapa pruebas.

Para moverse sobre el mapa sin cambiar la inclinación, el usuario lo hace igual que un mapa web tradicional, mantiene presionado click y arrastra. Para cambiar la inclinación de la cámara sin mover el mapa, el usuario debe primero mantener presionado Shift y arrastran hacia arriba y hacia abajo para cambiarla. Si lo que se quiere es que la cámara el norte o punto de vista, también se presiona Shift y se arrastra a la derecha o izquierda.

La siguiente figura muestra dos capturas de como se ve el mismo marcador de waypoint desde una perspectiva completamente cenital (2D) y otra con inclinación (estructuras 3D):

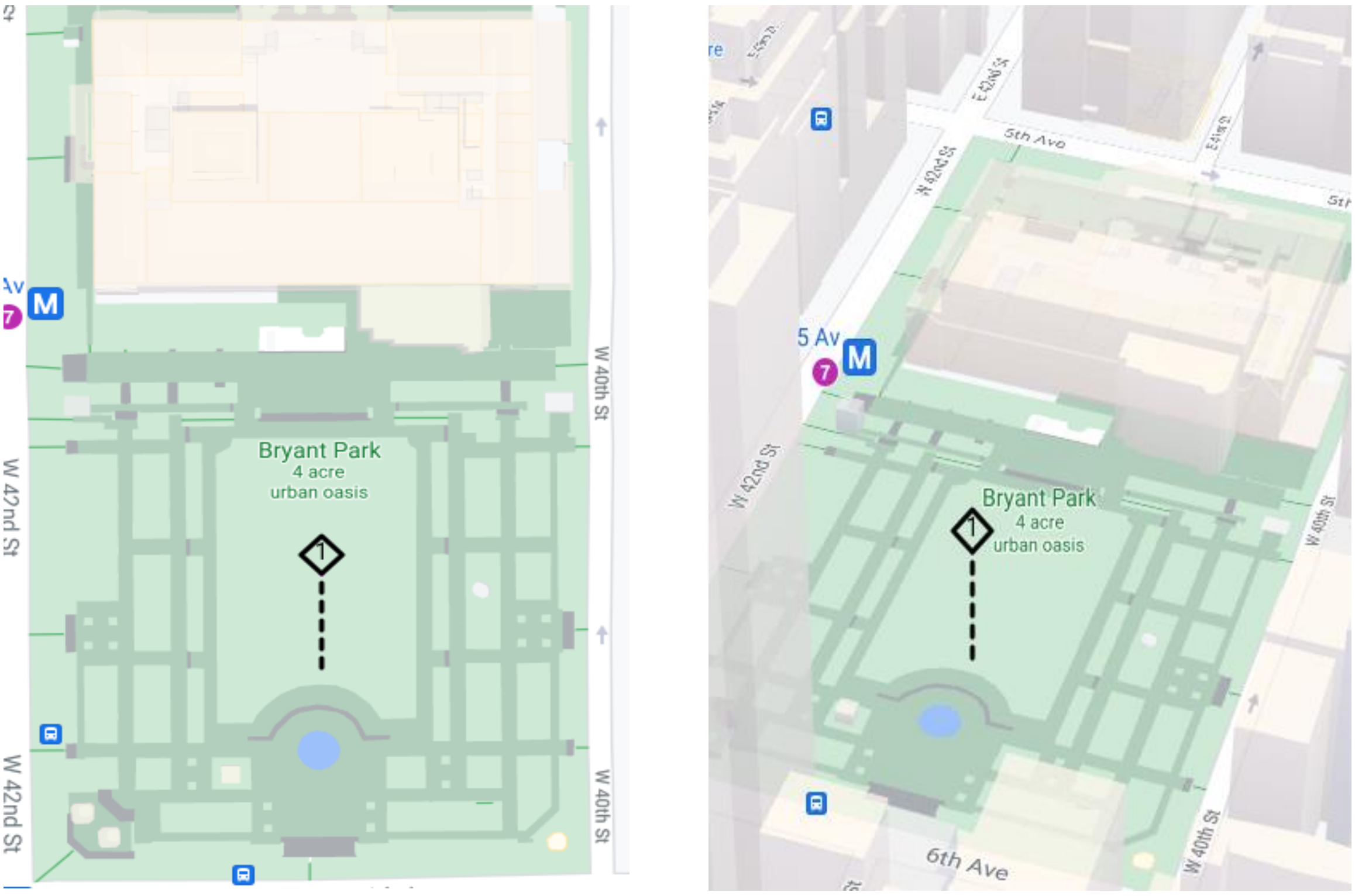

*Figura 61. RF16 cambiar inclinación de cámara*

## R17 y R18: Cambiar a vista satelital y vista Mapa 3D

En la esquina superior izquierda del mapa se habilito la opción de cambiar el modo de visualización para tener varias formas de medir y estimar la ubicación de los puntos. En modo satélite se tiene una forma más precisa de ubicar los puntos con relación a puntos de referencia reales que fueron fotografiados. Por otro lado, la opción de Map, habilita la visualización clásica del mapa además de tener representación en 3D de estructuras arquitectónicas como edificios y monumentos, estando unas ciudades mejores y más modeladas que otras:

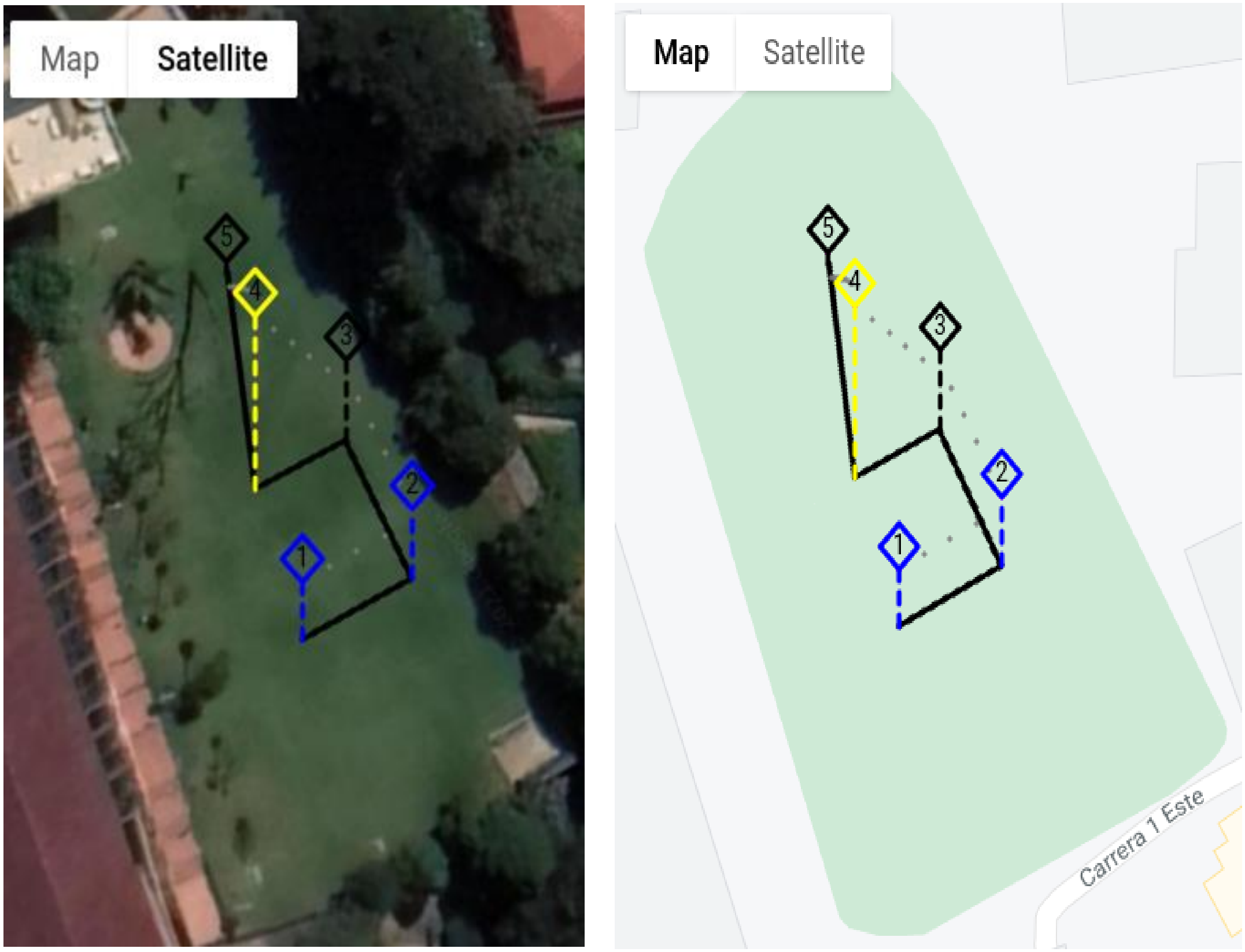

*Figura 62. RF17 y RF18 cambiar tipo de mapa*

# 8. Diseño y resultado de pruebas

## 8.1 Diseño de pruebas

Para probar la integración exitosa de los nuevos componentes desarrollados a la arquitectura desarrollada en proyectos anteriores, se definieron escenarios de prueba para comprobar la correcta ejecución de las rutas planeadas en diferentes ambientes y medir la precisión con que estas se completan. En total se definieron dos escenarios en entornos físicos diferentes y, por cada escenario, se plantearon tres rutas similares, cada una planeada en un ambiente de planeación diferente. Esto quiere decir que, por cada escenario, se planeó y ejecutó una ruta en ambiente de realidad virtual, una en la aplicación web y una en la aplicación móvil.

Estas tres rutas serán ejecutadas en cada escenario y por cada una se calculará la diferencia entre la posición de los waypoints planeados y los ejecutados por el dron. De esta forma es posible conocer la precisión de las rutas planeadas en cada ambiente distinto y la efectividad de estos ambientes a la hora de planear y ejecutar rutas de vuelo.

### 8.1.1 Escenario 1:

Se escogió la cancha de fútbol del centro deportivo de la Universidad de los Andes como entorno físico para el primer escenario (Fig. 63). Este entorno es lo suficientemente amplio como para planear rutas grandes que contengan waypoints lejanos en diferentes direcciones. Así mismo, es un espacio libre de obstáculos que puedan impedir la correcta ejecución de la ruta planeada.

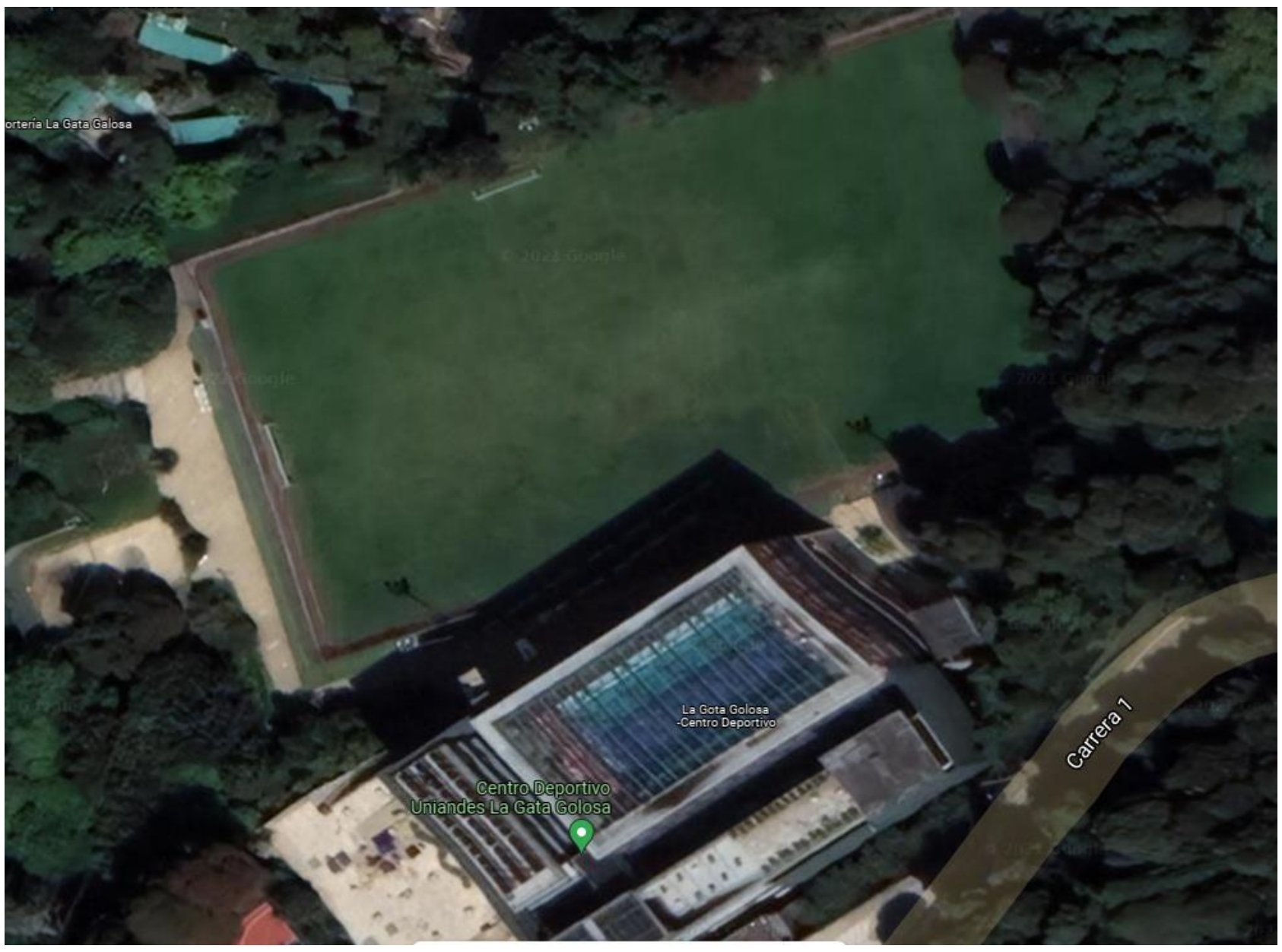

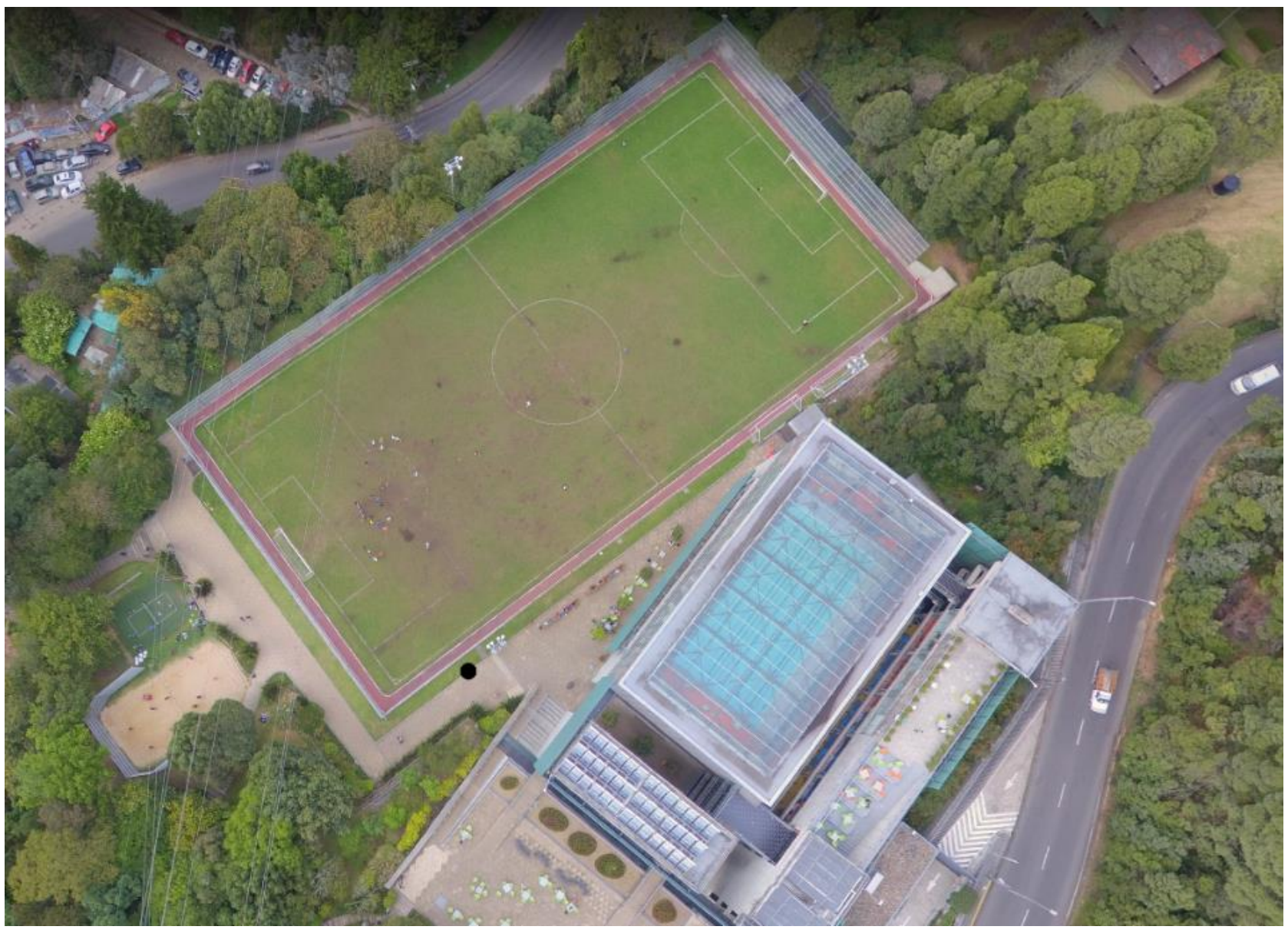

*Figura 63. Vista satelital (arriba) y StreetView (abajo) del centro deportivo (Escenario 1)*

Para este escenario se eligió la ruta 96 como ruta planeada en ambiente de realidad virtual. Esta ruta fue planeada y probada en este entorno en el proyecto de Múnera [2], por lo que se consideró la ruta más óptima para realizar las pruebas. La geometría de esta ruta se puede apreciar en la siguiente imagen:

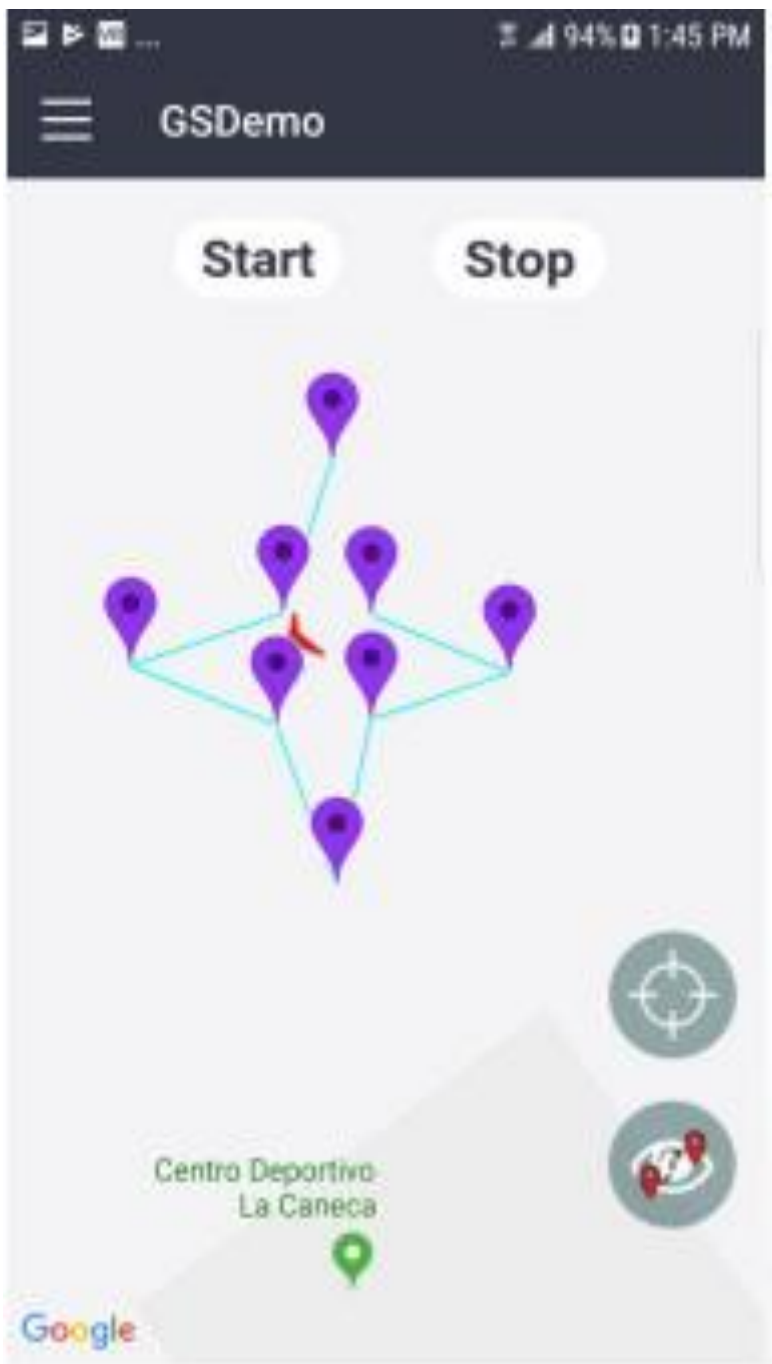

*Figura 64 . Geometría de la ruta 96 (planeada en ambiente RV)*

Con base a esta ruta, se planeó una nueva ruta con forma similar en el ambiente web para ser ejecutada en este escenario, la cual tiene el ID número 103. El resultado de la planeación se puede apreciar en la figura a continuación:

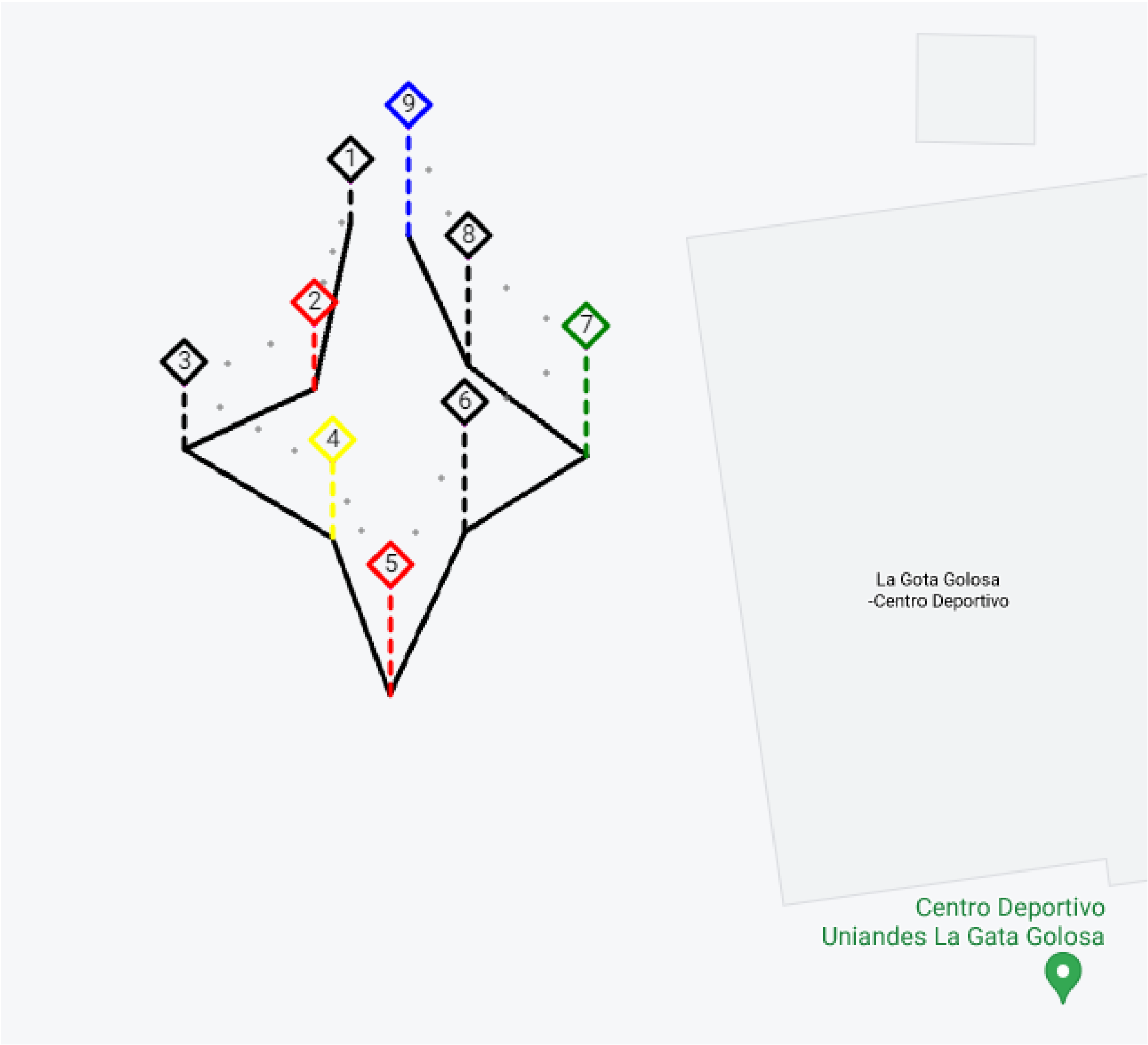

*Figura 65. Geometría de la ruta 103 (planeada en ambiente web)*

Por último, se creó una nueva ruta con ID 107 por medio de la aplicación móvil para este escenario. La geometría de esta ruta se puede apreciar en la siguiente imagen:

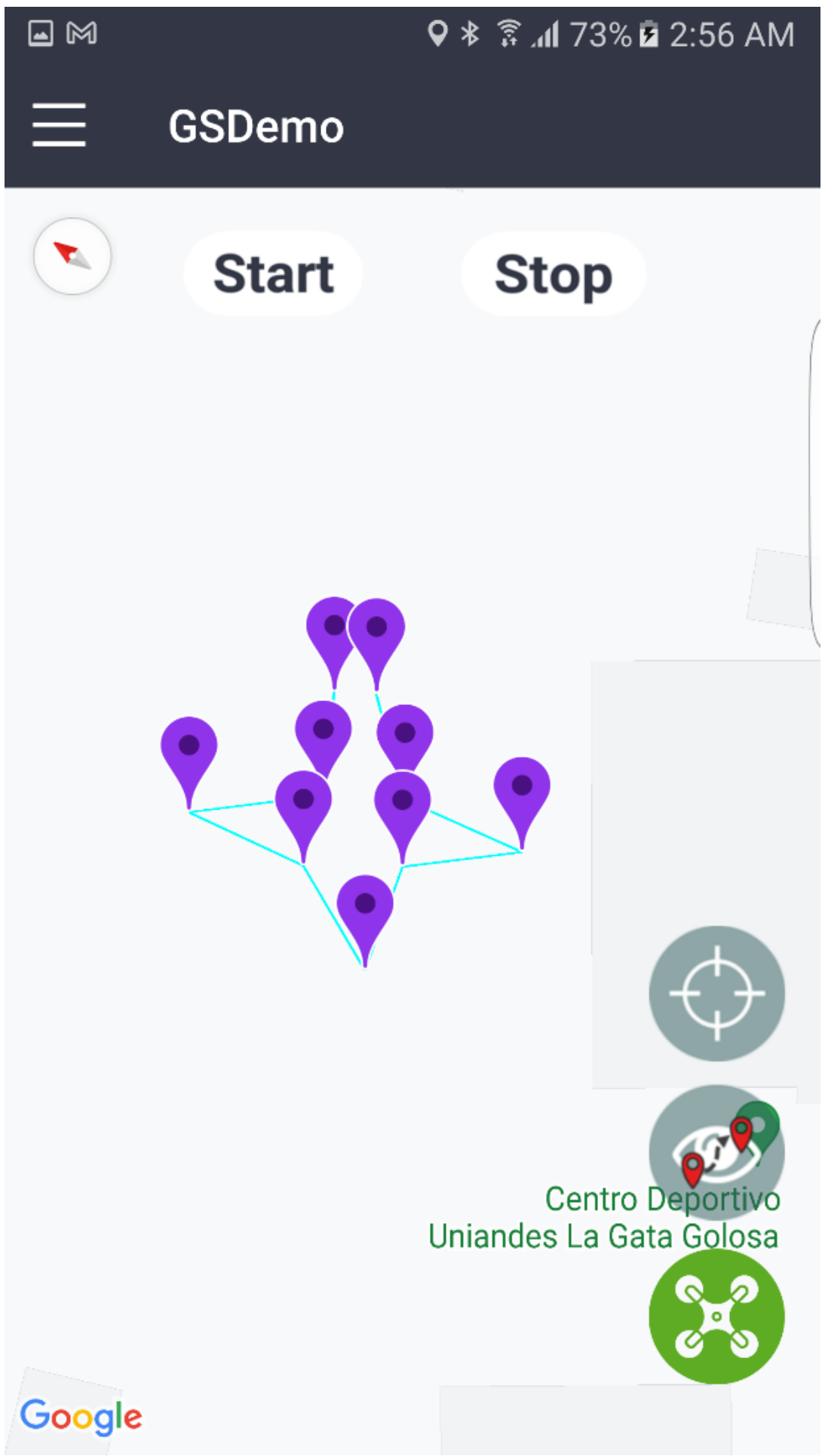

*Figura 66. Geometría de la ruta 107 (planeada en ambiente móvil)*

### 8.1.2 Escenario 2:

Para el segundo escenario se escogió como entorno físico la zona verde donde se encuentra la estatua de San Alberto Magno dentro del campus de la Universidad de los Andes, popularmente conocido como “el bobo” (Fig. 67). Se eligió este entorno debido a que es un espacio relativamente amplio sin muchos obstáculos internos, y es de fácil acceso para realizar múltiples pruebas. Además, este espacio fue utilizado por Sánchez [1] en la ejecución de pruebas de su proyecto.

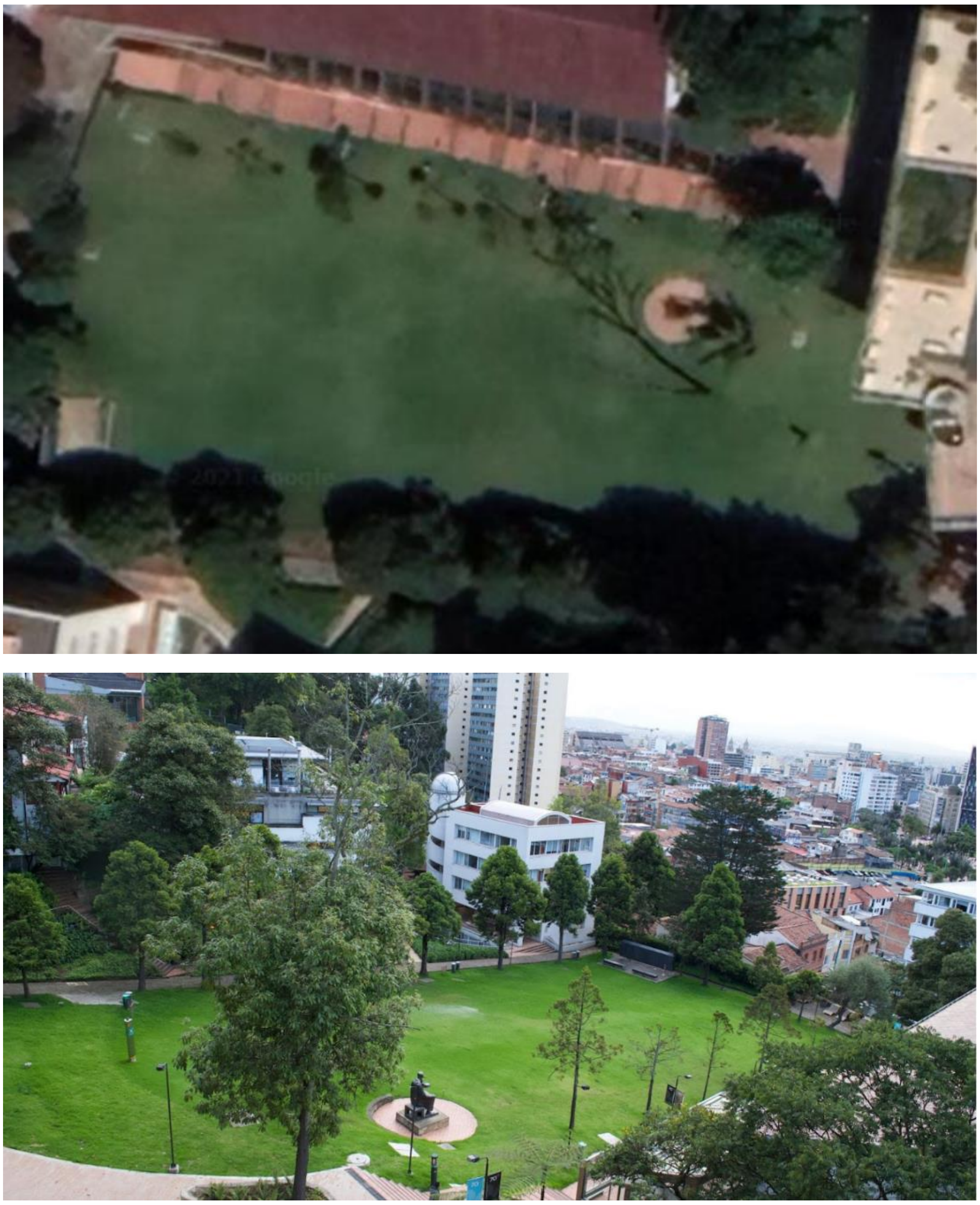

*Figura 67 . Vista satelital (arriba) y StreetView (abajo) estatua de San Alberto Magno (Escenario 2)*

Para este escenario se eligió la ruta 3 como ruta planeada en ambiente de realidad virtual. Esta ruta fue planeada en este entorno en el proyecto de Múnera [2]. La geometría de esta ruta se puede apreciar en la siguiente imagen:

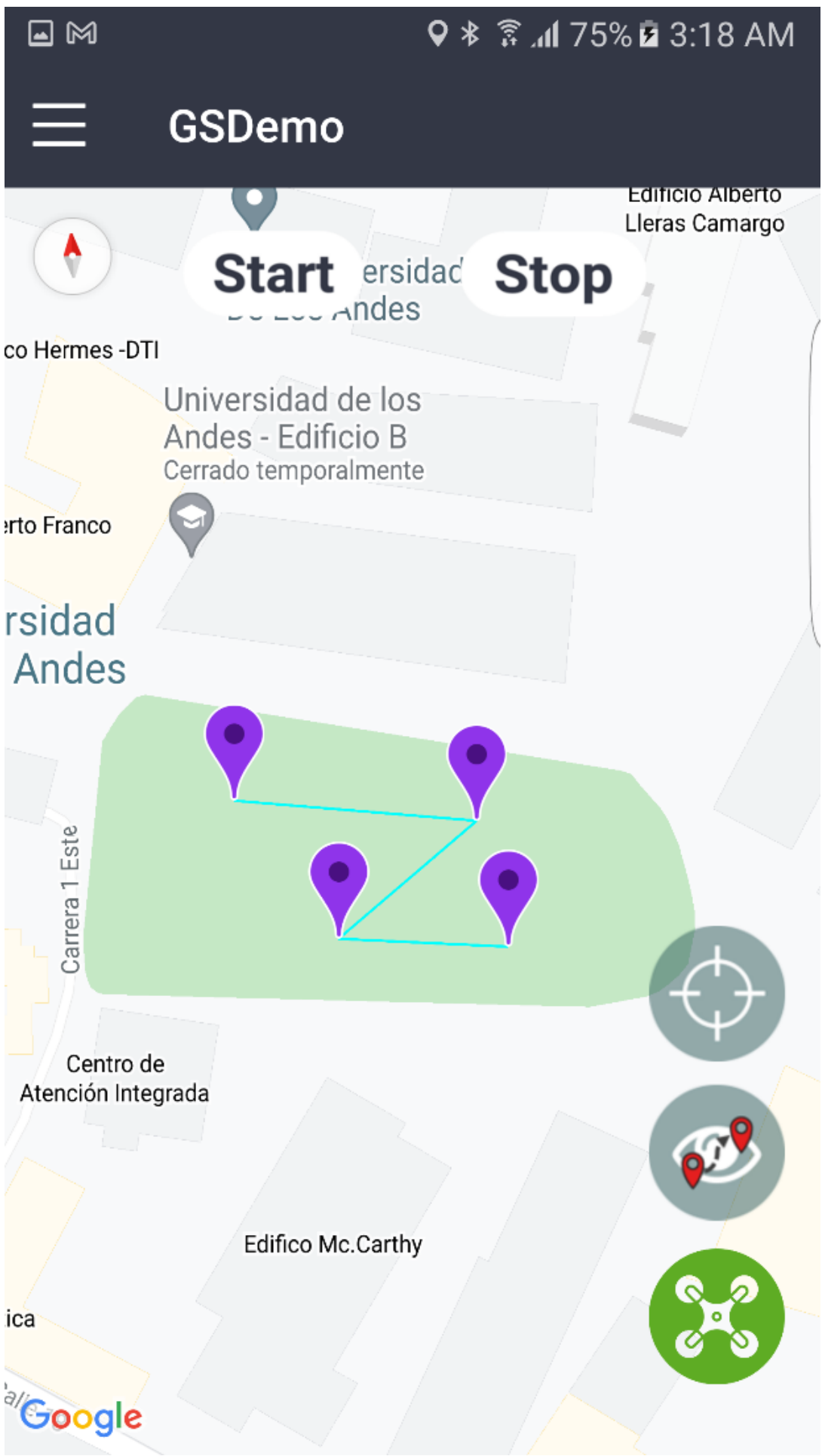

*Figura 68. Geometría de la ruta 3 (planeada en ambiente RV)*

Con base a esta ruta, se planeó una nueva ruta con forma similar en el ambiente web para ser ejecutada en este escenario, la cual tiene el ID número 97. El resultado de la planeación se puede apreciar en la figura a continuación:

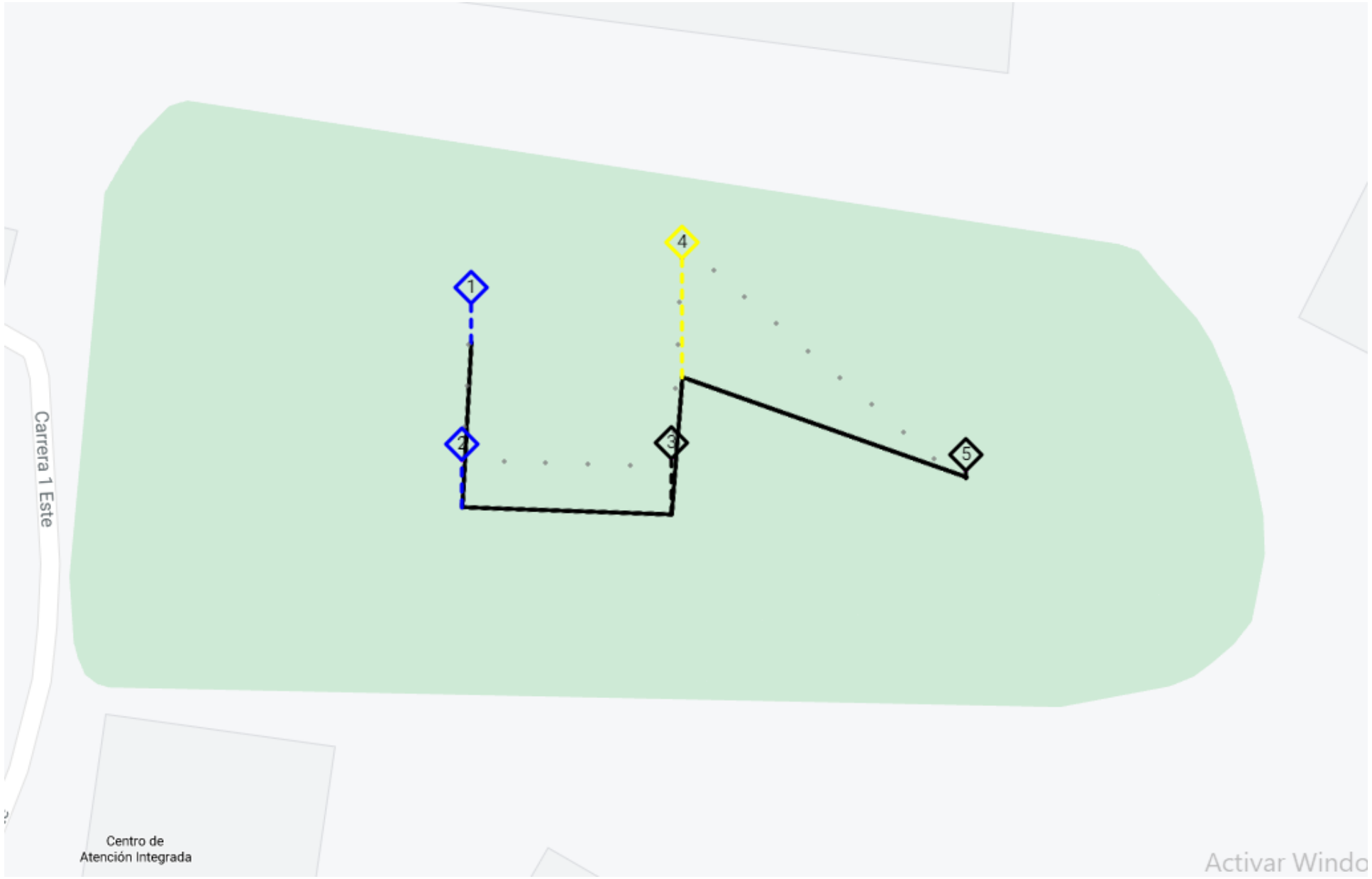

*Figura 69. Geometría de la ruta 97 (planeada en ambiente web)*

Por último, se creó una nueva ruta con ID 108 por medio de la aplicación móvil para este escenario. La geometría de esta ruta se puede apreciar en la siguiente imagen:

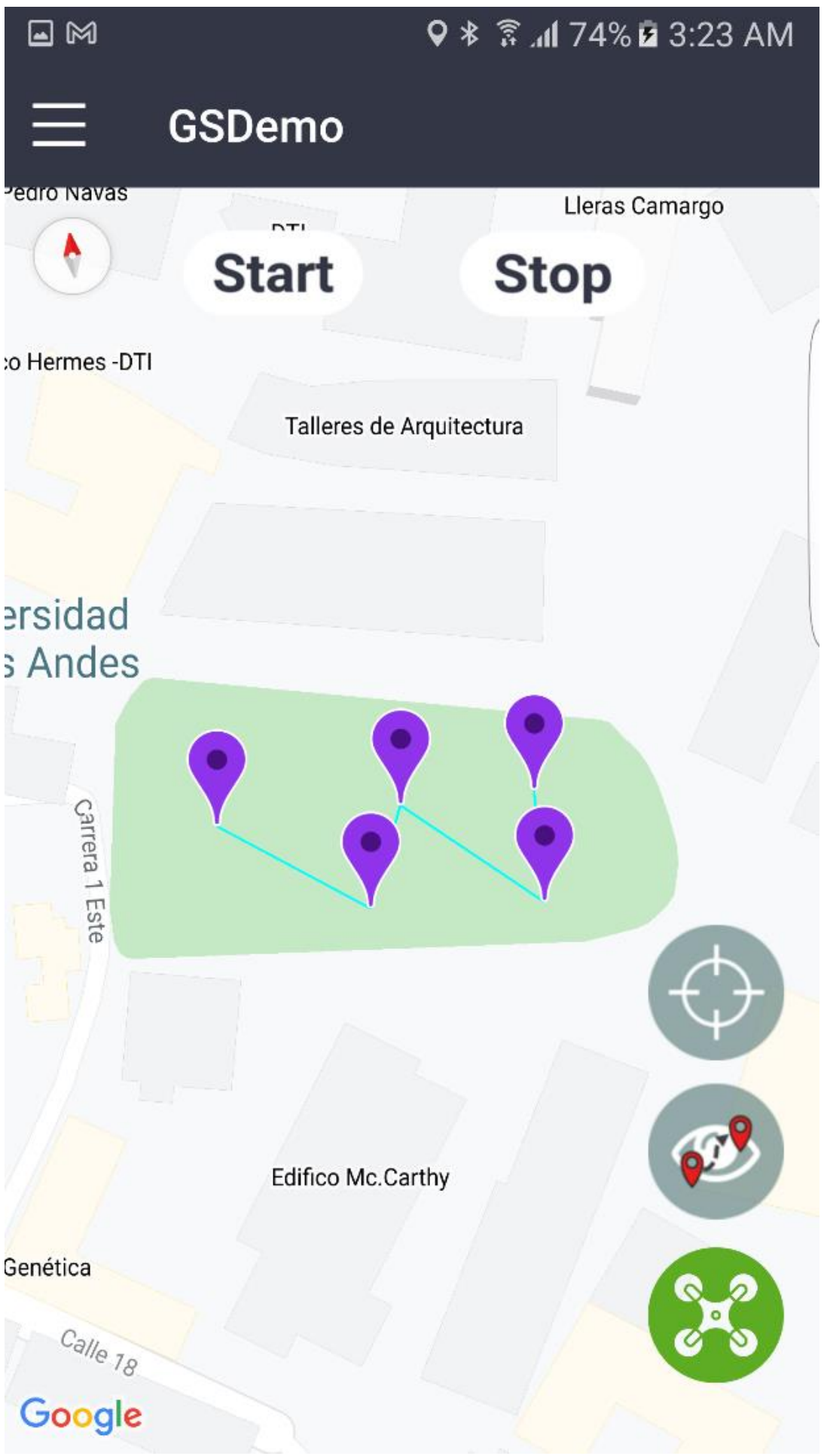

*Figura 70. Geometría de la ruta 108 (planeada en ambiente móvil)*

Tras la ejecución de cada ruta se recogieron los datos de planeación y vuelo para su posterior análisis. Los datos de planeación corresponden a la longitud y latitud de cada punto planeado por el usuario, los cuales se encuentran en la base de datos al planear una ruta. Los datos de vuelo corresponden a la longitud y latitud de cada punto recorrido por el dron en la ejecución de la ruta. Siguiendo la metodología empleada por Múnera [2] en sus pruebas, a partir de los datos recopilados se puede medir la precisión con que cada ruta es ejecutada. De esta forma, es posible analizar la efectividad de cada ambiente de planeación y compararlos entre sí.

## 8.2 Resultados de Pruebas

Para analizar correctamente las diferencias entre la ruta planeada y la ruta ejecutada, es necesario calcular la distancia de cada punto de la ruta con respecto al punto inicial (donde despega el dron). Para esto se utilizó la fórmula propuesta por Múnera [2]:

$$ZLatitude = \frac{\pi \cdot R_{tierra} \cdot (NuevaLatitud - LatitudHome)}{180}$$

$$XLongitude = \frac{R_{tierra} \cdot (NuevaLongitud - LongitudHome)}{\left(\frac{180}{\pi \cdot \cos\left(\frac{LatitudHome \cdot 180}{\pi}\right)}\right)}$$

De esta forma se calcula la distancia en metros tanto en el eje X como en el eje Z. Esto permite comparar las distancias entre los puntos de la ruta planeada y la ruta ejecutada, y así conocer el error de ejecución de la ruta.

### 8.2.1 Escenario 1:

Los datos obtenidos y calculados para las rutas del escenario 1 se pueden encontrar en la siguiente tabla:

| | Ruta 96 | | Ruta 103 | |
|---|---|---|---|---|
| | Prueba | Original | Prueba | Original |
| Longitud | -74,0626736305917000 | -74,0626708986067000 | -74,0626879456687000 | -74,0626882025568000 |
| | -74,0627043779282000 | -74,0627031312638000 | -74,0627678030480000 | -74,0627691348150000 |
| | -74,0628044425615000 | -74,0628043166052000 | -74,0628277411323000 | -74,0628264077886000 |
| | -74,0627070165221000 | -74,0627076130313000 | -74,0628313533416000 | -74,0628310780028000 |
| | -74,0626677819614000 | -74,0626679093220000 | -74,0628868197358000 | -74,0628861249675000 |
| | -74,0626448838722000 | -74,0626449994063000 | -74,0627983743645000 | -74,0627973763958000 |
| | -74,0625532155282000 | -74,0625533370873000 | -74,0627362791421000 | -74,0627366131549000 |
| | -74,0626453693681000 | -74,0626450975082000 | -74,0627250051200000 | -74,0627240097524000 |
| | | | -74,0626795135204000 | -74,0626793523982000 |
| Latitud | 4,60076522102383000 | 4,60076652464770000 | 4,60049136956037000 | 4,60048916568499000 |
| | 4,60066692061588000 | 4,60066636721870000 | 4,60046891979560000 | 4,60046582170739000 |
| | 4,60063167059083000 | 4,60063163766860000 | 4,60051144431447000 | 4,60050902996986000 |
| | 4,60059299630795000 | 4,60059277158126000 | 4,60042194740016000 | 4,60042314938550000 |
| | 4,60048607531645000 | 4,60048664994933000 | 4,60036105432938000 | 4,60036147440865000 |
| | 4,60059497452543000 | 4,60059602057194000 | 4,60036652551038000 | 4,60036647559494000 |
| | 4,60062676146890000 | 4,60062709833645000 | 4,60033027428866000 | 4,60033075188694000 |
| | 4,60066517034147000 | 4,60066534790818000 | 4,60040591381873000 | 4,60040409905449000 |
| | | | 4,60046027538303000 | 4,60046108199943000 |
| Coord X | 2,4227383738184700 | 2,7134635029748300 | 9,7596917466104800 | 9,7324562147775900 |
| | - 0,8492500717723600 | - 0,7165858489117360 | 1,2931329800200500 | 1,1519377188220700 |
| | - 11,4976626268241000 | - 11,4842589430781000 | - 5,0615623208715000 | - 4,9201998969726600 |
| | - 1,1300369538571300 | - 1,1935146862025300 | - 5,4445323427293800 | - 5,4153406502207300 |
| | 3,0451223905990600 | 3,0315692675220100 | - 11,3251346134671000 | - 11,2514745885549000 |
| | 5,4818304708581200 | 5,4695358689392200 | - 1,9480683964456800 | - 1,8422627626825900 |
| | 15,2367489963043000 | 15,2238132429609000 | 4,6353288322676200 | 4,5999164634735000 |
| | 5,4301662564523200 | 5,4590963220899700 | 5,8306118665398200 | 5,9361417293059600 |
| | | | 10,6536764951868000 | 10,6707588297402000 |
| Coord Z | 16,5331470717679000 | 16,6782626999877000 | 9,0578234635772800 | 8,8124944481960700 |
| | 5,5906307601642700 | 5,5290281911492700 | 6,5587807606267600 | 6,2139105665018600 |
| | 1,6667002075204600 | 1,6630354003308100 | 11,2924868670156000 | 11,0237290274208000 |
| | - 2,6384087953768300 | - 2,6634247186978300 | 1,3299499346579300 | 1,4637514565464200 |
| | - 14,5405434653343000 | - 14,4765769997510000 | - 5,4484900959666000 | - 5,4017280899853800 |
| | - 2,4181993628880900 | - 2,3017564992677600 | - 4,8394540950783300 | - 4,8450105371635400 |
| | 1,1202309922044700 | 1,1577301109336100 | - 8,8748349572990400 | - 8,8216701019475200 |
| | 5,3957952891942100 | 5,4155615003288400 | - 0,4548618477697420 | - 0,6568761396156520 |
| | | | 5,5965098248184900 | 5,6863000229695800 |

*Figura 71. Tabla con resultados de las rutas del escenario 1*

Esta tabla se encuentra dividida verticalmente por las rutas que se ejecutaron en el escenario. Cada ruta contiene una columna con los datos de la ruta planeada y una columna con los datos obtenidos de la ejecución de la ruta. Horizontalmente se encuentra dividida por longitud y latitud, los cuales representan las coordenadas planeadas y recorridas de cada punto de la ruta. Debajo de esto se encuentran las variables calculadas a partir de las coordenadas de cada punto. Estas representan las distancias en el eje X y Z con respecto al punto de despegue del dron. Con estas distancias fue posible calcular la diferencia entre puntos planeados y recorridos, y obtener el error de cada punto de la ruta. Finalmente, se calculó el promedio de los errores en ambos ejes de los puntos por cada ruta. El resultado de estos cálculos se puede apreciar en la siguiente tabla:

| | Ruta 96 | Ruta 103 |
|---|---|---|
| Error X | 0,290725129 | 0,027235532 |
| | 0,132664223 | 0,141195261 |
| | 0,013403684 | 0,141362424 |
| | 0,063477732 | 0,029191693 |
| | 0,013553123 | 0,073660025 |
| | 0,012294602 | 0,105805634 |
| | 0,012935753 | 0,035412369 |
| | 0,028930066 | 0,105529863 |
| | | 0,017082335 |
| Promedio | 0,070998039 | 0,075163904 |
| Error Z | 0,145115628 | 0,245329015 |
| | 0,061602569 | 0,344870194 |
| | 0,003664807 | 0,26875784 |
| | 0,025015923 | 0,133801522 |
| | 0,063966466 | 0,046762006 |
| | 0,116442864 | 0,005556442 |
| | 0,037499119 | 0,053164855 |
| | 0,019766211 | 0,202014292 |
| | | 0,089790198 |
| Promedio | 0,059134198 | 0,154449596 |

*Figura 72. Tabla con los errores de cada punto y el promedio de estos*

Para la ruta 96 (ambiente RV), en el eje X, el error más alto fue de 0,3 metros y el promedio de error fue de 0,1 metros. Para el eje Z, el error más alto fue de 0,2 metros y el promedio de error también fue de 0,1 metros. Para la ruta 103 (ambiente web), en el eje X, el error más alto fue de 0,2 metros y el promedio de error fue de 0,1 metros. Para el eje Z, el error más alto fue de 0,4 metros y el promedio de error fue de 0,2 metros. En las figuras 78 y 79, se pueden apreciar las gráficas del recorrido de cada ruta de este escenario. El punto (0,0) representa el punto de despegue del dron, la línea verde representa la ruta planeada y la línea naranja representa la ruta ejecutada por el dron.

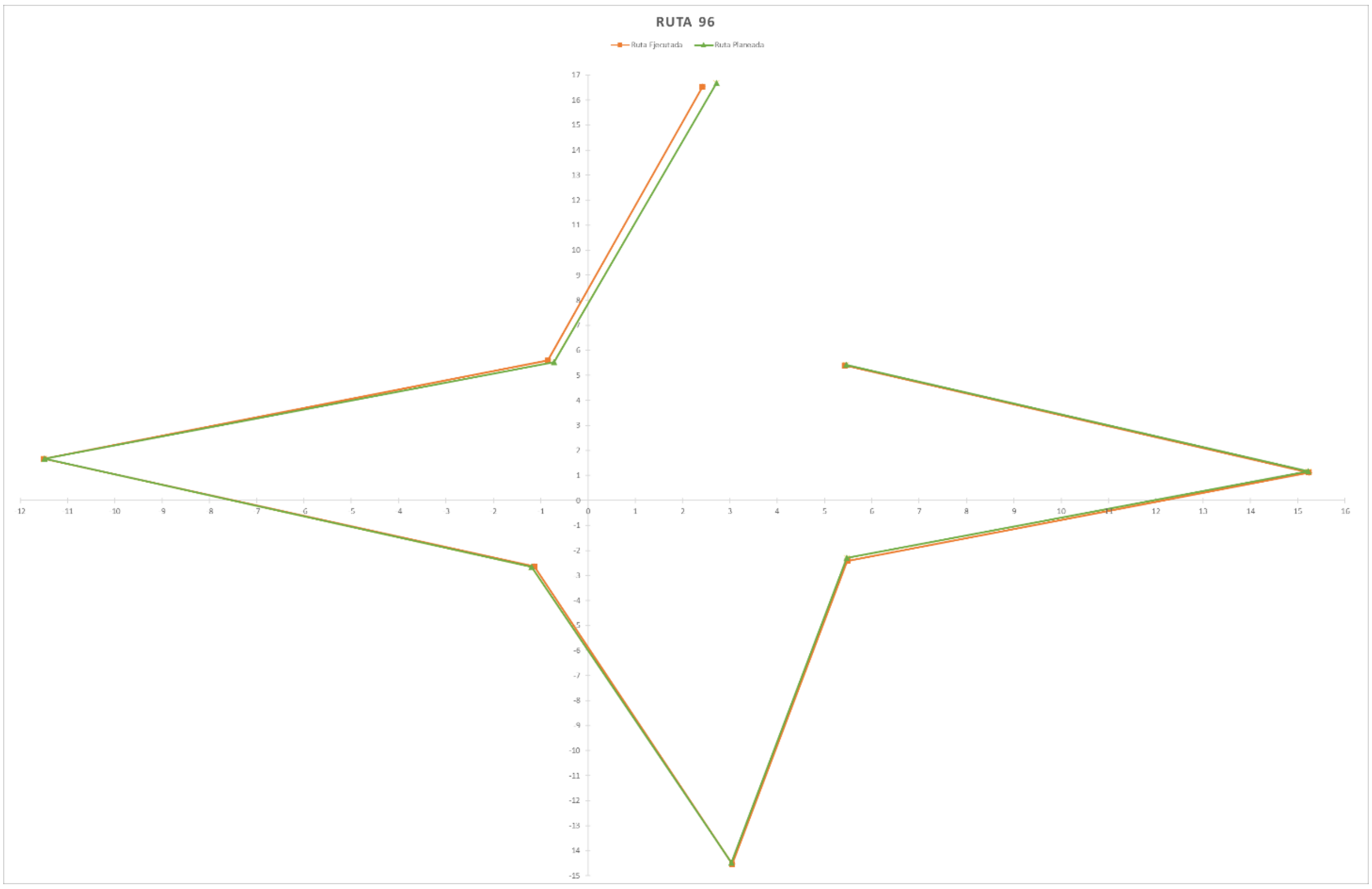

*Figura 73. Gráfica de recorrido y planeación de la ruta 96*

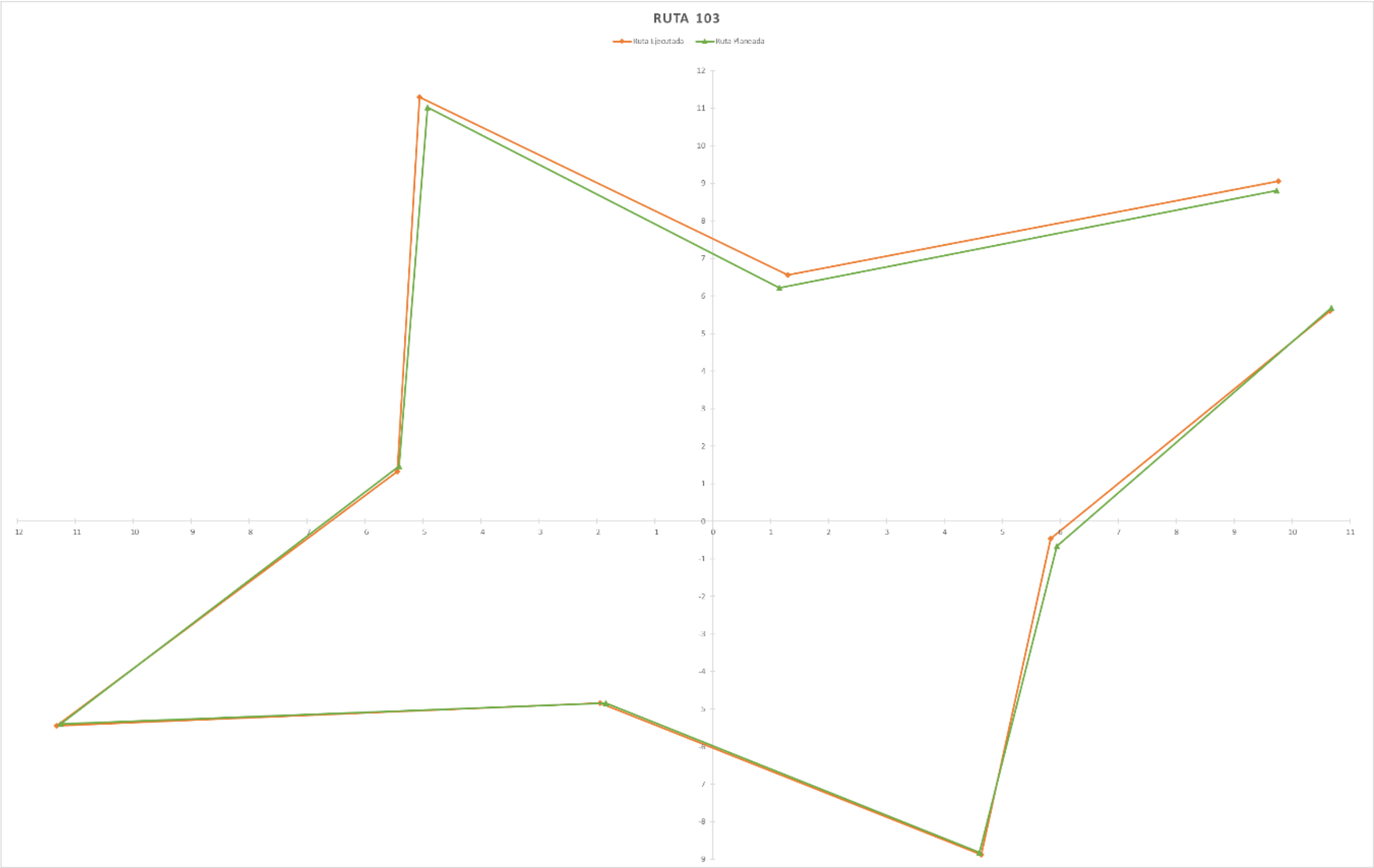

*Figura 74. Gráfica de recorrido y planeación de la ruta 103*

### 8.2.2 Escenario 2:

Así como el anterior escenario, los datos obtenidos y calculados para las rutas del escenario 2 se pueden encontrar en la siguiente tabla:

| | Ruta 3 | | Ruta 97 | |
|---|---|---|---|---|
| | Prueba | Original | Prueba | Original |
| Longitud | -74,0655593028419000 | -74,0655563975354000 | -74,0658266994795000 | -74,0658289531006000 |
| | -74,0656766731362000 | -74,0656773467640000 | -74,0658402154936000 | -74,0658404804359000 |
| | -74,0656049477381000 | -74,0656048735453000 | -74,0657236662742000 | -74,0657182451011000 |
| | -74,0657452019068000 | -74,0657460919554000 | -74,0657071880709000 | -74,0657071958649000 |
| | | | -74,0655389238865000 | -74,0655447877314000 |
| Latitud | 4,60115649740042000 | 4,60115086851309000 | 4,60123223280595000 | 4,60123740980174000 |
| | 4,60114841663791000 | 4,60114872020155000 | 4,60114148123823000 | 4,60114194874273000 |
| | 4,60121419585572000 | 4,60121419307994000 | 4,60112465500192000 | 4,60112977963185000 |
| | 4,60120692818426000 | 4,60120777889504000 | 4,60120856697050000 | 4,60120913953186000 |
| | | | 4,60113967542392000 | 4,60114026400182000 |
| Coord X | 7,9115218882956100 | 8,2236407489153800 | -2,8696817856375500 | -3,1119028160122600 |
| | -4,6976408622977900 | -4,7700091113088400 | -4,3223939798437500 | -4,3508701939673500 |
| | 3,0078626764826200 | 3,0158332542728900 | 8,2044109704468800 | 8,7870830670308400 |
| | -12,0597287465146000 | -12,1553472150335000 | 9,9755016705058300 | 9,9746639650309400 |
| | | | 28,0606731449415000 | 27,4304223240325000 |
| Coord Z | -3,7132680780109500 | -4,3398594900337900 | 1,3617204794980400 | 1,9380086359906400 |
| | -4,6127951239032200 | -4,5790032999261000 | -8,7404808316352200 | -8,6884395866020200 |
| | 2,7095566226819300 | 2,7092476309467900 | -10,6135286564058000 | -10,0430697155816000 |
| | 1,9005405142169000 | 1,9952391709281300 | -1,2726916846374200 | -1,2089558146351000 |
| | | | -8,9414988431671300 | -8,8759800583744200 |

*Figura 75. Tabla con resultados de las rutas del escenario 1*

De igual manera, esta tabla contiene información sobre cada ruta del escenario. Esta información representa la longitud y latitud de cada punto de la ruta planeada y ejecutada, así como las variables de distancia calculadas para el eje X y Z. Con estas distancias fue posible calcular la diferencia entre puntos planeados y recorridos, y obtener el error de cada punto de la ruta. Finalmente, se calculó el promedio de los errores en ambos ejes de los puntos por cada ruta. El resultado de estos cálculos se puede apreciar en la siguiente tabla:

| | Ruta 3 | Ruta 97 |
|---|---|---|
| Error X | 0,31211886061977300 | 0,24222103037470900 |
| | 0,07236824901104380 | 0,02847621412359570 |
| | 0,00797057779026478 | 0,58267209658396100 |
| | 0,09561846851884150 | 0,00083770547489515 |
| | | 0,63025082090898400 |
| Promedio | 0,12201903898498100 | 0,29689157349322900 |
| Error Z | 0,62659141202283700 | 0,57628815649260000 |
| | 0,03379182397712200 | 0,05204124503319820 |
| | 0,00030899173514332 | 0,57045894082420300 |
| | 0,09469865671123200 | 0,06373587000232290 |
| | | 0,06551878479271340 |
| Promedio | 0,18884772111158300 | 0,26560859942900800 |

*Figura 76. Tabla con los errores de cada punto y el promedio de estos*

Para la ruta 3 (ambiente RV), en el eje X, el error más alto fue de 0,4 metros y el promedio de error fue de 0,2 metros. Para el eje Z, el error más alto fue de 0,7 metros y el promedio de error también fue de 0,2 metros. Para la ruta 97 (ambiente web), en el eje X, el error más alto fue de 0,7 metros y el promedio de error fue de 0,3 metros. Para el eje Z, el error más alto fue de 0,6 metros y el promedio de error también fue de 0,3 metros. En las figuras 82 y 83, se pueden apreciar las gráficas del recorrido de cada ruta de este escenario. El punto (0,0) representa el punto de despegue del dron, la línea verde representa la ruta planeada y la línea naranja representa la ruta ejecutada por el dron.

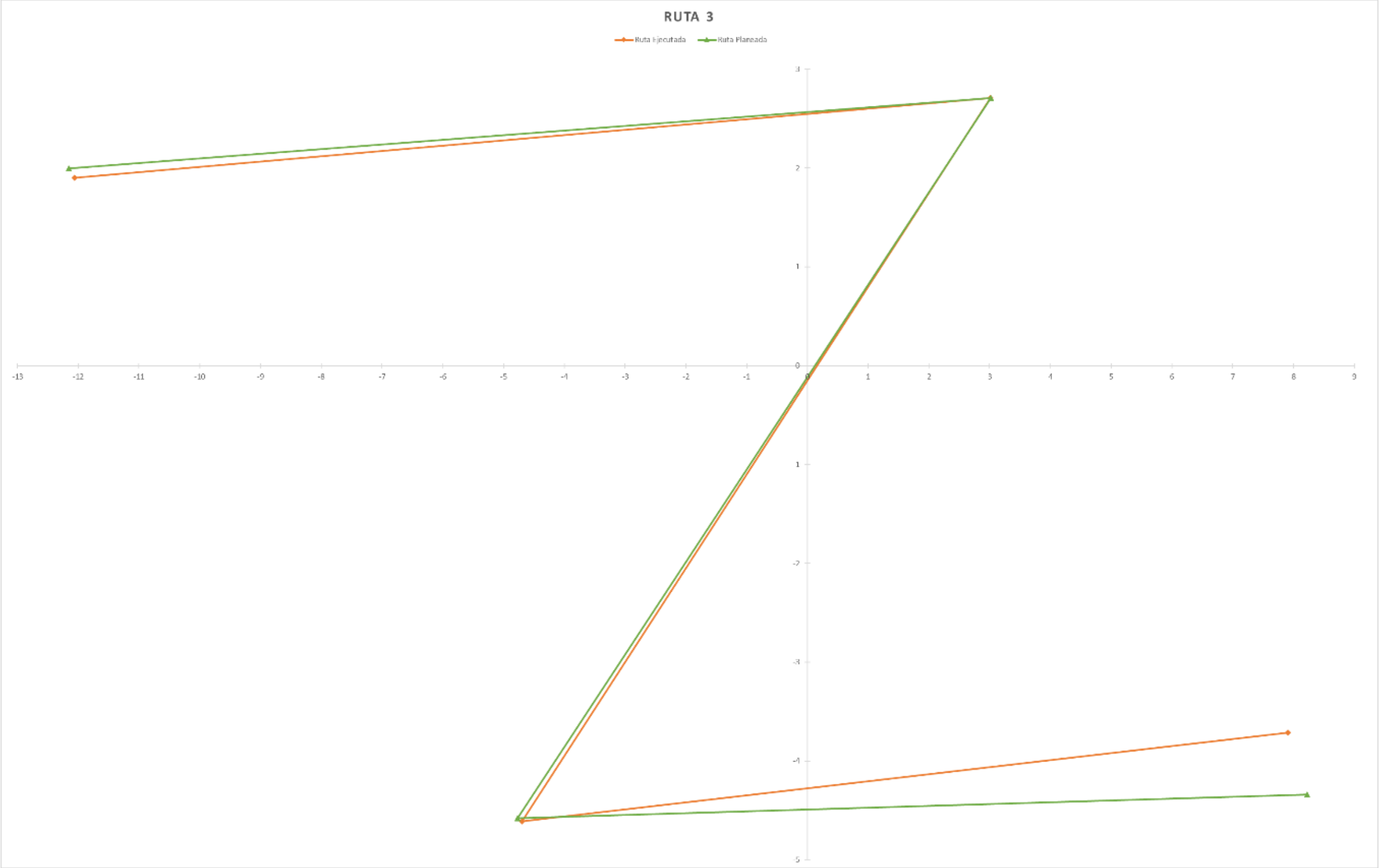

*Figura 77. Gráfica de recorrido y planeación de la ruta 3*

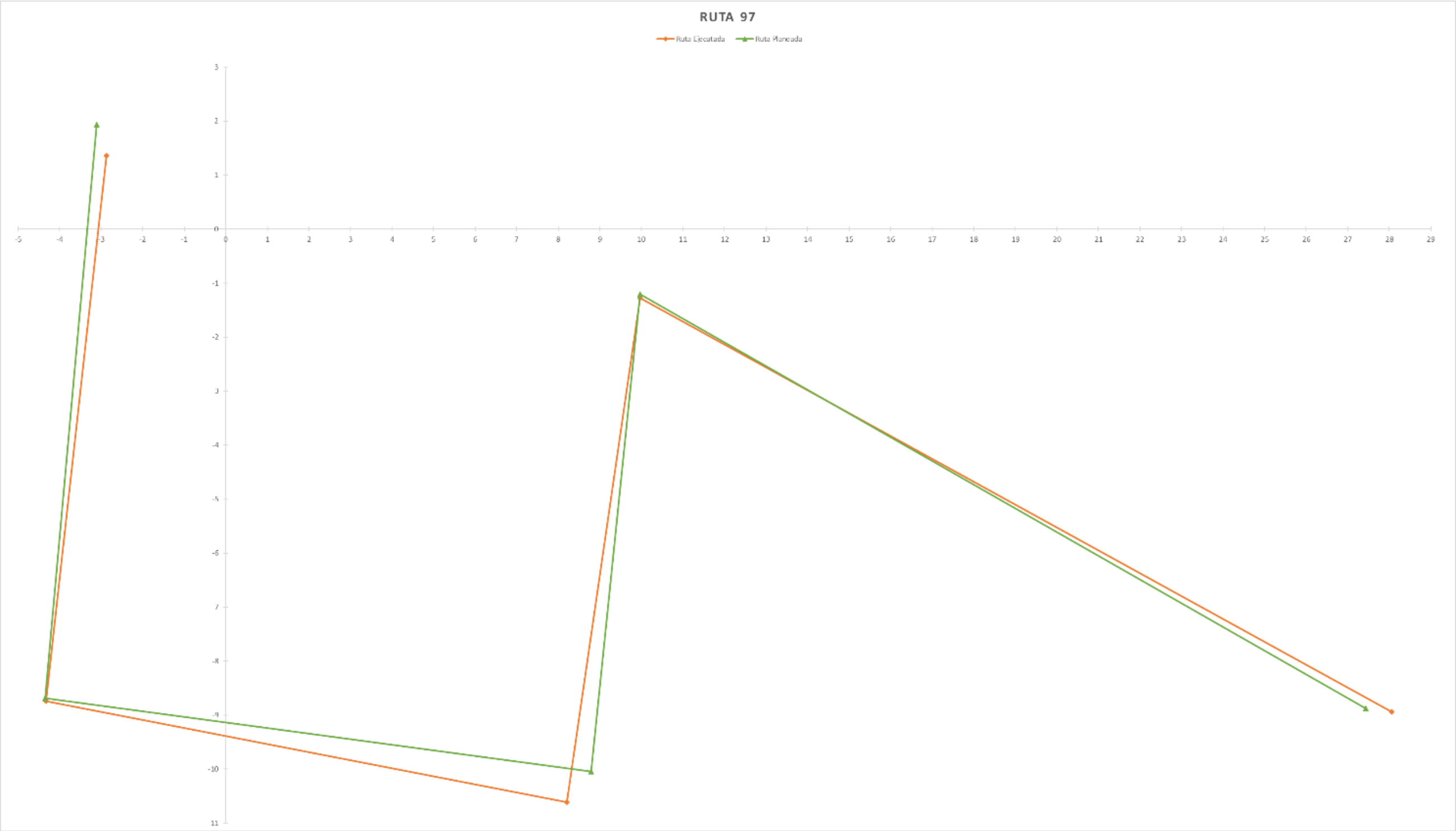

*Figura 78. Gráfica de recorrido y planeación de la ruta 97*

# 9. Conclusiones y Trabajos Futuros

## 9.1 Conclusiones

El proyecto fue desarrollado con el objetivo de extender la arquitectura desarrollada por Sánchez [1] y Múnera [2] basada en la planeación de rutas de dron en un ambiente de Realidad Virtual y su ejecución y monitoreo por medio de una aplicación web. A partir de esta arquitectura, se diseñaron componentes nuevos que añadieran funcionalidades y/o pudieran sustituir elementos de esta, para reducir la dependencia de hardware especializado.

Se desarrollo más a fondo la aplicación móvil con el objetivo de aprovechar al máximo las capacidades del dron, haciendo uso de las funcionalidades de cámara e integrando estas a la arquitectura de planeación de rutas. Para esto, se extendió el esquema de almacenamiento que se manejaba previamente y se añadió la posibilidad de persistir información sobre tareas de cámara.

Así mismo, se integró un nuevo componente de planeación en un ambiente web. Este permite planear y simular rutas 3D en un entorno que no depende de hardware especializado, como es la Realidad Virtual. De igual manera, en la aplicación móvil se implementó la funcionalidad de definir rutas. Sin embargo, en este ambiente esta definición pierde detalle pues es en 2D.

Con base en los resultados obtenidos, se pudo comparar el desempeño de los ambientes de definición de rutas. Si bien, en las pruebas realizadas el ambiente de Realidad Virtual tuvo errores más pequeños que los de web, la diferencia no es muy significativa en distancias largas por lo que ambas opciones son viables a la hora de planear rutas de vuelo.

Finalmente, después del diseño, desarrollo y prueba de los componentes integrados, se puede concluir que el trabajo de este proyecto contribuyó al enriquecimiento de la arquitectura de rutas de dron que el grupo IMAGINE de la Universidad de los Andes ha venido desarrollando en los últimos años. Los esfuerzos realizados en este proyecto facilitan el uso y estudio de la arquitectura al reducir la dependencia existente anteriormente, y abren la puerta a nuevos caminos de desarrollo e investigación al incluir las funcionalidades de cámara.

## 9.2 Trabajos Futuros

Como trabajo futuro se plantean varias propuestas:

Por un lado, al explorar el API de mapas de Google, se reconoce su gran potencial para mejorar la arquitectura existente aún más. Esta herramienta puede ser integrada al entorno de Realidad Virtual en Unity para facilitar la generación del entorno virtual de planeación.

Por otra parte, se reconoce el potencial de ROS como una herramienta para el control y operación de drones y otros sistemas robóticos que no cuenten con un SDK de fabricante que soporte las tareas que se quieran llevar a cabo. Así mismo, puede ser usado para llevar la operación de la arquitectura desarrollada a otros UAVs e integrarlos.

Por último, uno de los mayores impedimentos a la hora de extender la aplicación móvil fue la versión de Android sobre la cual se desarrolló. Se recomienda fuertemente actualizar el equipo móvil con el que se cuenta para poder aprovechar al máximo las funcionalidades que ofrece el SDK del fabricante (DJI). Así mismo, se recomienda actualizar el código fuente de la app al lenguaje Kotlin y hacer uso de Android Jetpack. Esto abre las puertas a desarrollos muy interesantes, como el análisis de *computer vision* que ofrecen librerías como OpenCV, las cuales requieren una capacidad de cómputo que el hardware actual no soporta.

# 10. Bibliografía